\documentclass[longauth]{aa}  
\usepackage{rotating} 
\usepackage{graphicx}
\usepackage{txfonts}
\usepackage{xcolor}
\usepackage{longtable}
\usepackage{subcaption}
\usepackage{caption}
\usepackage{ulem}
\usepackage{booktabs}
\usepackage{natbib}
\usepackage{comment}
\usepackage{lipsum, babel}
\usepackage{orcidlink}
\usepackage{hyperref}
\usepackage{silence}
\usepackage{array}
\usepackage{bm}
\usepackage{geometry}

\usepackage{svg}
\newcommand{\orcid}[1]{\href{https://orcid.org/#1}{\includesvg[width=10pt]{orcid_logo}}}

\bibpunct{(}{)}{;}{a}{}{,} 
\DeclareUnicodeCharacter{0303}{\~{}} 
\renewcommand{\appendix}{\par \setcounter{section}{0} \renewcommand{\thesection}{\Alph{section}}}

\begin{document} 

\authorrunning{C. E. Ferreira Lopes et al.}
\titlerunning{Stellar atmospheric parameters and chemical abundances from S-PLUS }
\title{Stellar atmospheric parameters and chemical abundances of $\sim5$ million stars from S-PLUS multi-band photometry}

\author{
  C. E. Ferreira Lopes\inst{1,2},
  L. A. Guti\'{e}rrez-Soto\inst{3},
  V. S. Ferreira Alberice\inst{4,5},
  N. Monsalves\inst{6},
  D. Hazarika\inst{1,2},
  M. Catelan\inst{7,8,2},
  V. M. Placco\inst{9},
  G. Limberg\inst{10},
  F. Almeida-Fernandes\inst{5,11},
  H. D. Perottoni\inst{5},
  A. V. Smith Castelli\inst{3,12},
  S. Akras\inst{13},
  J. Alonso-Garc\'{i}a\inst{14,2},
  V. Cordeiro\inst{15},
  M. Jaque Arancibia\inst{6,16},
  S. Daflon\inst{15},
  B. Dias\inst{17},
  D. R. Gon\c{c}alves\inst{11},
  E. Machado-Pereira\inst{9,15},
  A. R. Lopes\inst{3},
  C. R. Bom\inst{18},
  R. C. Thom de Souza\inst{19,20},  
  N. G. de Isídio\inst{21},
  A. Alvarez-Candal\inst{22,23},
   M. E. De Rossi\inst{24,25},
   C. J. Bonatto\inst{26},
   B. Cubillos Palma\inst{27},
   M. Borges Fernandes\inst{15},
    P. K. Humire\inst{5},
    G. B. Oliveira Schwarz\inst{4,5},
   W. Schoenell\inst{28},
   A. Kanaan\inst{29},
  C. Mendes de Oliveira\inst{5}
  \thanks{Thanks to the S-PLUS team for their efforts in data reduction, precise calibration, and photometry.}
}

\institute{
  Instituto de Astronom\'{i}a y Ciencias Planetarias, Universidad de Atacama, Copayapu 485, Copiap\'{o}, Chile\\
  \email{ferreiralopes1011@gmail.com}
  \and
  Millennium Institute of Astrophysics, Nuncio Monse\~{n}or Sotero Sanz 100, Of. 104, Providencia, Santiago, Chile
  \and
  Instituto de Astrof\'{i}sica de La Plata (CCT La Plata - CONICET - UNLP), B1900FWA, La Plata, Argentina
  \and
  Universidade Presbiteriana Mackenzie, Rua da Consolação, 930, Consolação, S\~{a}o Paulo, 01302-907, Brazil
  \and
  Universidade de S\~{a}o Paulo, Instituto de Astronomia, Geof\'{i}sica e Ciências Atmosf\'{e}ricas, Departamento de Astronomia, Rua do Mat\~{a}o, 1226, 05509-090, S\~{a}o Paulo, Brazil
  \and
  Departamento de Astronomía, Universidad de La Serena, Avenida Juan Cisternas 1200, La Serena, Chile
  \and
  Instituto de Astrof\'{i}sica, Pontificia Universidad Cat\'{o}lica de Chile, Av. Vicu\~{n}a Mackenna 4860, 7820436 Macul, Santiago, Chile
  \and
  Centro de Astro-Ingenier\'{i}a, Pontificia Universidad Cat\'{o}lica de Chile, Av. Vicu\~{n}a Mackenna 4860, 7820436 Macul, Santiago, Chile
  \and
  NSF NOIRLab, Tucson, AZ 85719, USA
  \and
  Kavli Institute for Cosmological Physics, University of Chicago, 5640 S Ellis Avenue, Chicago, IL 60637, USA
  \and
  Observat\'{o}rio do Valongo, Universidade Federal do Rio de Janeiro, Ladeira Pedro Antonio 43, 20080-090, Rio de Janeiro, Brazil
  \and
  Facultad de Cs. Astron\'{o}micas y Geof\'{i}sicas, Universidad Nacional de La Plata, Paseo del Bosque S/N, B1900FWA, La Plata, Argentina
  \and
  Institute for Astronomy, Astrophysics, Space Applications and Remote Sensing, National Observatory of Athens, GR 15236 Penteli, Greece
  \and
  Centro de Astronom\'{i}a (CITEVA), Universidad de Antofagasta, Av. Angamos 601, Antofagasta, Chile
  \and
  Observat\'{o}rio Nacional (ON), MCTI, Rua Gal. Jos\'{e} Cristino 77, Rio de Janeiro, 20921-400, RJ, Brazil
  \and
  Instituto de Investigaci\'{o}n Multidisciplinar en Ciencia y Tecnolog\'{i}a, Universidad de La Serena, Ra\'{u}l Bitr\'{a}n 1305, La Serena, Chile
  \and
  Instituto de Astrof\'{i}sica, Departamento de Ciencias F\'{i}sicas, Facultad de Ciencias Exactas, Universidad Andres Bello, Fernández Concha 700, Las Condes, Santiago, Chile
  \and
  Centro Brasileiro de Pesquisas F\'{i}sicas, Rua Dr. Xavier Sigaud 150, CEP 22290-180, Rio de Janeiro, RJ, Brazil
  \and
  Campus Avançado em Jandaia do Sul, Universidade Federal do Paran\'{a}, Jandaia do Sul, PR, 86900-000, Brazil
  \and
  Programa de P\'{o}s-graduaç\~{a}o em Ci\^{e}ncia da Computaç\~{a}o, Universidade Estadual de Maring\'{a}, Maring\'{a}, PR, 87020-900, Brazil
  \and
  European Southern Observatory, Karl Schwarzschild strasse 2, 85748, Garching bei München, Germany
  \and 
  Instituto de Astrof\'{i}sica de Andaluc\'ia, CSIC, Apt 3004, E18080 Granada, Spain
  \and 
  Instituto de F\'{i}sica Aplicada a las Ciencias y las Tecnolog\'ias, Universidad de Alicante, San Vicent del Raspeig, E03080, Alicante, Spain
  \and
  Universidad de Buenos Aires, Facultad de Ciencias Exactas y Naturales y Ciclo B\'asico Com\'un. Buenos Aires, Argentina
  \and
  CONICET-Universidad de Buenos Aires, Instituto de Astronom\'ia y F\'isica del Espacio (IAFE). Buenos Aires, Argentina
  \and
  Departamento de Astronomia, Instituto de Física, Universidade Federal do Rio Grande do Sul, Porto Alegre, RS, Brazil
  \and
  Departamento de Astronomía, Universidad de La Serena, Ra\'{u}l Bitr\'{a}n 1720256, La Serena, Coquimbo
  \and
  The Observatories of the Carnegie Institution for Science, 813 Santa Barbara St, Pasadena, CA 91101, USA
  \and
  Departamento de F\'{i}sica, Universidade Federal de Santa Catarina, Florian\'{o}polis, SC, 88040-900, Brazil
}


\date{Received xxx, 0000; accepted yyy, 0000}

  \abstract
{The APOGEE, GALAH, and LAMOST spectroscopic surveys have substantially contributed to our understanding of the Milky Way by providing a wide range of stellar parameters and chemical abundances. Complementing these efforts, photometric surveys that include narrow/medium-band filters, such as the Southern Photometric Local Universe Survey (S-PLUS), provide a unique opportunity to estimate atmospheric parameters and elemental abundances for a much larger number of sources compared to spectroscopic surveys.}
{Establish methodologies for extracting stellar atmospheric parameters and selected chemical abundances from S-PLUS photometric data, which cover approximately $3000$ square degrees, by applying seven narrowband and five broad-band filters.}
{We used all 66 S-PLUS colors to estimate parameters based on three different training samples from the LAMOST, APOGEE, and GALAH surveys, applying Cost-Sensitive Neural Network (NN) and Random Forest (RF) algorithms. We kept stellar abundances that lacked corresponding absorption features in the S-PLUS filters to test for spurious correlations in our method. Furthermore, we evaluated the effectiveness of the NN and RF algorithms by using estimated $T_\mathrm{eff}$ and \(\log g\) as input features to determine other stellar parameters and abundances. The NN approach consistently outperforms the RF technique on all parameters tested. Moreover, incorporating $T_\mathrm{eff}$ and \(\log g\) leads to an improvement in the estimation accuracy by approximately $3\%$. We kept only parameters with a goodness-of-fit higher than $50\%$.}
{Our methodology allowed reliable estimates for fundamental stellar parameters ($T_\mathrm{eff}$, \(\log g\), and [Fe/H]) and elemental abundance ratios such as [$\alpha$/Fe], [Al/Fe], [C/Fe], [Li/Fe], and [Mg/Fe] for approximately 5 million stars across the Milky Way, with goodness-of-fit above $60\%$. We also obtained additional abundance ratios, including [Cu/Fe], [O/Fe], and [Si/Fe]. However, these ratios should be used cautiously due to their low accuracy or lack of a clear relationship with the S-PLUS filters. Validation of our estimations and methods was performed using star clusters, TESS (Transiting Exoplanet Survey Satellite) data, and J-PLUS (Javalambre Photometric Local Universe Survey) photometry, further demonstrating the robustness and accuracy of our approach.}
{By leveraging S-PLUS photometric data and advanced machine-learning techniques, we have established a robust framework for extracting fundamental stellar parameters and chemical abundances from medium- and narrowband photometric observations. This approach offers a cost-effective alternative to high-resolution spectroscopy, and the estimated parameters hold significant potential for future studies, particularly in classifying objects within our Milky Way or gaining insights into its various stellar populations.}
\keywords{Stars: fundamental parameters -- Stars: abundances -- Techniques: photometric -- Surveys}

\maketitle
\section{Introduction}\label{sec:introduction}

Over the last century, a worldwide campaign involving ground-based observatories has allowed the accumulation of a vast and heterogeneous dataset comprising chemical abundance measurements for approximately two million stars. However, this dataset exhibits significant spatial heterogeneity due to fragmented sky coverage \citep[][]{Recio-Blanco-2023}. This scenario was revolutionized by the third {\it Gaia} data release \citep[{\it Gaia} DR3 -][]{GaiaCollaborationDR3-2021}, which contains the parametrization of Radial Velocity Spectrometer (RVS) data performed by the General Stellar Parametrizer-spectroscopy module, GSP-Spec. GSP-Spec estimates the chemo-physical parameters from combined RVS spectra of single stars, without additional inputs from astrometric, photometric, or spectro-photometric BP/RP data \citep[e.g.,][]{GaiaCollaboration2018,GaiaCollaborationDR3-2021}. Individual chemical abundances of N, Mg, Si, S, Ca, Ti, Cr, Fe I, Fe II, Ni, Zr, Ce, and Nd have recently been derived from Gaia data \citep[][]{Recio-Blanco-2023}. With about $5.6$ million stars, the Gaia DR3 GSP-Spec all-sky catalog is the most extensive compilation of stellar chemo-physical parameters ever published and the first from space-based data. The inclusion of Gaia’s extensive spectral capabilities highlights the importance of these methods as a complementary approach, improving the accuracy and broadening the scope of stellar chemical abundance studies. The homogeneity and quality of the estimated parameters allow chemo-dynamical studies of Galactic stellar populations, interstellar extinction studies from individual spectra, and explicit constraints on stellar evolution models.\\

Estimating stellar parameters and abundances using photometry represents a fundamental frontier in contemporary astrophysics. The technological advances, unprecedented data availability, and the growing need for comprehensive knowledge of stellar populations drive this scenario. While spectroscopic surveys, exemplified by instruments such as the Large Sky Area Multi-Object Fiber Spectroscopy Telescope (LAMOST; \citealp[][]{Cui-2012}), can gather around 4000 spectra in a single exposure, photometric surveys provide a comprehensive view of the universe, capturing the luminous signatures of millions of astrophysical objects in the sky. Photometric surveys maintain a significantly higher observational efficiency, yielding more data in less time. This abundance of data offers the potential to estimate stellar parameters for a vast number of stars, as highlighted by recent surveys of critical astronomical phenomena. For instance, \citet{Gore:2024} emphasized the importance of targeting M and K stars known to host exoplanets. Other examples of investigations highlighting the synergy between spectroscopic observations and photometric analyses include \citet{Borucki-2010}, who precisely calculated stellar metallicities and parameters, including systems with planets identified by missions such as Kepler or COnvection, internal ROtation and transiting planets \citep[CoRoT][]{Baglin-2007,Auvergne-2009}; and Transiting Exoplanet Survey Satellite (TESS) \citep[e.g.,][]{Fischer-2005,Huber2013,Buchhave-2014,Buchhave2015}. These efforts to combine spectroscopic and photometric observations allow us to confirm the presence of exoplanets and accurately determine stellar and exoplanet parameters.\\

\citet[][]{Schonhut:2024} highlighted the importance of estimating stellar parameters in large samples. By combining spectroscopic observations from the Apache Point Observatory Galactic Evolution Experiment (APOGEE; \citealp[][]{Wilson-2019}) with asteroseismology from {\it Kepler} 2 \citep[K2 -][]{Howell-2014}, fundamental stellar properties for approximately 7500 evolved stars were obtained. This comprehensive catalog furnishes temperatures, metallicities, and global asteroseismic parameters, shedding light on stellar evolution and Galactic dynamics. It includes the kinematic properties of low-metallicity stars, highlighting their potential role in Galactic mergers, and playing a critical role in the investigation of past Galactic mergers that gave rise to the present-day Galaxy \citep[][]{Yates-2012}. Characterizing the selection function for the APO-K2 sample allows robust statistical inferences, highlighting the significance of large-scale surveys in advancing our understanding of the universe \citep[][]{Fallows-2022}.\\

On the other hand, large-scale ground-based photometric surveys, including the Sloan Digital Sky Survey \citep[SDSS;][]{York-2000}, the SkyMapper Southern Survey \citep[SMSS - ][]{Keller-2007}, the UKIRT Infrared Deep Sky Survey \citep[UKIDSS-][]{Lawrence:2007}, the All-Sky Automated Survey for Supernovae \citep[ASAS-SN -][]{Kochanek:2017}, Stellar Abundance and Galactic Evolution \citep[SAGE - ][]{Gordon-2011}, the Javalambre-Physics of the Accelerated Universe Astrophysical Survey \citep[J-PAS - ][]{Benitez-2014, Bonoli-2021}, Panoramic Survey Telescope and Rapid Response System 1 survey \citep[Pan-STARRS1 - ][]{Chambers-2016}, and the Dark Energy Survey \citep[DES - ][]{DESI-2016}. As the first all-sky survey, the SDSS has set a benchmark in astrostatistics and the astro-photosurvey era, offering a comprehensive database that has been crucial for numerous studies. Its contributions range from mapping the large-scale structure of the universe to identifying and classifying a myriad of celestial objects \citep[e.g.,][]{Eisenstein-2011}. The Javalambre Photometric Local Universe Survey \citep[J-PLUS - ][]{Cenarro-2019} and the Southern Photometric Local Universe Survey \citep[S-PLUS - ][]{MendesOliveira-2019} now provide vast amounts of valuable photometric data for millions of astronomical objects. Characterized by the use of medium- and narrowband filters optimized for precise measurements of key stellar properties, these surveys herald a new era in the determination of stellar types, surface gravity (\(\log g\)), effective temperatures ($T_\mathrm{eff}$), and metallicity indicators such as [M/H] and [Fe/H] \citep[e.g.,][]{Arnadottir-2010, BailerJones-2013, Placco-2022, Bellazzini-2023}. \\

Several empirical and theoretical approaches have emerged to extract stellar labels from photometric data, fueled by the wealth of high-quality spectroscopic datasets from surveys like the Medium-Resolution LAMOST, APOGEE, and the Galactic Archaeology with HERMES \citep[GALAH -][]{Sheinis-2015}. These methods, ranging from empirical correlations of metallicity-dependent stellar parameter to machine learning algorithms, now provide stellar labels that can be accurately determined and combined with the precision of spectroscopy for high-quality photometry. Examples include using stellar loci in color-magnitude diagrams and applying machine learning to large photometric datasets \citep[e.g.,][]{Fallows-2022,BailerJones-2019}. \\

\begin{figure*}
   \centering
   \includegraphics[width=0.99\textwidth]{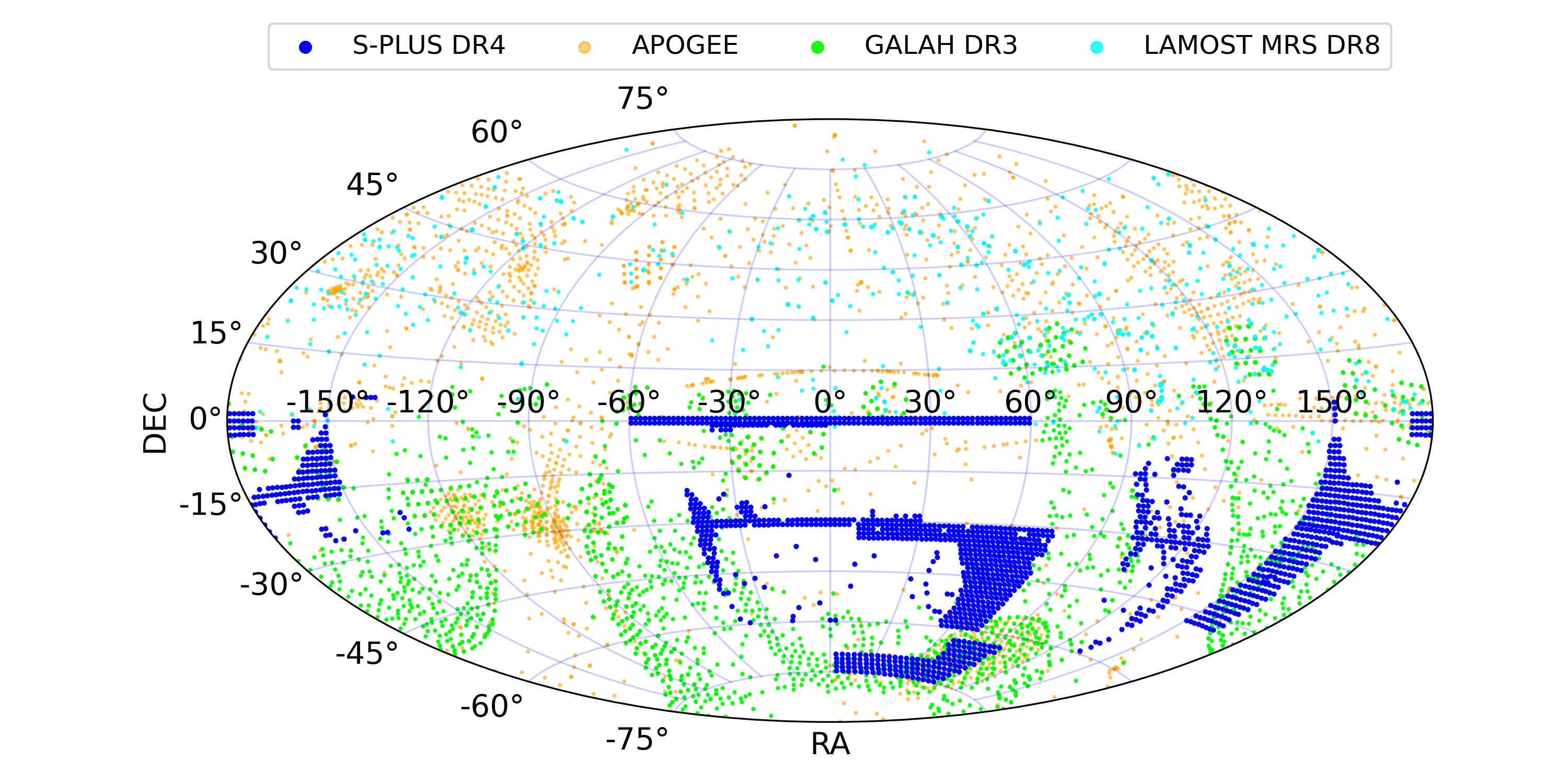}

   \caption{Sky footprints of S-PLUS DR4, APOGEE, GALAH DR3, and LAMOST MRS DR8 surveys across the celestial sphere. Each survey's distinctive coverage and observed regions are depicted, along with their overlapping regions.}
    \label{Fig:Aitoff-plot}%
\end{figure*}

For instance, the estimated elemental abundances in the current work can be utilized to undertake a more nuanced exploration of the impact of narrowband filters in identifying metal-poor stars \citep[e.g.,][]{Almeida-Fernandes-2023,Placco-2021,Placco-2022,Placco:2023}. The utilization of these filters has significantly enhanced the precision in locating these rare stellar specimens, representing a notable advancement in astronomical research. For instance, the groundbreaking discovery reported by \citet{Placco-2021} exemplifies the efficacy of narrowband photometry in detecting ultra-metal-poor star candidates. Their findings, initially made possible through the analysis of data from the S-PLUS Data Release 1, underscore the potential of such surveys in identifying targets with low metallicity. This achievement not only validates the effectiveness of narrowband photometric techniques but also charts new pathways for exploration in astrophysics. Building on this, we now delve into specific catalogs and studies that illustrate the impact of narrowband filters in identifying metal-poor stars. For example, the research conducted by \citet{Galarza:2022}, \citet{Placco-2022}, and \citet{Placco:2023} offers valuable insights into how the application of narrowband filters has contributed to the identification and characterization of these stellar populations. In a parallel study, \citet{Whitten-2021} provided photometric estimates of key stellar parameters for more than 700\,000 stars within S-PLUS Data Release 2, identifying 364 candidate carbon-enhanced and metal-poor stars. This further emphasizes the significance of narrowband photometry in delineating specific stellar populations within large datasets.\\

The S-PLUS narrowband filters are strategically centered around important spectral features: [O II] for J0378; Ca H + K for $J$0395; H$\delta$ for $J$0410; CH G-band for $J$0430; Mgb triplet for $J$0515; H$\alpha$ for $J$0660; and Ca triplet for $J$0861. For instance, the Ca line, measured by the J0395 filter, is particularly sensitive to metallicity. In this wavelength, narrowband photometry around this feature is typically employed by surveys searching for metal-poor stars \citep[][]{Keller-2007,Starkenburg-2017}. Similarly, the Mgb triplet, measured by the J0515 filter, is well known for its surface gravity sensitivity \citet{Geisler-1984,Majewski-2000}, and narrowband photometry has been used by surveys such as APOGEE to search for giant stars \citet{Zasowski-2013, Majewski-2016}. Through a combination of broad- and narrowband filters that serve to identify the main stellar spectral features (absorption lines and continuum), this photometric system was designed for the optimal classification of stars \citep{Gruel-2012,MarinFranch:2012}. \\

Recently, \citet{Almeida-Fernandes-2023} explored the chemodynamical properties and ages of 522 metal-poor candidates, initially selected by \citet{Placco-2022} from the S-PLUS DR3. As noted in \citet{Placco-2022}, approximately $92\%$ of these stars were confirmed to be metal-poor based on medium-resolution spectroscopic analysis. Together, these studies highlight the crucial role of spectroscopic estimates in identifying precise targets for more detailed investigations.\\

\citet{Gutierrez:2020} leveraged the strong H$\alpha$ emission lines characteristic of Planetary Nebulae (PNe) to explore the S-PLUS survey within the context of PNe. Similarly, \citet{FabianodeSouza-2024} examined age and metallicity determinations of the Small Magellanic Cloud, demonstrating a preference for younger clusters in the center and older ones predominantly located in outer regions. Additionally, \citet{Quispe-Huaynasi-2024} achieved highly accurate estimates of magnesium abundances for approximately 64 high-velocity stars, exemplifying the vast potential of spectral parameter determination from multi-wavelength data. \citet[][]{Yang-2022} used J-PLUS data to determine metallicity ([Fe/H]), [$\alpha$/Fe], and the abundances of four elements ([C/Fe], [N/Fe], [Mg/Fe], and [Ca/Fe]). Similar investigations using S-PLUS data are currently absent. \\

In this research, we propose a methodology to extract atmospheric parameters directly from S-PLUS photometric data. By utilizing all possible S-PLUS color combinations (66), we conducted estimations based on three distinct training samples obtained through the LAMOST, APOGEE, and GALAH surveys. The analysis employed the Cost-Sensitive Neural Network \citep[NN - ][]{Whitten-2019,Ksoll-2020,Whitten-2021,Yang-2022} and Random Forest \citep[e.g.,][]{Breiman-2001} methods. Figure~\ref{Fig:Aitoff-plot} exhibits the regions in the sky where these surveys and their respective data releases overlap. We use this overlapping data to build predictive models using deep/machine learning algorithms.\\

The following sections of this paper will delve into database construction (see Sect. \ref{sec:splus}), detailing the acquisition of training data. Sect. \ref{Sec:Methodology} will outline the methodology and constraints employed to enhance outputs. Next, Sect. \ref{Sec:Results} will present the primary findings and provide a catalog description of the sources studied in this paper. In Sect. \ref{Sec:conclusion}, our final remarks will be presented.

\section{Data construction}
\label{sec:splus}
\begin{figure*}
    \centering
    \includegraphics[width=0.8\linewidth]{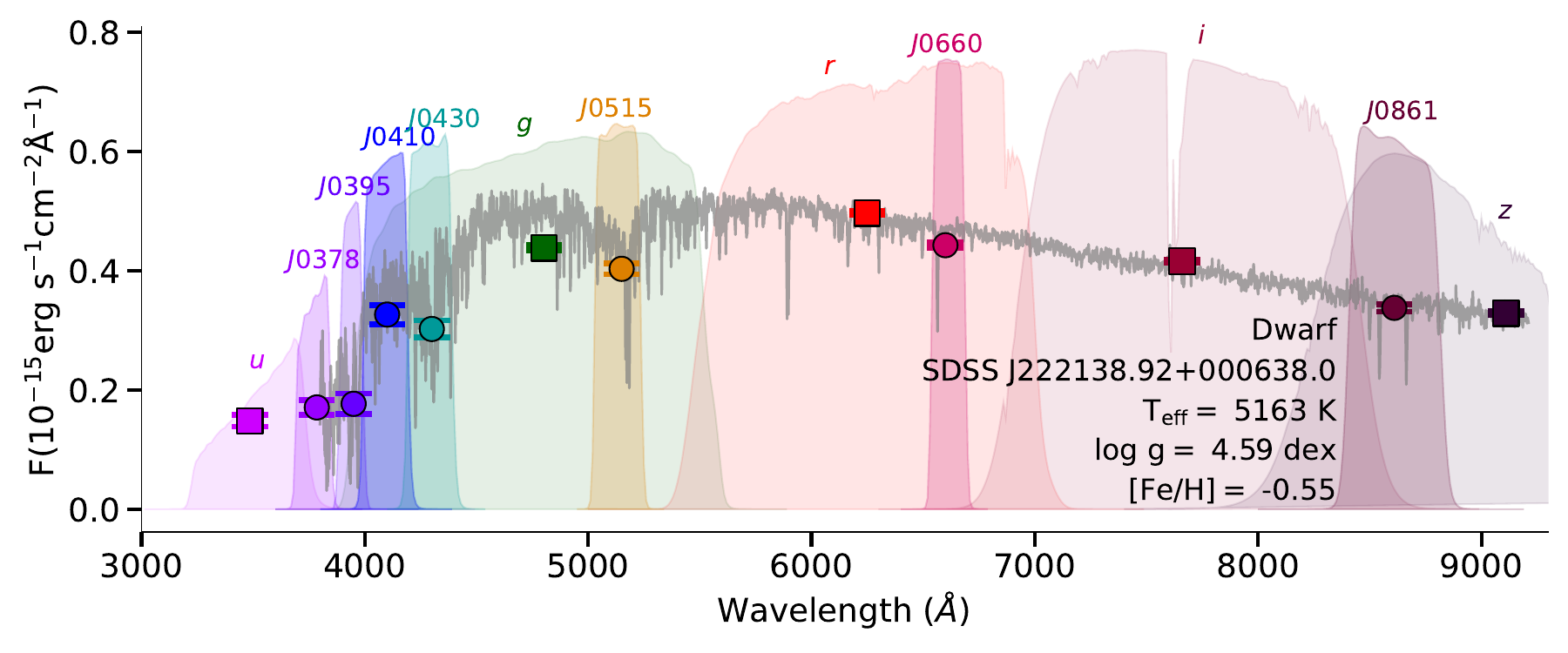} 
    \includegraphics[width=0.8\linewidth]{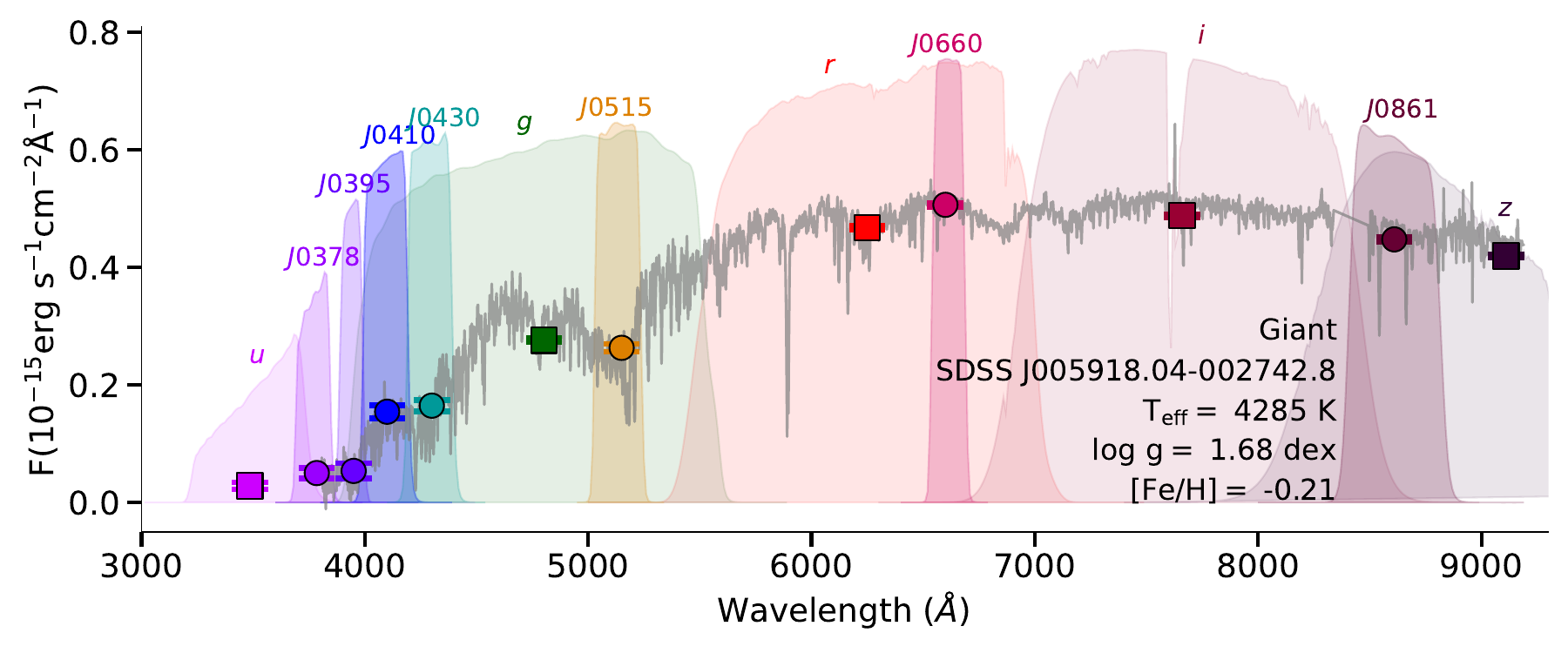}
    \caption{S-PLUS photometry and SDSS spectra of a dwarf star (top panel) and a giant star (bottom panel). The colored symbols represent S-PLUS photometry in flux units for the following filters: $u$, $J0378$, $J0395$, $J0410$, $J0430$, $g$, $J0515$, $r$, $J0660$, $i$, $J0861$, and $z$ (from left to right). Square symbols denote broad-band filters, while circular symbols indicate narrowband filters. The filter responses are shown accordingly.}
\label{fig:splus-photometry}
\end{figure*}

S-PLUS observes the sky in 12 filters covering the visible spectrum from 3000\text{\AA} to 10\,000\text{\AA}, comprising five wide bands ($ugriz$), akin to those used in the Sloan Digital Sky Survey \citep[SDSS - ][]{York-2000}, and seven narrowbands ($J$0378, $J$0395, $J$0410, $J$0430, $J$0515, $J$0660, $J$0861) unique to the Javalambre system \citep{MarinFranch:2012}. Figure \ref{fig:splus-photometry} illustrates spectral energy distributions (SEDs) constructed from S-PLUS photometry, alongside SDSS spectra, for both a dwarf star and a giant star. The S-PLUS survey, conducted with the T80-South (T80S) telescope, a Ritchey-Chrétien Cassegrain telescope with an 86 cm primary mirror, located at the Inter-American Observatory of Cerro Tololo, Chile. It uses the T80SCam camera, which features a detector with dimensions of $9200\times9200$ pixels, a plate scale of $0.55$ arcsec/pixel, and covering a field of view of $1.4\times1.4$ degrees \citep[][]{MendesOliveira-2019}. \\

\begin{table*}
\centering
\caption{Summary of training datasets used for Neural Network (NN) and Random Forest (RF) models.}
\resizebox{\textwidth}{!}{%
\begin{tabular}{l | ccc | ccc | ccc | ccc | ccc | ccc}
\toprule \midrule
& \multicolumn{9}{c|}{Dwarfs} & \multicolumn{9}{c}{Giants} \\
\midrule 
& \multicolumn{3}{c|}{APOGEE} & \multicolumn{3}{c|}{GALAH} & \multicolumn{3}{c|}{LAMOST} & \multicolumn{3}{c|}{APOGEE} & \multicolumn{3}{c|}{GALAH} & \multicolumn{3}{c}{LAMOST} \\
\midrule
Parameter & Min & Max & N & Min & Max & N & Min & Max & N & Min & Max & N & Min & Max & N & Min & Max & N \\
\midrule
$T_{\rm eff}$ & $3296$ & $8186$ & $8885$ & $3156$ & $7965$ & $24740$ & $4025$ & $6830$ & $2863$ & $3651$ & $5540$ & $5500$ & $3813$ & $5721$ & $4840$ & $4158$ & $5849$ & $190$ \\
$\log g$ & $2.40$ & $5.18$ & $8805$ & $2.91$ & $4.99$ & $24700$ & $1.96$ & $4.90$ & $2852$ & $-0.18$ & $3.66$ & $5500$ & $0.93$ & $3.68$ & $4842$ & $0.31$ & $4.53$ & $190$ \\
$\mathrm{[Fe/H]}$ & $-1.42$ & $0.46$ & $8673$ & $-1.76$ & $1.00$ & $24479$ & $-1.61$ & $0.42$ & $2856$ & $-2.44$ & $0.47$ & $5484$ & $-2.33$ & $1.00$ & $4811$ & $-1.92$ & $0.33$ & $190$ \\
$[\alpha/\mathrm{M}]$ & $-0.30$ & $0.41$ & $8692$ & $-0.41$ & $0.55$ & $23819$ & $-0.14$ & $0.30$ & $2146$ & $-0.31$ & $0.71$ & $5479$ & $-0.40$ & $0.66$ & $4724$ & $-0.02$ & $0.33$ & $107$  \\
$\mathrm{[C/Fe]}$ & $-0.49$ & $0.47$ & $8667$ & $-1.14$ & $1.55$ & $18116$ & $-0.31$ & $0.31$ & $2145$ & $-1.18$ & $1.03$ & $5461$ & $-0.42$ & $1.43$ & $332$ & $-0.18$ & $0.38$ & $106$  \\
$\mathrm{[Ca/Fe]}$ & $-0.43$ & $0.48$ & $8403$ & $-0.63$ & $0.77$ & $23328$ & $-0.31$ & $0.41$ & $2146$ & $-0.45$ & $0.68$ & $5425$ & $-0.53$ & $0.75$ & $4624$ & $-0.09$ & $0.29$ & $107$  \\
$\mathrm{[N/Fe]}$ & $-1.00$ & $1.24$ & $6532$ & $...$ & $...$ & $...$ & $-0.26$ & $0.50$ & $2145$ & $-0.66$ & $1.26$ & $5460$ & $...$ & $...$ & $...$ & $-0.15$ & $0.47$ & $107$  \\
$\mathrm{[Ni/Fe]}$ & $-0.35$ & $0.35$ & $8649$ & $-0.58$ & $0.62$ & $19570$ & $-0.13$ & $0.14$ & $2143$ & $-0.45$ & $0.49$ & $5446$ & $-0.62$ & $0.72$ & $4623$ & $-0.06$ & $0.11$ & $106$  \\
$\mathrm{[Mg/Fe]}$ & $-0.55$ & $0.51$ & $8698$ & $-0.61$ & $0.74$ & $23385$ & $-0.27$ & $0.39$ & $2146$ & $-0.47$ & $0.61$ & $5482$ & $-0.58$ & $0.93$ & $4677$ & $-0.02$ & $0.38$ & $107$  \\
$\mathrm{[Si/Fe]}$ & $-0.37$ & $0.49$ & $8692$ & $-0.54$ & $0.65$ & $23695$ & $-0.15$ & $0.29$ & $2146$ & $-0.39$ & $0.66$ & $5480$ & $-0.51$ & $0.86$ & $4706$ & $-0.04$ & $0.35$ & $107$  \\
$\mathrm{[C I/Fe]}$ & $-0.61$ & $0.57$ & $8609$ & $...$ & $...$ & $...$ & $...$ & $...$ & $...$ & $-1.24$ & $0.96$ & $5425$ & $...$ & $...$ & $...$ & $...$ & $...$ & $...$  \\
$\mathrm{[O/Fe]}$ & $-0.65$ & $0.77$ & $8645$ & $-1.19$ & $1.46$ & $23745$ & $...$ & $...$ & $...$ & $-0.61$ & $0.99$ & $5468$ & $-1.04$ & $1.78$ & $4678$ & $...$ & $...$ & $...$  \\
$\mathrm{[Na/Fe]}$ & $-2.37$ & $1.68$ & $6678$ & $-0.49$ & $0.65$ & $23943$ & $...$ & $...$ & $...$ & $-1.75$ & $1.66$ & $5254$ & $-0.66$ & $0.78$ & $4713$ & $...$ & $...$ & $...$  \\
$\mathrm{[Al/Fe]}$ & $-0.51$ & $0.69$ & $6910$ & $-0.80$ & $1.07$ & $23683$ & $...$ & $...$ & $...$ & $-1.11$ & $0.90$ & $5409$ & $-0.59$ & $1.09$ & $4644$ & $...$ & $...$ & $...$  \\
$\mathrm{[K/Fe]}$ & $-0.69$ & $0.88$ & $8629$ & $-0.97$ & $1.11$ & $23362$ & $...$ & $...$ & $...$ & $-0.85$ & $1.12$ & $5384$ & $-0.90$ & $1.08$ & $4520$ & $...$ & $...$ & $...$ \\
$\mathrm{[Ti/Fe]}$ & $-0.88$ & $0.91$ & $8182$ & $-0.86$ & $0.98$ & $23344$ & $...$ & $...$ & $...$ & $-0.94$ & $0.86$ & $5332$ & $-0.80$ & $0.97$ & $4730$ & $...$ & $...$ & $...$ \\
$\mathrm{[Ti II/Fe]}$ & $-1.44$ & $0.94$ & $669$ & $...$ & $...$ & $...$ & $...$ & $...$ & $...$ & $-0.82$ & $0.95$ & $5243$ & $...$ & $...$ & $...$ & $...$ & $...$ & $...$  \\
$\mathrm{[V/Fe]}$ & $-1.74$ & $1.56$ & $5405$ & $...$ & $...$ & $...$ & $...$ & $...$ & $...$ & $-1.70$ & $1.66$ & $5325$ & $...$ & $...$ & $...$ & $...$ & $...$ & $...$  \\
$\mathrm{[Mn/Fe]}$ & $-1.55$ & $1.32$ & $6853$ & $-0.66$ & $0.59$ & $23769$ & $...$ & $...$ & $...$ & $-1.27$ & $1.18$ & $5348$ & $-0.98$ & $0.76$ & $4604$ & $...$ & $...$ & $...$  \\
$\mathrm{[Co/Fe]}$ & $-1.88$ & $2.21$ & $685$ & $-0.46$ & $2.70$ & $17923$ & $...$ & $...$ & $...$ & $-1.60$ & $1.58$ & $5360$ & $-0.90$ & $1.42$ & $4406$ & $...$ & $...$ & $...$  \\
$\mathrm{[Ce/Fe]}$ & $-1.60$ & $1.98$ & $709$ & $...$ & $...$ & $...$ & $...$ & $...$ & $...$ & $-1.76$ & $1.36$ & $5152$ & $...$ & $...$ & $...$ & $...$ & $...$ & $...$ \\
$\mathrm{[Li/Fe]}$ & $...$ & $...$ & $...$ & $-1.65$ & $4.09$ & $20626$ & $...$ & $...$ & $...$ & $...$ & $...$ & $...$ & $-1.06$ & $2.65$ & $3092$ & $...$ & $...$ & $...$ \\
$\mathrm{[Sc/Fe]}$ & $...$ & $...$ & $...$ & $-0.54$ & $0.70$ & $23896$ & $...$ & $...$ & $...$ & $...$ & $...$ & $...$ & $-0.44$ & $0.59$ & $4732$ & $...$ & $...$ & $...$ \\
$\mathrm{[Cr/Fe]}$ & $...$ & $...$ & $...$ & $-0.66$ & $0.59$ & $23640$ & $...$ & $...$ & $...$ & $...$ & $...$ & $...$ & $-0.64$ & $0.62$ & $4720$ & $...$ & $...$ & $...$ \\
$\mathrm{[Cu/Fe]}$ & $...$ & $...$ & $...$ & $-0.68$ & $0.76$ & $21796$ & $...$ & $...$ & $...$ & $...$ & $...$ & $...$ & $-1.03$ & $1.07$ & $4709$ & $...$ & $...$ & $...$ \\
$\mathrm{[Zn/Fe]}$ & $...$ & $...$ & $...$ & $-0.79$ & $0.86$ & $23005$ & $...$ & $...$ & $...$ & $...$ & $...$ & $...$ & $-1.13$ & $1.29$ & $4042$ & $...$ & $...$ & $...$ \\
$\mathrm{[Y/Fe]}$ & $...$ & $...$ & $...$ & $-1.21$ & $1.21$ & $23591$ & $...$ & $...$ & $...$ & $...$ & $...$ & $...$ & $-1.12$ & $1.39$ & $4504$ & $...$ & $...$ & $...$ \\
\midrule \bottomrule
\end{tabular}}
\tablefoot{
The table includes data from APOGEE, LAMOST, and GALAH surveys, organized into two categories: Dwarfs and Giants. Each category is divided by dataset sources, detailing the parameters used, source counts (N), and value ranges (Max, Min).
}
\label{Table:SpectralLimits}
\end{table*}

\begin{figure*}
   \centering
   \includegraphics[width=0.95\textwidth]{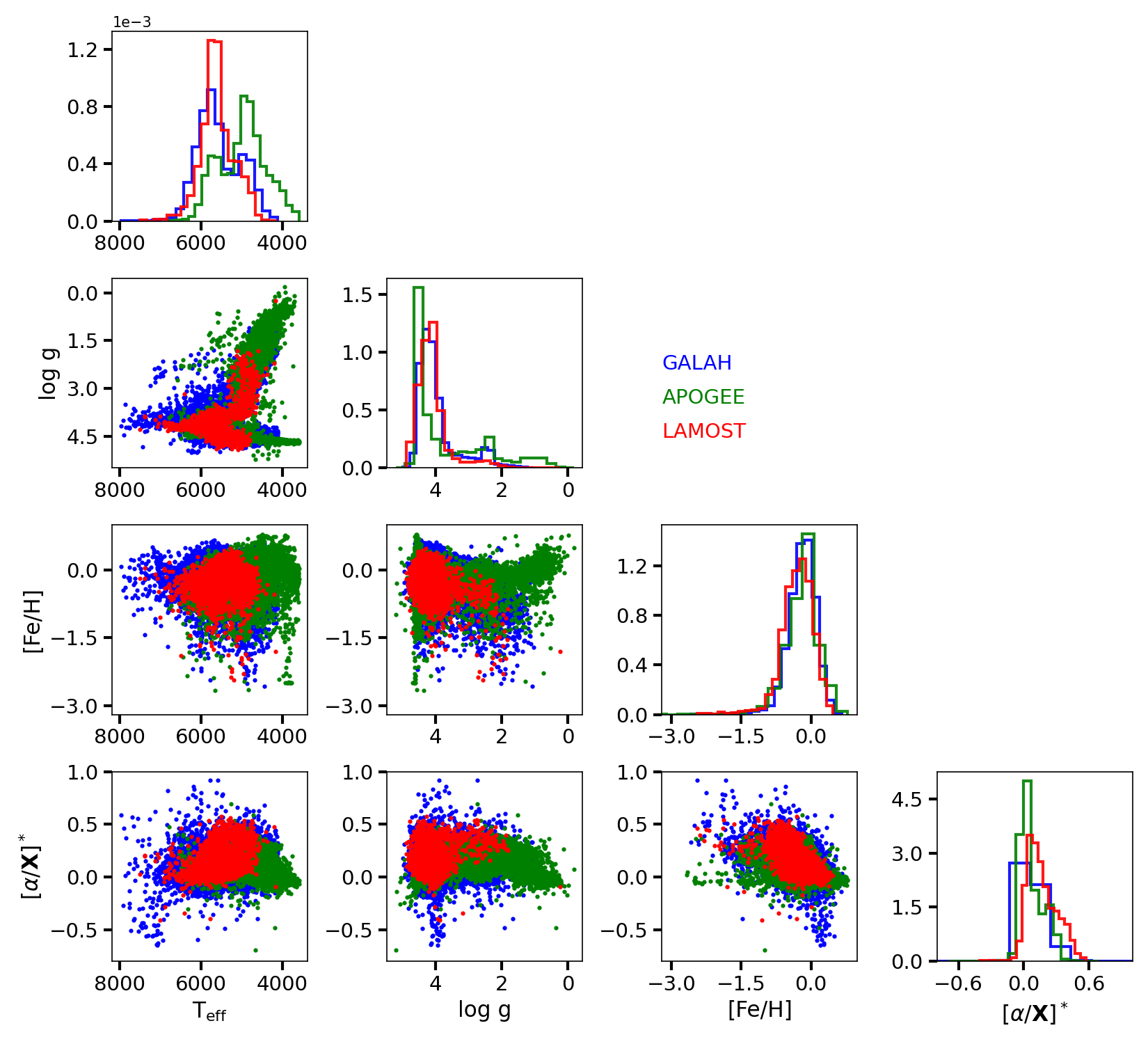}

   \caption{Corner plot for the APOGEE, LAMOST, and GALAH training sets, that coincide with S-PLUS data, indicated in green, red, and blue, respectively, showing the distribution of $T_\mathrm{eff}$, \(\log g\), [Fe/H], and [$\alpha$/X], where X denotes [M] for LAMOST and APOGEE data, and [Fe] for GALAH data. The histograms along the leading diagonal illustrate the parameter distributions for $T_\mathrm{eff}$, \(\log g\), [Fe/H], and [$\alpha$/X]  across the three trainning datasets.}
    \label{Fig:SPLUS-TrainsetAll}%
\end{figure*}

\begin{figure*}
   \centering
   \includegraphics[width=0.95\textwidth,height=1\textwidth]{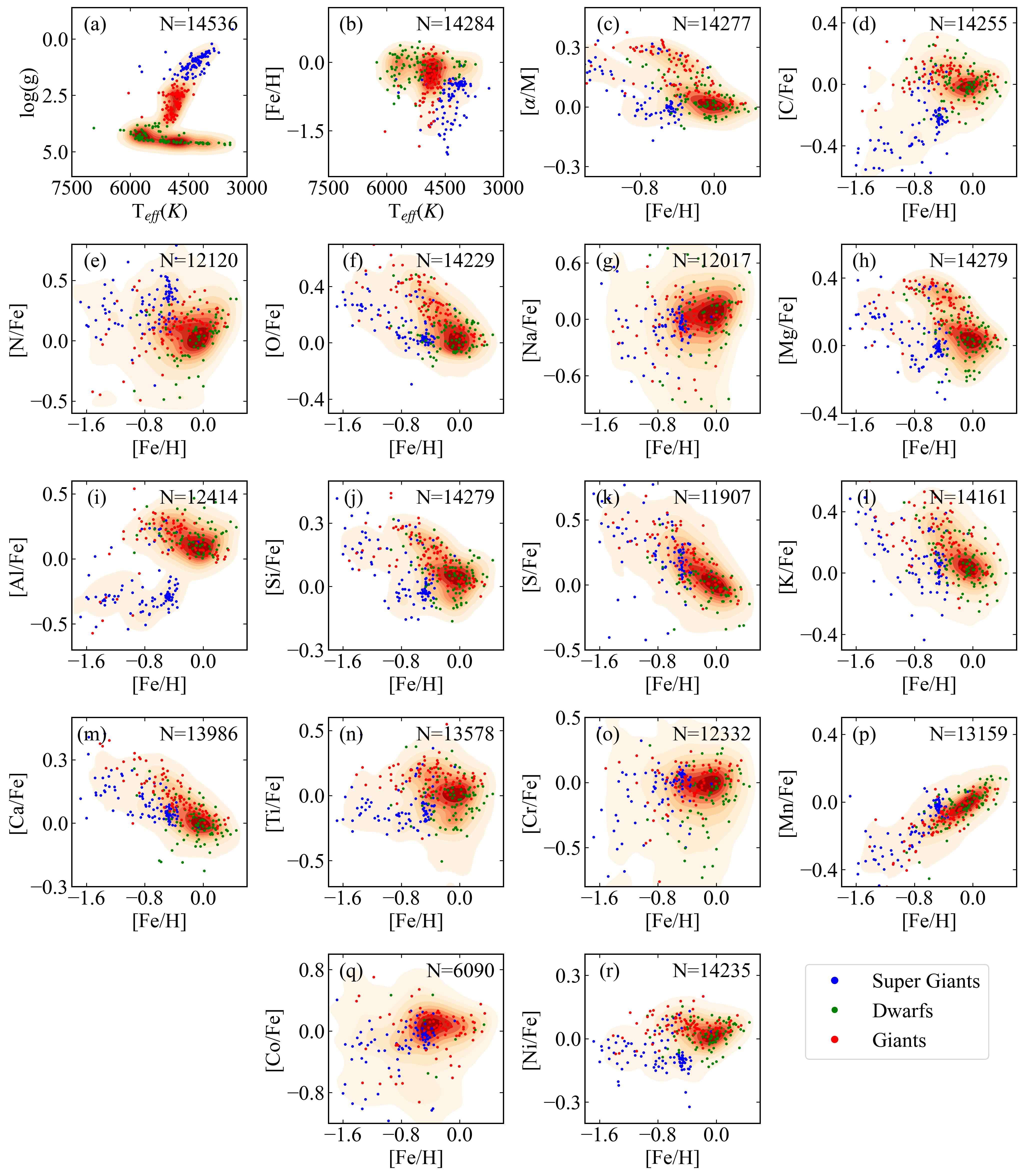}

   \caption{Density plots illustrating the distribution of \(\log g\), $T_{\rm eff}$, and different elemental abundances of the APOGEE sample used in the training. In the top-right corner of each panel the number of sources (N) in each distribution is shown. A randomly selected sample of 100 sources for dwarfs, giants, and supergiants is scatter plotted to avoid crowding. }
    \label{Fig:SPLUS-TrainsetAPOGEE}%
\end{figure*} 

The starting point for our analysis was the S-PLUS DR4 \citep{Herpich:2024}, which provides both aperture and point spread function (PSF) photometries for objects in a 3022.7 sq deg patch of sky visible from the Southern Hemisphere in 12 different optical bands. We utilized aperture photometry measured in a 3-arcsec aperture and corrected for the amount of missing flux (referred to as PStotal magnitudes in the S-PLUS survey, \citealp{Almeida-Fernandes-2022}) for all cases when this aperture is available. In more crowded regions, such as those at low Galactic latitudes or in the central areas of the Magellanic Clouds, PSF photometry was preferred to account for blending and contamination issues. \\

We used PSF photometry, which is less affected by the contamination of neighboring sources. Furthermore, we imposed a sample selection criterion, restricting our dataset to sources having measurements in all 12 available filters, with signal-to-noise ratios (SNR) greater than 5 and error bars smaller than $0.1$. These selection criteria serve two purposes: they enable us to retain more spectral information and establish the $66$ S-PLUS colors, while also ensuring the reliability of photometric measurements. Subsequently, we performed a crossmatch with {\it Gaia}  Data Release 3 \citep[{\it Gaia}  DR3 - ][]{GaiaCollaborationDR3-2021} to obtain preliminary estimates of parameters such as \(\log g\) and $T_\mathrm{eff}$. This resulted in an initial sample of approximately 5 million sources. \\ 

We corrected the magnitudes for interstellar medium extinctions using the maps from \citet[][commonly referred to as SFD]{Schlegel+1998} for the source of the $E_{B-V}^{\rm SFD}$ excess-color parameter, while also adopting the correction proposed by \citet{Schlafly-2011} for this extinction map: $E_{B-V}^{\rm adopted} = 0.86 \times E_{B-V}^{\rm SFD}$. We utilized the extinction coefficients $A_\lambda / A_V$ provided by the Spanish Virtual Observatory Filter Profile Service\footnote{\url{http://svo2.cab.inta-csic.es/theory/fps/}} for the S-PLUS filters, and adopted $A_V / E_{B-V} = 3.1$. \\

\citet[][]{Yang-2022} employed a similar methodology to estimate a comprehensive set of stellar parameters, including $T_\mathrm{eff}$, surface gravity (\(\log g\)), metallicity ([Fe/H]), as well as [$\alpha$/Fe] and four elemental abundances ([C/Fe], [N/Fe], [Mg/Fe], and [Ca/Fe]). They utilized data from the first data release (DR1) of J-PLUS, integrating it with {\it Gaia}  DR2 information and spectroscopic labels from LAMOST to train a series of cost-sensitive neural networks. Furthermore, the authors conducted a thorough comparison of their parameter and abundance estimates with those derived from other spectroscopic catalogs such as APOGEE and GALAH. The results demonstrated an overall agreement between their estimates and those from the aforementioned catalogs. We similarly adhere to the prescriptions outlined by \citet[][]{Yang-2022}.\\

In this study, we adopted a Cost-Sensitive NN approach proposed by \citet[][]{Yang-2022}, which consists of a combination of a traditional seven-layer feed-forward NN architecture and a 2D cost-sensitive learning algorithm to estimate stellar parameters and chemical abundances. The input (first) layer imports the stellar colors, the following five hidden layers extract nonlinear features from these colors, and finally, the output layer generates a stellar label through a weighted sum of the features learned in the last hidden layer. The cost-sensitive learning algorithm used to train the NN achieves better measurement precision for less frequent target values in an imbalanced training set. By adjusting the cost function to assign higher weights to less frequent cases, this approach mitigates bias. This adjustment enables the neural network to learn more effectively, thereby enhancing its ability to accurately predict these minority cases.\\

The separation between giant and dwarf stars in our training sample was based on the \(\log g\) and \(T_{\mathrm{eff}}\) values obtained from the \citet[{\it Gaia} DR3]{GaiaCollaborationDR3-2021}. We followed the selection criteria proposed by \citet{Thomas-2019} for the \(\log g\) vs \(T_{\mathrm{eff}}\) distribution (Kiel diagram) to differentiate between dwarfs and giants. For the selection of giants, \citet{Thomas-2019} proposed a range of \(5000 \le T_{\mathrm{eff}} \le 6000\) and \(3.2 \le \log g \le 3.9\), based on a machine learning algorithm trained on multiband photometric and spectroscopic data, applicable to southern hemisphere targets. Their work did not consider the subgiant class, and the specified limits exclude the majority of asymptotic giant branch (AGB) stars. To include more AGB stars, red giant branch stars (RGB), and subgiants in our giant selection and to further distinguish a supergiant class, we adjusted the limits as follows (see Fig. \ref{Fig:SPLUS-TrainsetAPOGEE}): for giants: \(3000 \le T_{\mathrm{eff}} \le 7000\) and \(1.8 \le \log g \le 3.9\); for supergiants: \(3000 \le T_{\mathrm{eff}} \le 7000\) and \(\log g \le 1.8\).\\

Defining precise boundaries between dwarfs, giants, and supergiants on the Kiel diagram presents significant challenges, particularly when accounting for uncertainties in \(\log g\), which tend to obscure the delineations between these stellar classifications. The inclusion of a supergiant category further complicates the classification algorithm and does not resolve the inherent ambiguity among these stellar classes, a challenge also noted by \citet{Thomas-2019}. Therefore, for model training and subsequent analysis, we have simplified the classification into two primary categories: dwarfs and giants. Within this framework, the giants category encompasses supergiants, RGB stars, AGB stars, and subgiants, if present. This classification criterion has been applied consistently across all three training sample sets. Table \ref{Table:SpectralLimits} displays the number of sources and the limits for each parameter used in our training sets, categorized by survey type for both dwarfs and giants. The GALAH data offer the widest range of parameters for both dwarf and giant class, while the APOGEE dataset includes a highest number of giant stars among all the three surveys. On the other hand, the LAMOST survey has the smallest number of metal-poor stars, providing a limited sample for estimating stellar parameters and elemental abundances for these stars (see Table \ref{table:MetalityInterval}).

\begin{table}
\centering
\scriptsize
\caption{Summary of the training datasets categorized by [Fe/H] intervals.}  \resizebox{0.9\columnwidth}{!}{%
  \begin{tabular}{c c c c c c}
    \toprule \midrule
     Interval & APOGEE &  GALAH & LAMOST  \\
       \midrule
     $\mathrm{[Fe/H]} > 0$      & 7755  & 11528  & 1200 \\
     $0 > \mathrm{[Fe/H]} > -0.5$  & 10514 & 26072  & 5309 \\
     $-0.5 > \mathrm{[Fe/H]} > -1$ & 2651  & 5189   & 2479 \\
     $-1 > \mathrm{[Fe/H]} > -2$   & 591   & 882    & 244 \\
     $-2 > \mathrm{[Fe/H]} > -3$   & 66    & 121    & 9 \\
     $\mathrm{[Fe/H]} < -3$      & 14    & 34     & 4 \\    
\midrule \bottomrule
    \label{table:MetalityInterval}
  \end{tabular}}
\end{table}


The initial sample was subjected to cross-matching with LAMOST Medium-Resolution, APOGEE, and GALAH. LAMOST uses fibers with a diameter of $3.3$ arcseconds, achieving a spectral resolving power of approximately $R \simeq 7500$. APOGEE, on the other hand, utilizes fibers with a sky diameter of $2.0$ arcseconds and offers a higher spectral resolving power, approximately $R \simeq 22\,500$. Meanwhile, GALAH utilizes fibers with a diameter of $2.1$ arcseconds, providing a spectral resolution range from approximately $R \simeq 20\,000$ to $R \simeq 50\,000$. As a note of caution, we emphasize that the LAMOST data used here is limited to the medium-resolution spectra. \\

It is important to note that machine learning techniques may not successfully recover all elemental abundances from the LAMOST, GALAH, and APOGEE surveys if those abundances do not significantly impact the photometric data obtained from the 12 S-PLUS filters. The S-PLUS filters provide 66 photometric colors, capturing a broad range of elemental abundance information that may not be immediately visible. Our initial hypothesis is that the S-PLUS colors retain significant information about these abundances, particularly those detected by the narrowband filters. Nevertheless, these techniques have the potential to uncover intricate correlations that may not be immediately apparent. It is crucial to exercise caution when interpreting elemental abundances that lack corresponding features in S-PLUS filters. However, retaining parameters that cannot be determined directly from the input data enables us to thoroughly assess potential biases and explore unexpected correlations identified by our analytical approach. A detailed discussion of the parameters used in this work is provided in Sect. \ref{Sec:Results}. \\

To reduce contamination due to fiber size, we performed cross-matching individually between the S-PLUS data and each survey's catalog—LAMOST, APOGEE, and GALAH—using a search radius of 2 arcseconds. For each survey, we accounted for the specific fiber size by removing crossmatched sources that had multiple counterparts within the fiber aperture. This approach allowed us to handle potential overlaps and ensure that only the most accurate matches were considered. The accuracy of these surveys varies significantly due to differences in resolution and fiber size. Therefore, we chose not to combine all the data acquired from these crossmatches, allowing for the consideration of any divergences among them in further studies. According to \citet{Hegedus:2023}, the parameters from APOGEE and GALAH surveys are generally in agreement within their respective uncertainties. However, offsets larger than $0.1$ dex in the absolute abundance scales are observed in certain regimes. Therefore, when combining the data, it is crucial to consider the peculiarities of each survey. Moreover, by independently estimating the parameters using three distinct datasets, we can effectively compare the results, which, in turn, aids in validating and enhancing our analytical methodologies. \\

The cross-matching resulted in the identification of numerous sources—designated as $\sim2877$, $\sim5916$, and $\sim573$ in the APOGEE, GALAH, and LAMOST datasets, respectively. It is important to note that the number of valid values for each spectroscopy parameter may exhibit variations due to data availability and completeness. To ensure data quality, a minimum SNR of 20 was adopted in the GALAH, LAMOST, and APOGEE datasets. Our parameter estimation approach was carried out separately for each survey, enabling a comparative assessment of the goodness-of-fit of our estimations. This separation led to the creation of dedicated matrices for each parameter, thereby maximizing the number of elements for our analysis. This approach acknowledges the complexities arising from the unique characteristics of each survey, ensuring a more accurate and robust analysis tailored to the specific attributes of the data.\\

In Figure \ref{Fig:SPLUS-TrainsetAll}, we present an overview of the parameter space, encompassing $T_\mathrm{eff}$, \(\log g\), metallicity ([Fe/H]), and $[\rm \alpha/M]$ abundances for the three distinct trainsets we are using: APOGEE (blue dots), LAMOST (red dots), and GALAH (green dots). Notably, these distributions cover an extensive range of $T_\mathrm{eff}$, \(\log g\), [Fe/H], and $[\alpha/\mathrm{M}]$, with values ranging from approximately $3000 \, \mathrm{K}$ to $7000 \, \mathrm{K}$ for $T_\mathrm{eff}$, $0.5$ to $4.5$ for \(\log g\), $-3$ to $1$ for [Fe/H], and $-0.4$ to $0.6$ for $[\alpha/\mathrm{M}]$. In summary, the three data samples consist of dwarf stars and giants stars, exhibiting similar distributions  (see the diagram of \(\log g\) as a function of $T_\mathrm{eff}$ in the second panel line of Fig. \ref{Fig:SPLUS-TrainsetAll}). Despite the clear separation of super-giant and giant stars in the plot, they are considered as a single class to maximize the number of training data (see top left panel in Fig. \ref{Fig:SPLUS-TrainsetAll}). Indeed, the GALAH sample is the largest, varying from about $19$ to $25$ thousand sources, with a primary focus on stars with temperatures exceeding $4000$K. Conversely, the LAMOST and GALAH datasets primarily include stars with higher temperatures than APOGEE; on the other hand, LAMOST has a wider range of \(\log g\) than does GALAH (for more details, see Table \ref{Table:SpectralLimits}). \\

We do not impose a typical cutoff in temperatures, so values for $T_\mathrm{eff} < 4000K$ or $T_\mathrm{eff} > 6500K$ should be interpreted with caution. The GALAH, LAMOST, and APOGEE datasets provided a wide array of elemental abundances, encompassing parameters such as [$\alpha$/Fe], [Fe/H],  [S/Fe], [Cr/Fe], [Mn/Fe], [Co/Fe],  [O/Fe], [Na/Fe], [K/Fe], [Ti/Fe], [C/Fe], [N/Fe], [Mg/Fe], and  [Ca/Fe]. This paper explores the possible linear or non-linear relationships between these parameters, using machine learning techniques, employing S-PLUS colors within our initial S-PLUS sample.\\

Figure \ref{Fig:SPLUS-TrainsetAPOGEE} shows the array of parameter planes within the APOGEE training sets. These planes include various parameters plotted against [Fe/H]. Incorporating these diverse parameters expands our ability to estimate stellar parameters across a wider range, broadening the scope of our analysis. Similar figures showing the distribution of available parameters for GALAH and LAMOST data can be found in Appendix \ref{appendixA}.\\

\section{Machine learning approaches and data normalization}
\label{Sec:Methodology}

Neural Networks (NN) and Random Forests (RF) were employed for a dual verification of the results, aiming to mitigate the dependency of the estimation on a single method. NN, with their capacity to model intricate, nonlinear data relationships, are highly adept at accommodating a wide variety of data types and capturing complex patterns. On the other hand, RF offers robustness against overfitting, demonstrates competence in managing high-dimensional data, and furnishes valuable insights through the importance of the input parameters. Moreover, the decision tree structure of RF enhances interpretability, a feature that can be more challenging to achieve with NN. A pragmatic approach combines NN and RF predictions for more robust and accurate parameter estimation \citep[][]{Cutler-2007,LeCun-2015}. This approach aligns with best practices in machine learning, which emphasize the importance of feature selection and engineering to enhance model performance.  \\

We harnessed the widely-used Keras library \citep[][]{Chollet-2018} and constructed a NN with multiple layers for processing. In this context, we explored various configurations, varying the number of layers from $1$ to $10$ and the number of neurons from $32$ to $2000$, i.e. typical parameters used to define the network's capacity to comprehend complex data structures. The optimal configuration was identified as using six layers with $1664$ neurons distributed among them. Indeed, this approach closely follows the methodology applied to J-PLUS observations \citep[]{Cenarro-2018,Yang-2022}.\\

\begin{figure}
   \centering
   \includegraphics[width=0.5\textwidth]{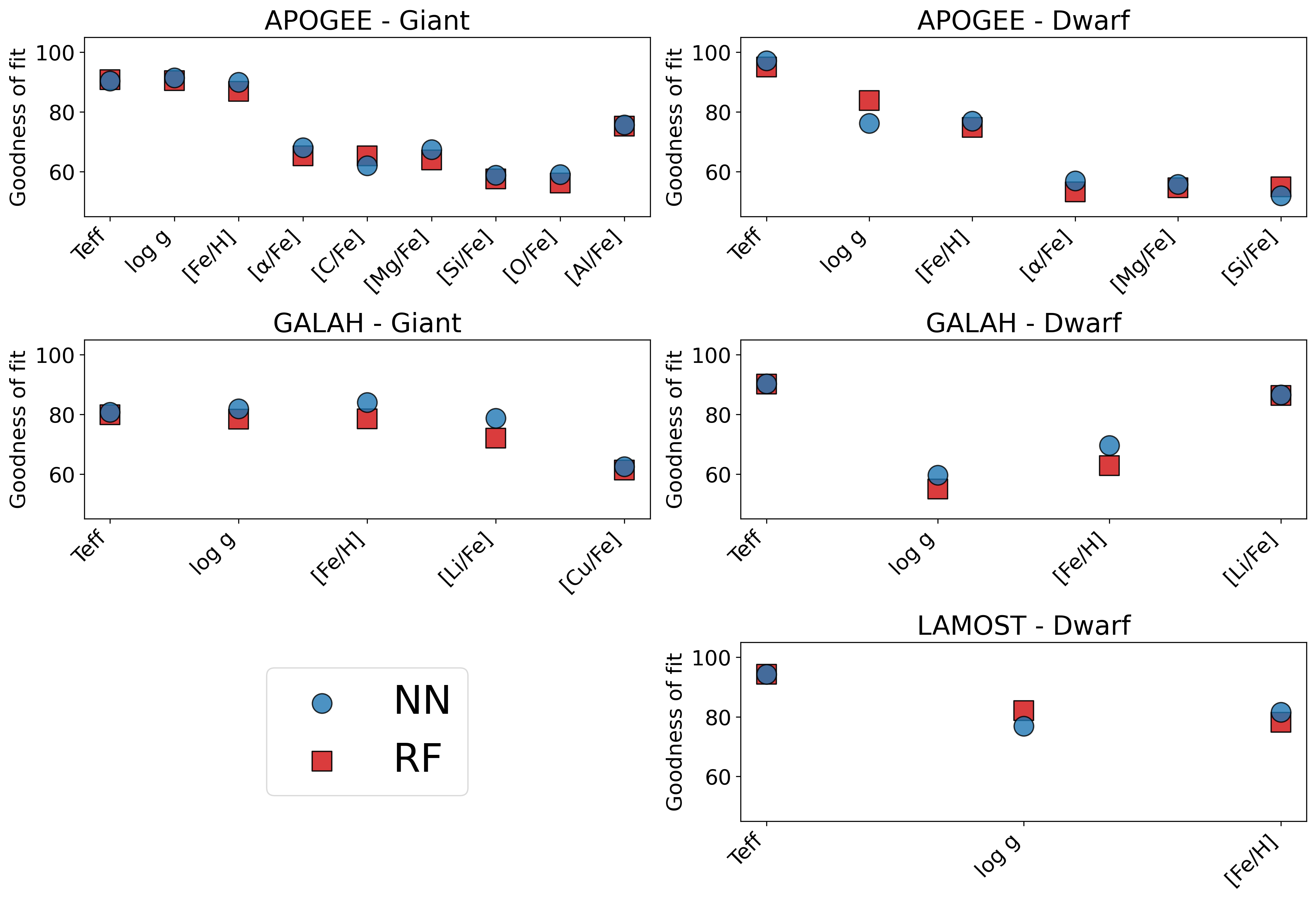}
\caption{Comparative analysis of RF and NN goodness-of-fit across diverse parameters in the APOGEE, GALAH, and LAMOST MRS Surveys. These results are based on models constructed using only S-PLUS color information. Only parameters with goodness-of-fit exceeding $50\%$  are included.}
    \label{Fig:accuracy}%
\end{figure} 

In the evaluation of RF, we conducted an extensive parameter-tuning exercise to identify the optimal configuration for our model. We systematically tested hyperparameters such as the number of trees, maximum depth of trees, and maximum features for each stellar parameter or abundance, exploring their impact on the model's predictive capability. By analyzing the mean square error associated with different hyperparameter configurations, we selected the most effective model for each stellar atmospheric parameter or abundance. This iterative process, which involved refining model parameters through repeated testing, enabled us to determine the optimal combination that yielded the best predictive results and assess the importance of each color in predicting stellar parameters or abundances, facilitating an astrophysical interpretation.\\

The available stellar atmospheric parameters from GALAH, LAMOST, and APOGEE were accessed using standardized input vectors—pre-processed data that ensure consistent scaling and units—along with the optimal NN and RF configurations. We partitioned the data into two segments: $80\%$ for training and $20\%$ for testing, adhering to standard cross-validation practices \citep[e.g.,][]{Belgiu-2016}. Each data sample was analyzed independently, and this partitioning strategy was employed to evaluate the model’s training efficacy. The results of this estimation are detailed in Section \ref{Sec:Results}.

The input variables, standardized to ensure uniform scales, form the foundation for training our robust NN and RF models. Initially, our analysis focuses on stars with all 12 photometric measurements. In future work, utilizing S-PLUS DR 5, we plan to extend our approach to include sources with incomplete photometric data. The 12 S-PLUS magnitudes result in a total of 66 colors, and we will employ three distinct approaches to leverage this data effectively: (A) using a 66-dimensional input vector composed of S-PLUS colors; (B) estimating the effective temperature ($T_\mathrm{eff}$) using the initial 66 S-PLUS colors. Subsequently, a matrix with 67 columns, comprising both the colors and $T_\mathrm{eff}$, is employed to estimate additional parameters; and (C) estimating both the effective temperature ($T_\mathrm{eff}$) and surface gravity ($\log g$) using the 66 S-PLUS colors. Following this, a matrix with 68 columns, including the colors, $T_\mathrm{eff}$, and $\log g$, is utilized to estimate the remaining parameters. The normalization is outlined as follows:

\begin{equation}
    y' = \frac{y - \mu}{\sigma},
\end{equation}

\noindent where $y$ and $y'$ are the original and standardized input vectors, respectively. Additionally, $\mu$ and $\sigma$ represent the mean and standard deviation of all the original input vectors, respectively.

\section{Results}
\label{Sec:Results}

This section presents the results obtained from two distinct methods, RF and NN, implemented using three different approaches (ABC, see Sect. \ref{Sec:Methodology}) and applied to three separate datasets. This results in a total of $18$ unique strategies for parameter estimation. To determine the most effective method, we utilize the \textit{r2\_score} function from scikit-learn \citep{scikit-learn}, which we have labeled as goodness-of-fit. This function provides a relative measure of goodness-of-fit, indicating the extent to which the model aligns with the observed data in comparison to a basic mean baseline. We chose to use the \textit{r2\_score} over Root Mean Square Error (RMSE) because \textit{r2\_score} offers a normalized measure that can indicate the proportion of variance explained by the model. The best possible score is 1.0, while a score of 0.0 indicates that the model performs no better than the mean baseline. Negative values indicate that the model performs worse than the mean baseline. This makes \textit{r2\_score} particularly useful for comparing the performance of different models across varying scales. On the other hand, RMSE provides an absolute measure of error, which can be more challenging to compare across different parameters or datasets. We only considered parameters with a goodness-of-fit greater than 50\% (see Appendix \ref{appendixB}) across all approaches used. Such a limit constrains our parameters to $T_\mathrm{eff}$, \(\log g\), $[\text{Fe/H}]$, $[\alpha/\text{Fe}]$, $[\text{Al}/\text{Fe}]$, $[\text{C}/\text{Fe}]$, $[\text{Cu}/\text{Fe}]$, $[\text{Li}/\text{Fe}]$, $[\text{Mg}/\text{Fe}]$, $[\text{O}/\text{Fe}]$, and $[\text{Si}/\text{Fe}]$. Figure \ref{Fig:accuracy} shows the goodness-of-fit for the parameters considered, all of which have scores greater than 50\%. However, not all parameters are available for both giants and dwarfs. Furthermore, the LAMOST sample for giant stars includes several parameters with a training set smaller than 100, and these were excluded from our analysis. \\

At this stage, two critical aspects require careful examination: the optimization of machine learning by adjusting input parameters to ensure alignment with S-PLUS filters, and the assessment of parameter reliability. These criteria culminate in the establishment of flags, which are described below.

The NN approach consistently outperforms the RF approach, showcasing goodness-of-fit improvements of up to $\sim6\%$ (see Appendix \ref{appendixB}). These results established it as the standard prediction method. However, according to our results, the RF approach performs better for [Si/Fe] and [C/Fe].

The RF approach offers better goodness-of-fit than NN for APOGEE [C/Fe] in giant stars, as well as for [Si/Fe] and $\log g$ in dwarfs. Additionally, LAMOST $\log g$ shows slightly better results with the RF approach.

The accuracies of APOGEE, GALAH, and LAMOST for the same parameters vary. For example, the goodness-of-fit of $T_\mathrm{eff}$ estimation for APOGEE is approximately 9\% better than that found for GALAH in the case of giant stars. In general, S-PLUS estimations based on GALAH input data exhibit smaller accuracies compared to those derived from APOGEE and LAMOST data. The near-infrared spectroscopy provided by APOGEE, along with the wide spectral coverage offered by LAMOST, appears to confer advantages for the parameters studied. However, a straightforward comparison of the performance of the different training sets should be approached carefully, as they do not equally cover the full range of values or sample all intervals uniformly (see Fig. \ref{Fig:SPLUS-TrainsetAll} and Table \ref{Table:SpectralLimits}). For example, the GALAH data tends to assume lower values of $T_\mathrm{eff}$ and [Fe/H], which may result in reduced in-sample accuracy for GALAH measurements.

Incorporating $T_\mathrm{eff}$ and surface gravity ($\log g$) as input parameters enhances the goodness-of-fit of parameter estimations by approximately $\sim1\%$ across several parameters. Specifically, we observe an increase in accuracy of about $2\%$ and $5\%$ for determining [Mg/Fe] and [Fe/H] ratios, respectively. This suggests that these particular parameters exhibit heightened sensitivity to variations in $T_\mathrm{eff}$ and $\log g$ compared to other parameters within the tested sample.

\begin{figure}[htb]
\centering

\includegraphics[width=0.45\textwidth]{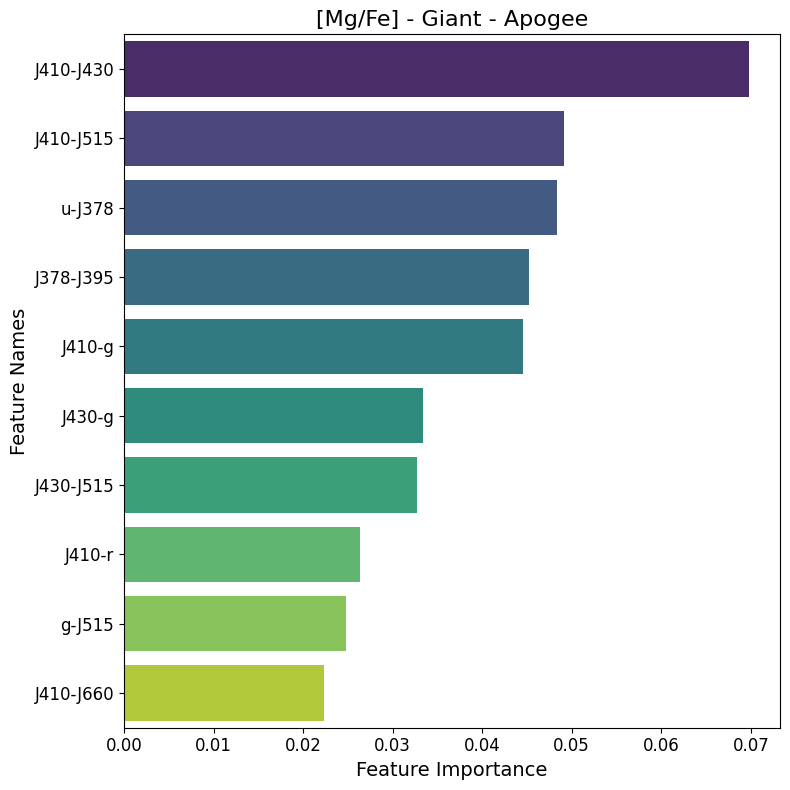}

\caption{The 10 primary features utilized for prediction are arranged in decreasing order of their importance in the RF model. At the top of each diagram, the star type (giant or dwarf), data sample (LAMOST, APOGEE, or GALAH), and the spectroscopic parameter is specified.}
\label{fig:FeaturesImportance}
\end{figure}

\begin{figure*}
\centering
\includegraphics[width=0.30\textwidth,height=0.23\textheight]{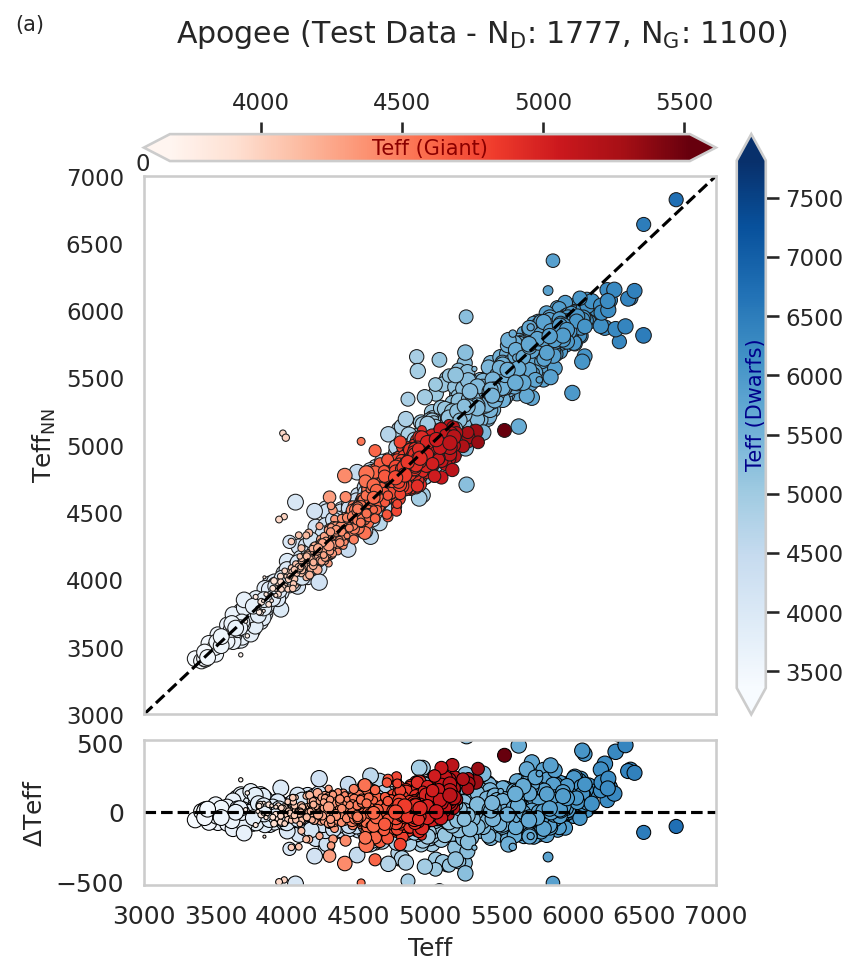} 
\includegraphics[width=0.30\textwidth,height=0.23\textheight]{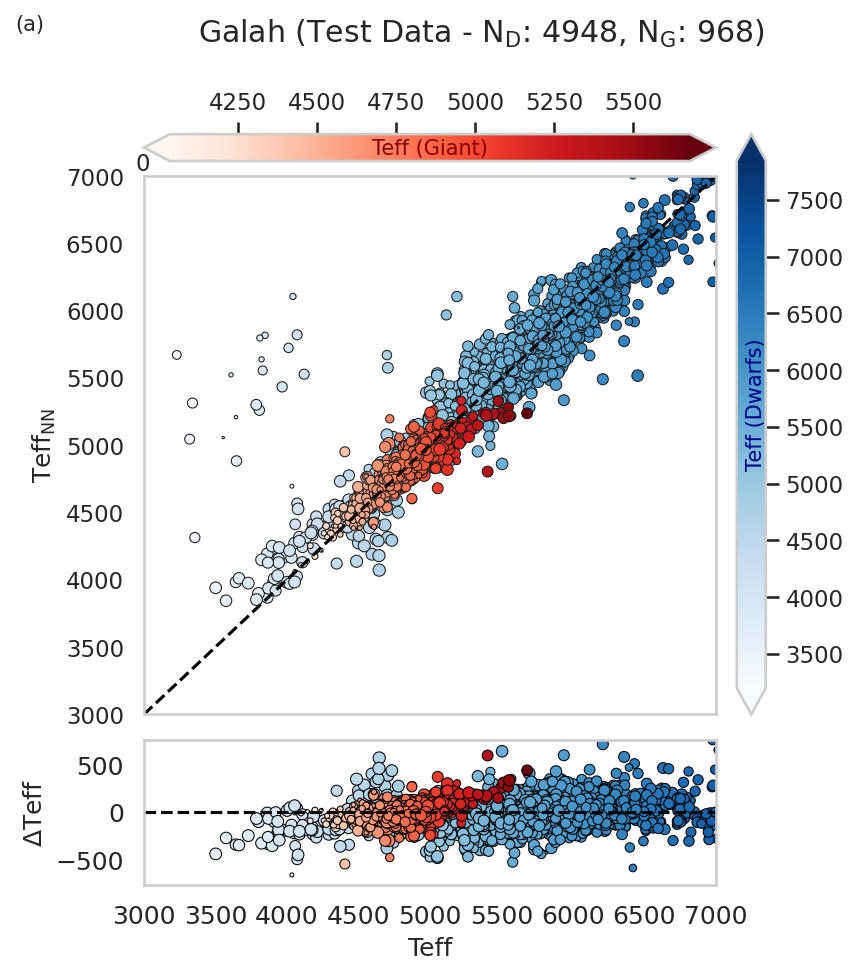} 
\includegraphics[width=0.30\textwidth,height=0.23\textheight]{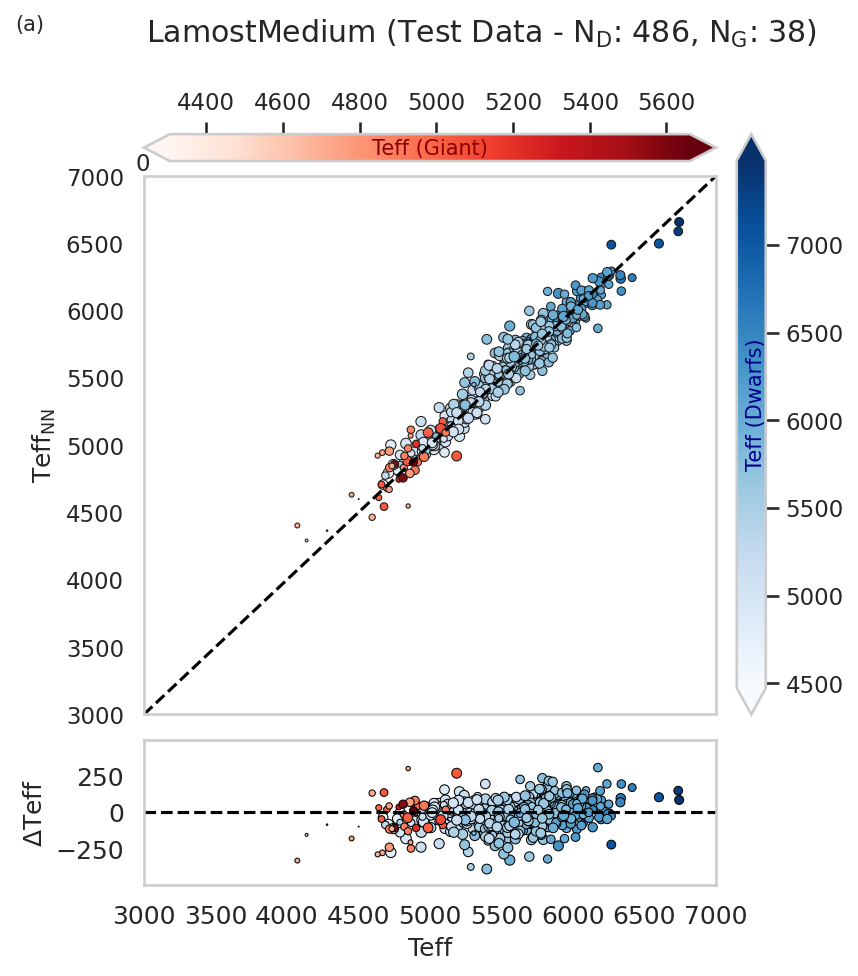} 

\includegraphics[width=0.30\textwidth,height=0.23\textheight]{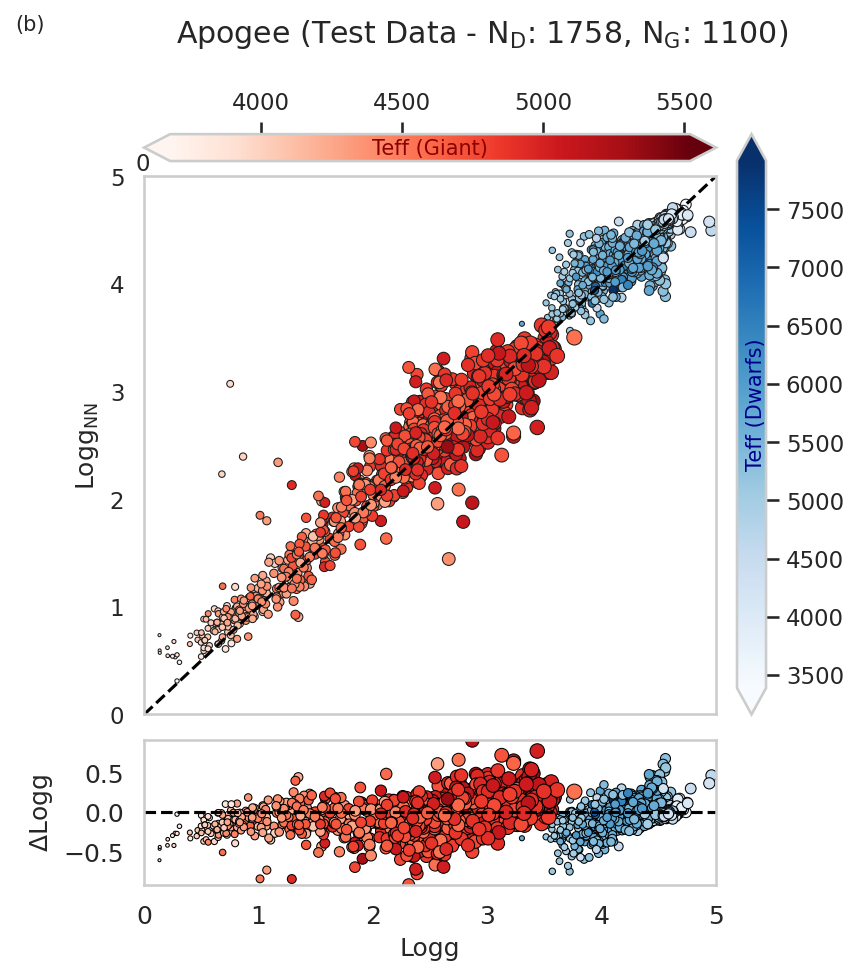} 
\includegraphics[width=0.30\textwidth,height=0.23\textheight]{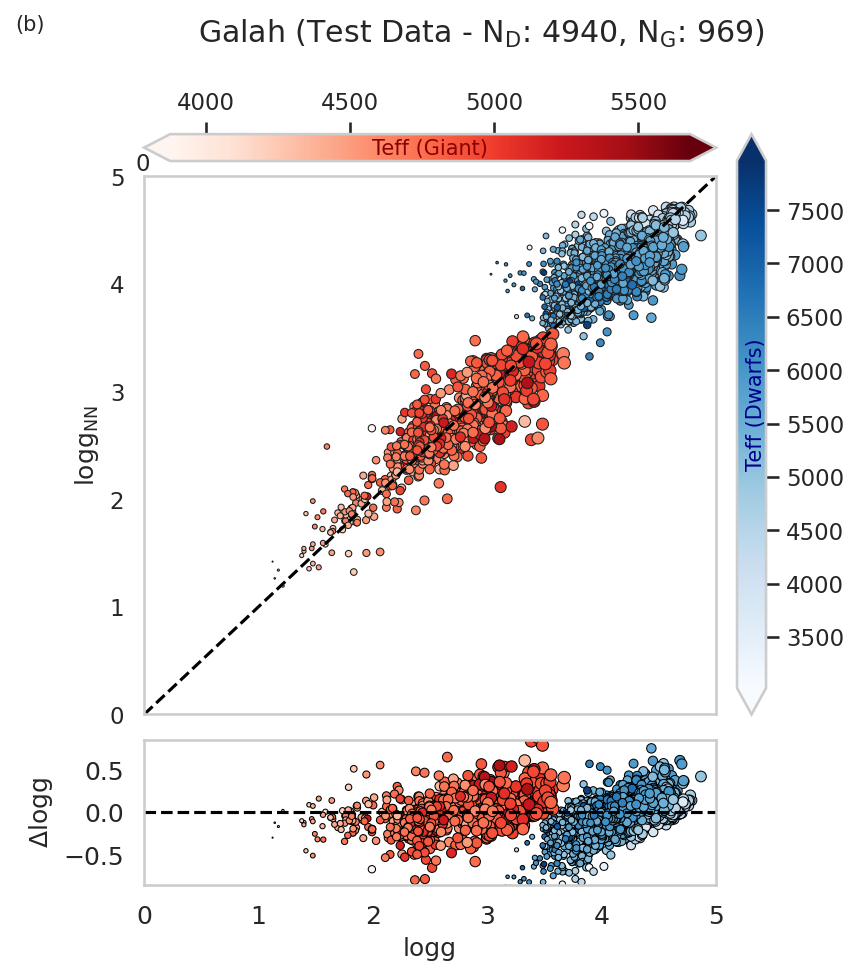} 
\includegraphics[width=0.30\textwidth,height=0.23\textheight]
{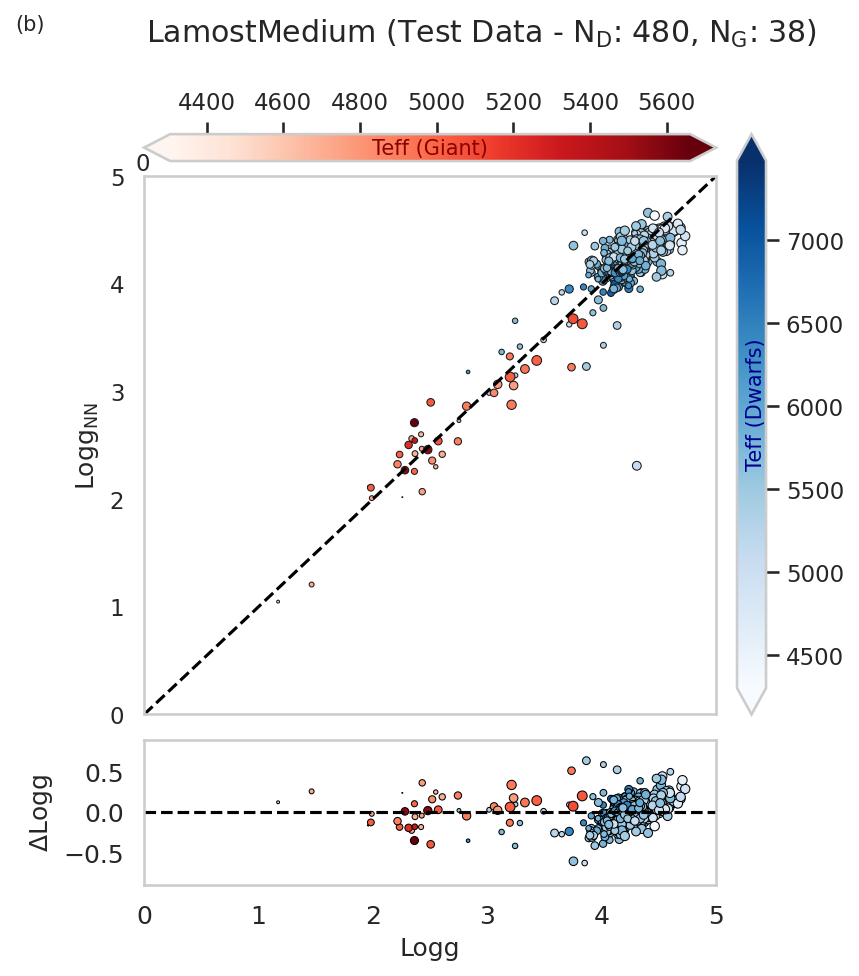} 

\includegraphics[width=0.30\textwidth,height=0.23\textheight]{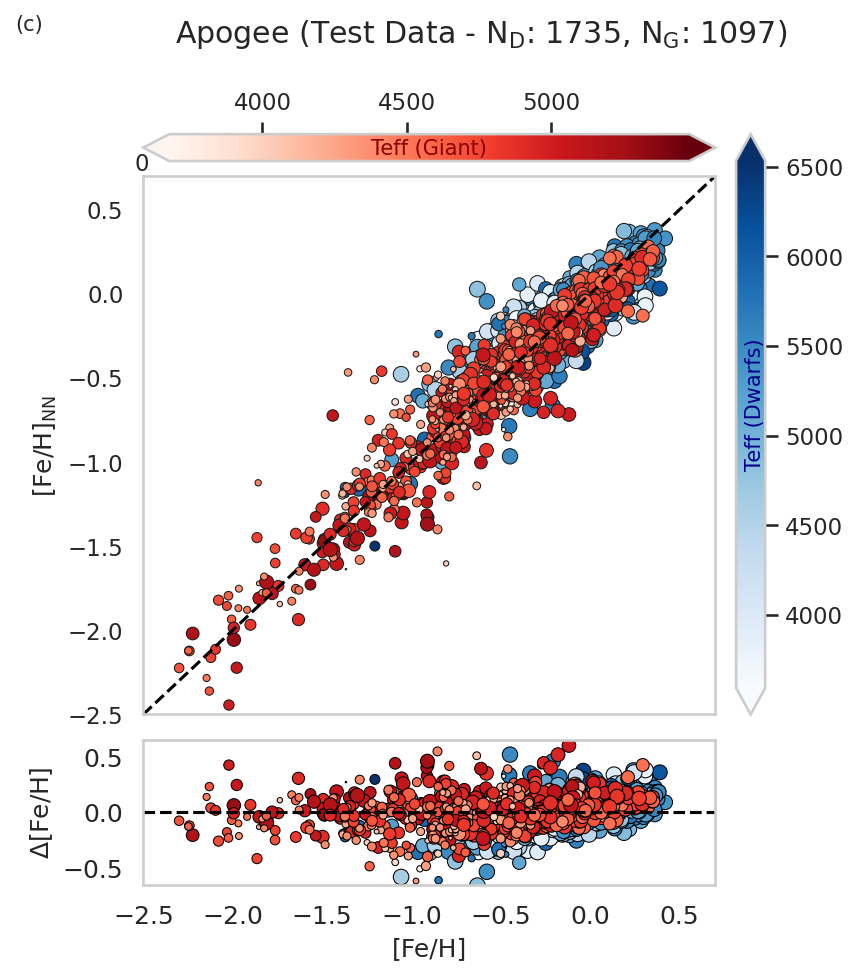} 
\includegraphics[width=0.30\textwidth,height=0.23\textheight]{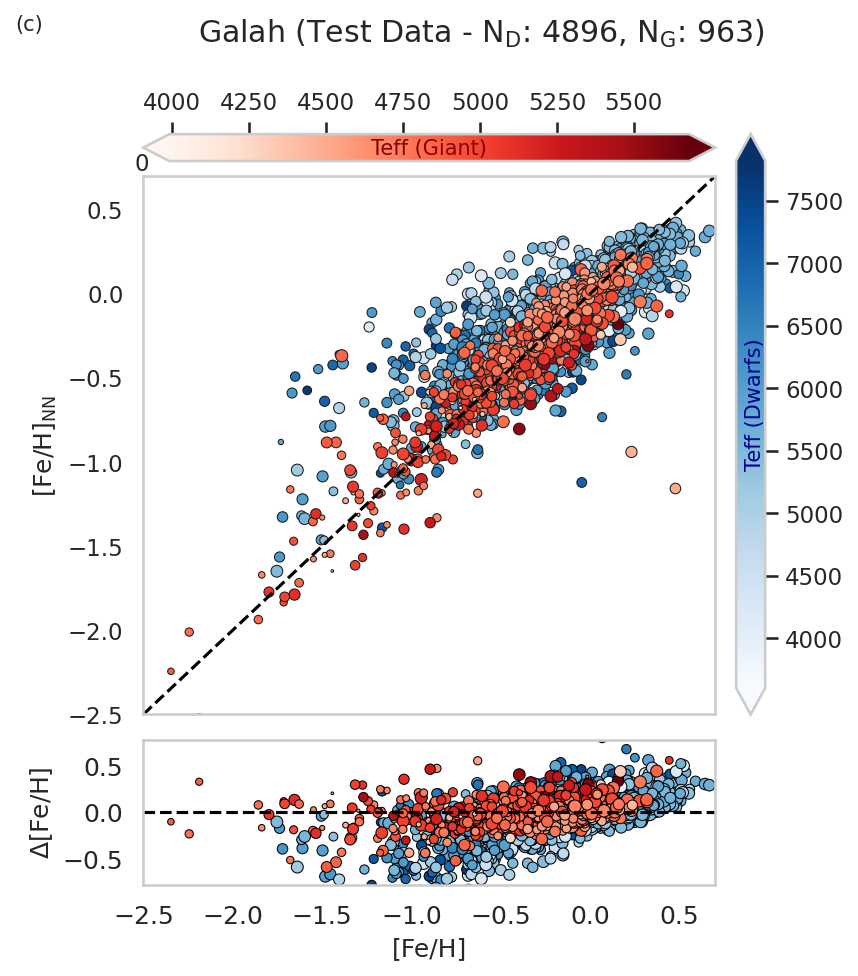} 
\includegraphics[width=0.30\textwidth,height=0.23\textheight]{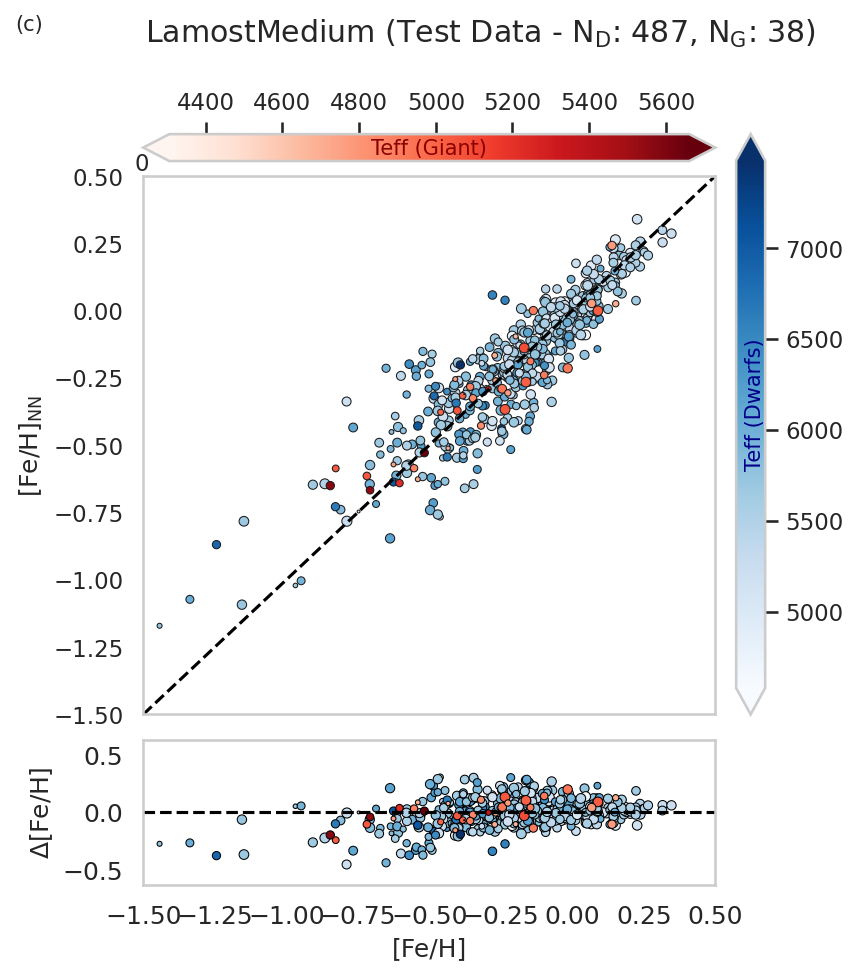} 

\includegraphics[width=0.30\textwidth,height=0.23\textheight]{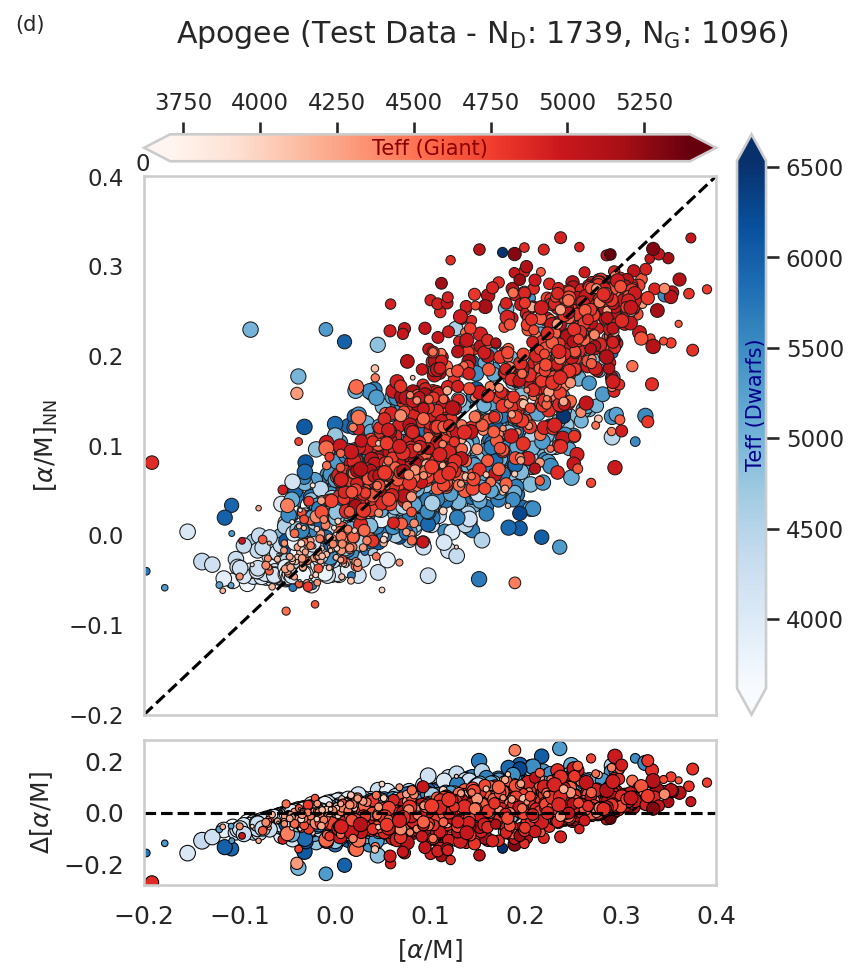} 
\includegraphics[width=0.30\textwidth,height=0.23\textheight]{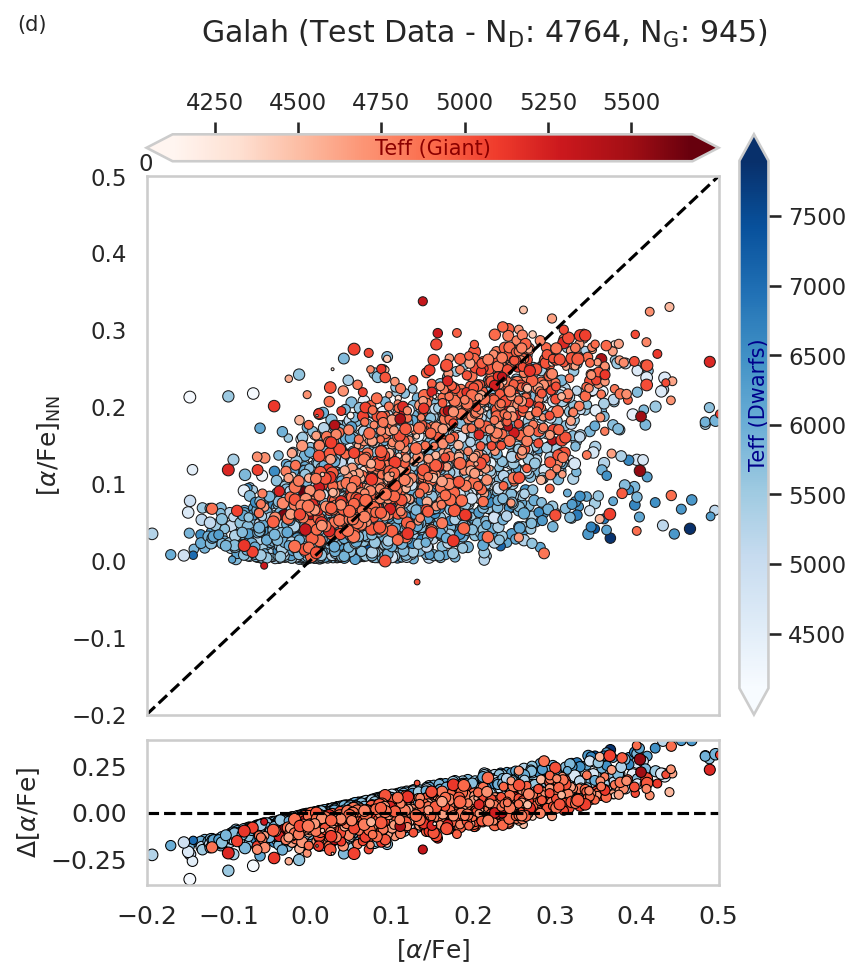} 
\includegraphics[width=0.30\textwidth,height=0.23\textheight]{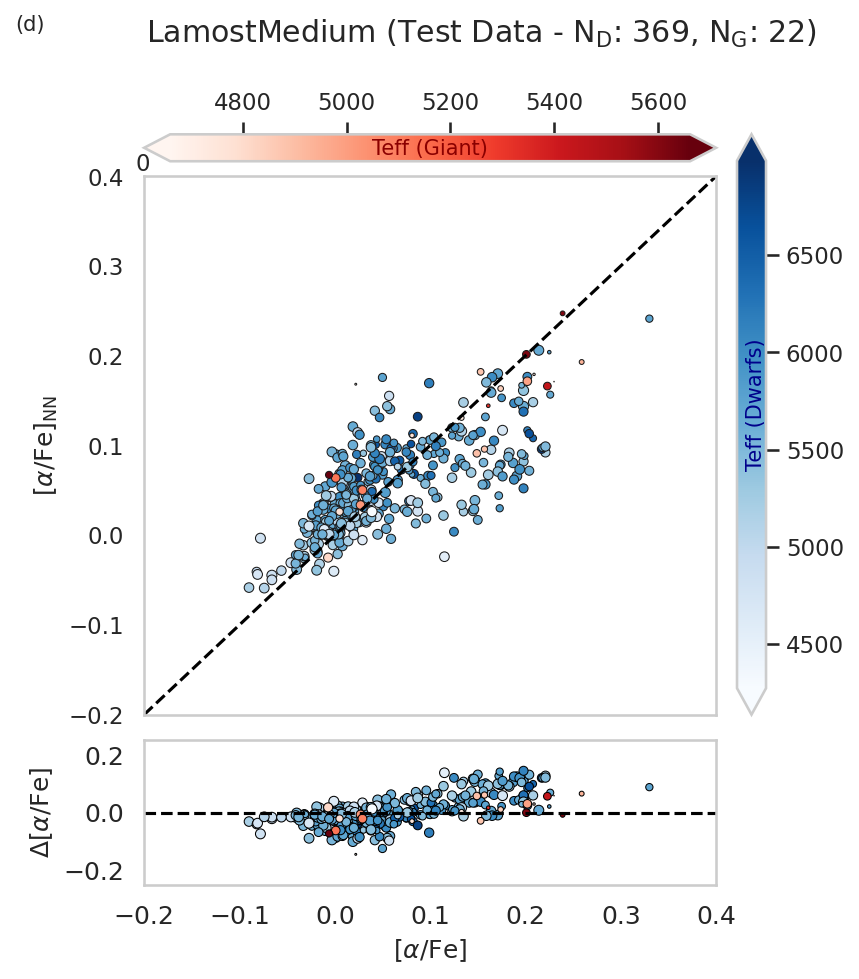} 
\caption{Results of the testing set for various astrophysical parameters ($T_\mathrm{eff}$, \(\log g\), [Fe/H], $\alpha$/M, [Mg/Fe]) obtained in this study using the NN model. Each plot features a one-to-one correlation between the parameters obtained from APOGEE, GALAH, and LAMOST surveys and those predicted by the NN, with identity line included for reference. Blue and red dots correspond to dwarfs and giants, respectively. For both dwarfs and giants, the colorbar indicates the $T_\mathrm{eff}$. The size of the symbols represents the value of \(\log g\).}
\label{fig:predict-NN-Dwarfs}
\end{figure*}

As $T_\mathrm{eff}$ increases, the thermal ionization and excitation rates of atoms change, leading $T_\mathrm{eff}$ as an important feature to estimate such parameters. Indeed, the dependence of spectral lines on $T_\mathrm{eff}$ and \(\log g\) are a fundamental aspect of stellar spectroscopy \citep[e.g.,][]{Payne-1925,Kurucz-1979,Cayrel-2001,Gray-2005}. The influence of $\log g$ on the studied chemical abundances is particularly notable in elements related to surface gravity-sensitive features. On the other hand, $T_\mathrm{eff}$ plays a crucial role in shaping the spectral features related to abundance ratios like [C/Fe], [O/Fe], and [Ti/Fe] which are known to be strongly influenced by changes in temperature.  Indeed, \citet{Beeson-2024} show that the use of photometric priors can improve the goodness-of-fit of abundances and $\log g$. [$\alpha$/Fe], [Al/Fe], and [Mg/Fe] are known to exhibit significant variations with changes in $\log g$; accordingly, our results also present $\log g$ as the main feature to estimate such parameters (see Fig. \ref{fig:FeaturesImportance}). The [C/Fe], [O/Fe], and [Cu/Fe] lines originate from atomic transitions that are sensitive to the thermal conditions in the stellar atmosphere. \\

Moreover, we have also observed that the narrowband filters $J$0378, $J$0395, $J$0410, and $J$0430, along with the $u$-band, play a crucial role in determining stellar parameters and abundances. These filters contribute significantly to the key features used in our analysis, as the feature importance for many of the outputs is strongly linked to these filters (see Fig. \ref{fig:FeaturesImportance}). Intriguingly, the importance of these filters varies across different surveys, suggesting that their efficacy may be influenced by the unique characteristics of each dataset (refer to Appendix \ref{appendixC}). Moreover, despite reddening correction, high extinction regions, such as those in the disk and bulge (observed with APOGEE-S), will lead to large uncertainties in colors, particularly when using bluer bands, potentially resulting in poorer outcomes. This should be considered when selecting reliable samples. \\

\begin{table*}
  \centering
  \scriptsize
\caption{Summary of the number of giant and dwarf stars with computed stellar parameters or abundances. }  \resizebox{\textwidth}{!}{%
  \begin{tabular}{c c | c c c c | c c c c}
    \toprule \midrule
    &  & \multicolumn{4}{|c|}{Giant} & \multicolumn{4}{|c}{Dwarf} \\
    \midrule
    Survey & Parameter & $100\%$ & $90-99\%$ & $80-89\%$ & $<= 70\%$ &  $100\%$ & $90-99\%$ & $80-89\%$ & $<= 70\%$ \\
    \midrule
APOGEE  &  $T_\mathrm{eff}$  &  146448  &  1907  &  60  &  41  &  4259122  &  246971  &  41770  &  243935   \\ 
APOGEE  &  $\log g$  &  146448  &  1907  &  60  &  41  &  4164703  &  284182  &  52371  &  287206   \\ 
APOGEE  &  [Fe/H]  &  146259  &  2091  &  65  &  41  &  3859132  &  465852  &  60370  &  403628   \\ 
APOGEE  &  [$\alpha$/M]  &  146259  &  2091  &  65  &  41  &  3859855  &  465290  &  60230  &  403623   \\ 
APOGEE  &  [Al/Fe]  &  146084  &  2199  &  100  &  59  &  ...  &  ...  &  ...  &  ...   \\ 
APOGEE  &  [C/Fe]  &  146259  &  2091  &  65  &  41  &  ...  &  ...  &  ...  &  ...   \\ 
APOGEE  &  [Mg/Fe]  &  146259  &  2091  &  65  &  41  &  4011971  &  420630  &  59487  &  294183   \\ 
APOGEE  &  [O/Fe]  &  146259  &  2091  &  65  &  41  &  ...  &  ...  &  ...  &  ...   \\ 
APOGEE  &  [Si/Fe]  &  146259  &  2091  &  65  &  41  &  3859855  &  465290  &  60230  &  403623   \\ 
GALAH  &  $T_\mathrm{eff}$  &  137731  &  7367  &  1119  &  1601  &  4294462  &  275018  &  40211  &  190765   \\ 
GALAH  &  $\log g$  &  137731  &  7367  &  1119  &  1601  &  4278260  &  274900  &  47118  &  198056   \\ 
GALAH  &  [Fe/H]  &  134852  &  10158  &  1181  &  1615  &  4183255  &  239017  &  45081  &  323317   \\ 
GALAH  &  [Cu/Fe]  &  133570  &  10650  &  1360  &  2167  &  ...  &  ...  &  ...  &  ...   \\ 
GALAH  &  [Li/Fe]  &  132732  &  11288  &  1437  &  2266  &  4169646  &  246914  &  46106  &  328053   \\ 
LAMOST  &  $T_\mathrm{eff}$  &  ...  &  ...  &  ...  &  ...  &  3237456  &  667860  &  122621  &  722699   \\ 
LAMOST  &  $\log g$  &  ...  &  ...  &  ...  &  ...  &  3215348  &  688342  &  123573  &  723106   \\ 
LAMOST  &  [Fe/H]  &  ...  &  ...  &  ...  &  ...  &  3208193  &  691026  &  125597  &  725318   \\ 
\midrule \bottomrule
    \label{table:IDR4Prediction}
  \end{tabular}}
  \tablefoot{
The Feature Flag (FF) represents the percentage of features meeting the training set constraints, categorized as $100\%$, $90$–$99\%$, $80$–$89\%$, and $\leq 70\%$.
}
\end{table*}

In summary, the results obtained using RF and NN with approaches ABC were compared, and they agree within the uncertainties. However, the NN implementation using approach C tends to yield superior results in terms of estimated accuracies (see Appendix \ref{appendixD}). Figure \ref{fig:predict-NN-Dwarfs} displays the predicted versus estimated values for $T_\mathrm{eff}$, $\log{g}$, $[\alpha/\text{Fe}]$, and $[\text{Fe/H}]$ across the APOGEE, LAMOST, and GALAH datasets. The results are largely consistent. However, there exists a distinct cluster of stars, primarily dwarf stars, in the GALAH dataset that falls outside the boundaries defined by the dashed line. We isolated this group of stars and found that their GALAH effective temperatures fall within the range of approximately $\sim3200$ to $\sim4500$K. However, the predicted temperatures using NN indicate a wider range, spanning from $\sim4300$ to $\sim6300$K. To investigate this discrepancy, we crossmatched these stars with the {\it Gaia} DR3 and the Transiting Exoplanet Survey Satellite \citep[TESS - ][]{Ricker-2015} catalogs, which present temperature intervals of approximately $\sim4300$ to $\sim6600$K and $\sim4300$ to $\sim6300$K, respectively. These $T_\mathrm{eff}$ values align more closely with our predictions, suggesting a possible underestimation in GALAH temperature measurements for this group.\\ 

Previous studies have also addressed similar aspects of GALAH temperature calibrations. For example, \citet{Giribaldi:2023} note differences in GALAH DR2 parameters and suggest that additional standard stars could enhance calibration. \citet{Hegedus:2023} discuss variations across high-resolution surveys and the need for further investigation into these discrepancies. Additionally, \citet{Wheeler:2020} highlights variations in GALAH's temperature and metallicity measurements, suggesting that further validation could be useful. These references underscore the importance of a thorough reassessment and the incorporation of complementary parameters such as photometric colors, variability, evolutionary stage indicators, and reddening characteristics to address potential differences in GALAH's temperature estimations more effectively.\\

For each parameter and training set, we obtained one NN and one RF model. Users can utilize these models to estimate stellar parameters and abundances for new data obtained by S-PLUS that will be available at the S-PLUS cloud\footnote{\url{https://splus.cloud/}}. Input requirements include the 12 magnitudes with their respective errors for the determination of $T_\mathrm{eff}$ and $\log g$. For the remaining parameters, users can employ the ABC approach. For the B approach, 66 colors plus $T_\mathrm{eff}$ values are required, while for the C approach, $\log g$ values are also necessary.\\

We also investigated the correlation of the abundance ratios determined in this study with temperature and $\log g$, but found no strong correlation between them. The ability of the NN and RF methods to uncover non-linear relationships among these parameters, coupled with the abundance ratios falling within the S-PLUS filters, enhances the reliability of the estimated elemental abundances. However, to provide a cautionary note regarding the parameter estimates, the latter were flagged as follows: Flag 00: Stellar parameters or elemental abundances with a goodness-of-fit higher than 60\% in all approaches and methods used in this work. Flag 01: Elemental abundances of [Mg/Fe] and [$\alpha$/M] within S-PLUS filters that have a goodness-of-fit below 60\%. These abundance ratios were also flagged as 00 using another sample. Flag 02: Elemental abundances of [Cu/Fe], [O/Fe], and [Si/Fe] with a goodness-of-fit above 50\% and below 60\% that were not flagged as 00 or 01. Such abundances do not match with higher accuracy in any other tests using different samples. Given the lower accuracy, they should be used with extra caution.

\subsection{S-PLUS Parameter Estimation}
\label{Sec:EstimationsSPLUS}

The initial sample comprises approximately $5$ million stars observed through the $12$ filters. Subsequently, this sample was crossmatched with {\it Gaia} DR3 data (using a search radius of two arcseconds) to obtain {\it Gaia} $\log g$ and $T_\mathrm{eff}$. This procedure facilitates the classification of stars as giants or dwarfs. Stars near the classification thresholds or parameter limits should be interpreted with caution. However, in addition to this step, we also determine $\log g$ and $T_\mathrm{eff}$ solely based on the S-PLUS data. \\

\begin{figure*}
\centering
\includegraphics[width=0.45\textwidth,height=0.23\textheight]{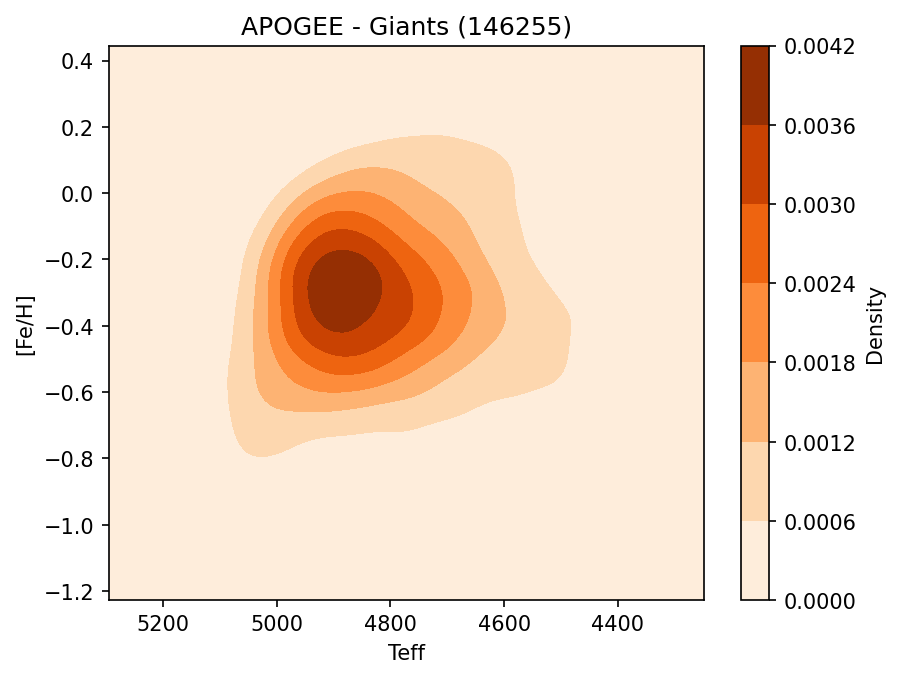} 
\includegraphics[width=0.45\textwidth,height=0.23\textheight]{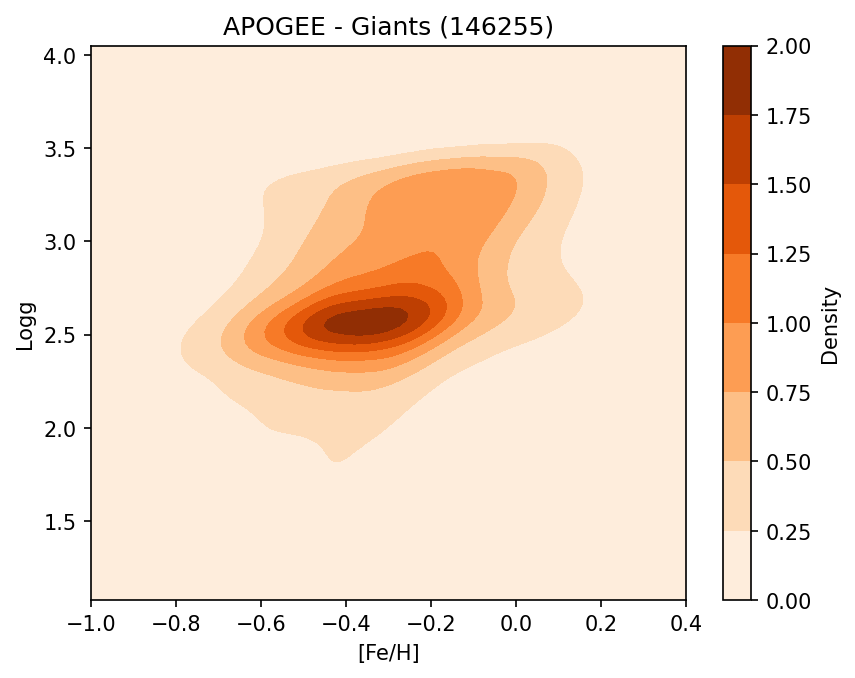}

\includegraphics[width=0.45\textwidth,height=0.23\textheight]{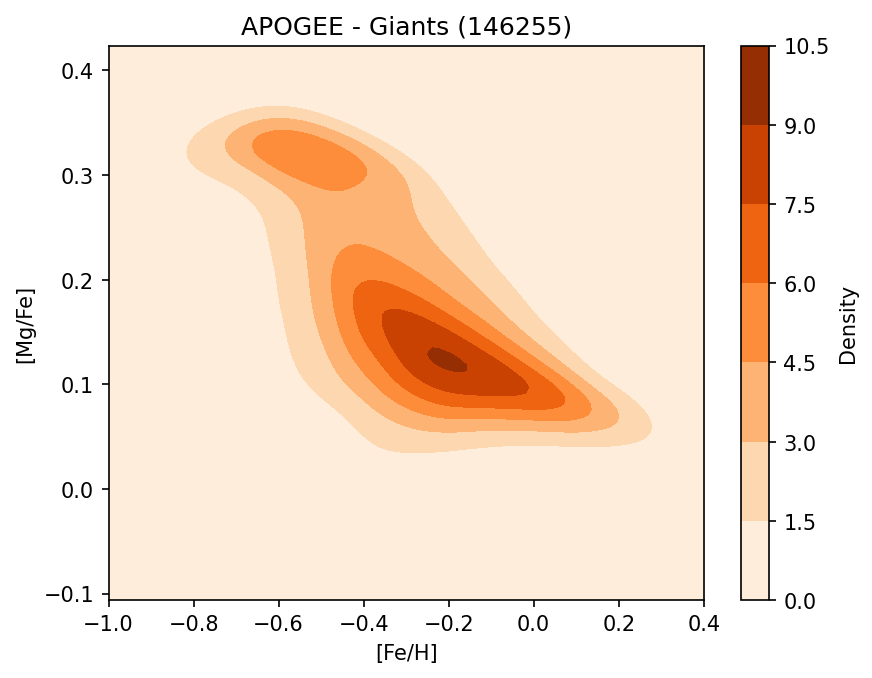} 
\includegraphics[width=0.45\textwidth,height=0.23\textheight]{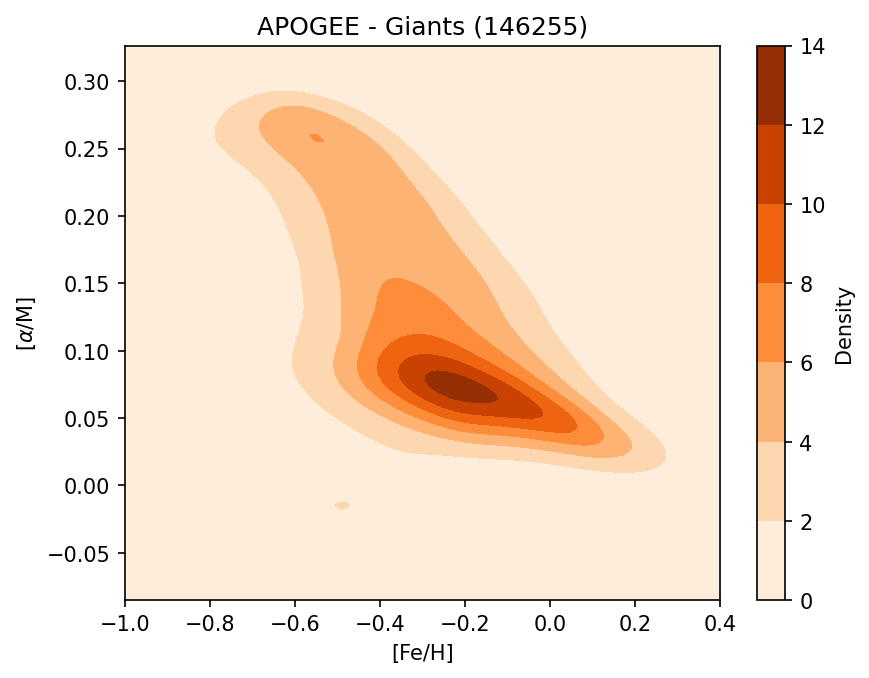} 

\includegraphics[width=0.45\textwidth,height=0.24\textheight]{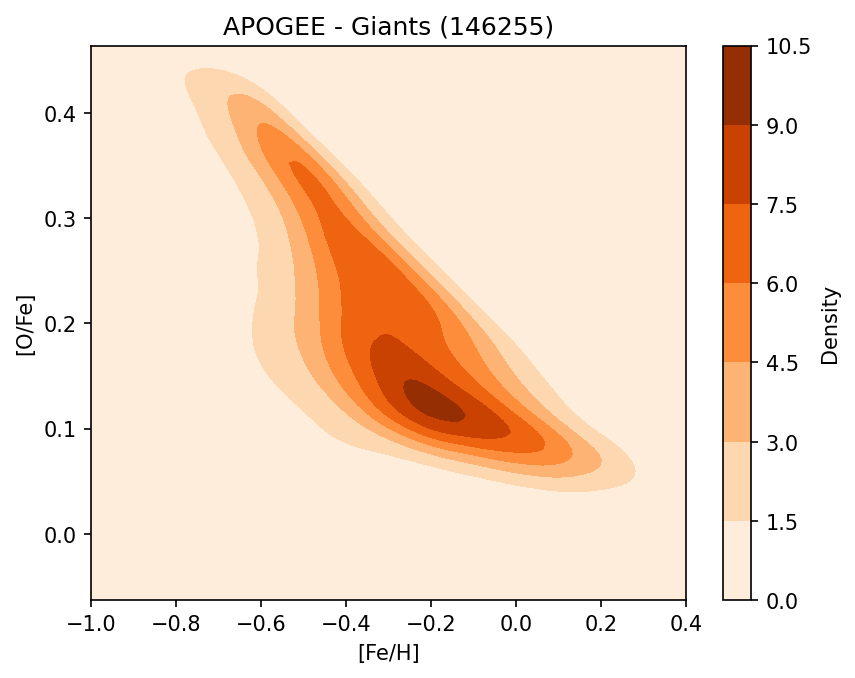} 
\includegraphics[width=0.45\textwidth,height=0.24\textheight]{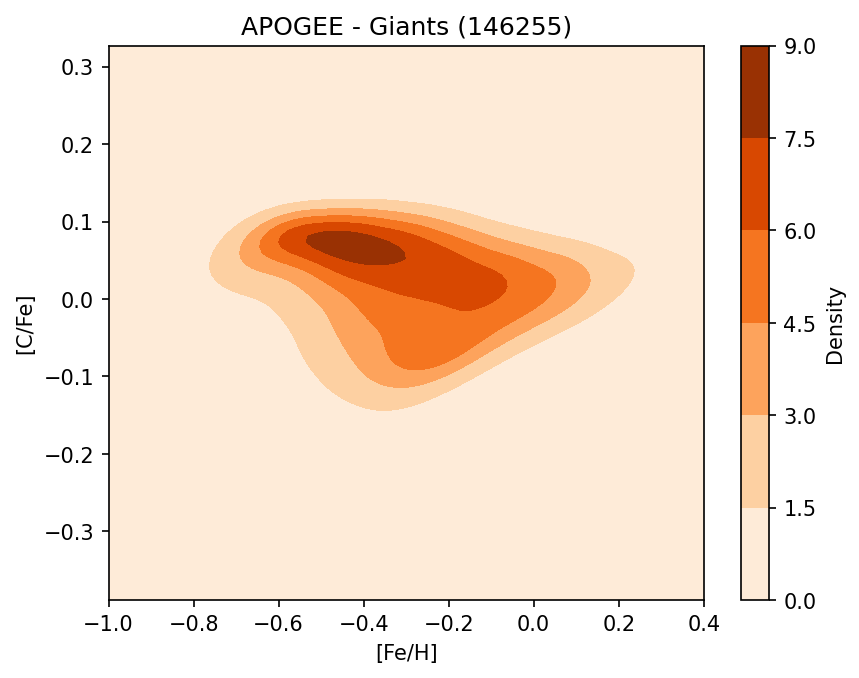} 

\includegraphics[width=0.45\textwidth,height=0.24\textheight]{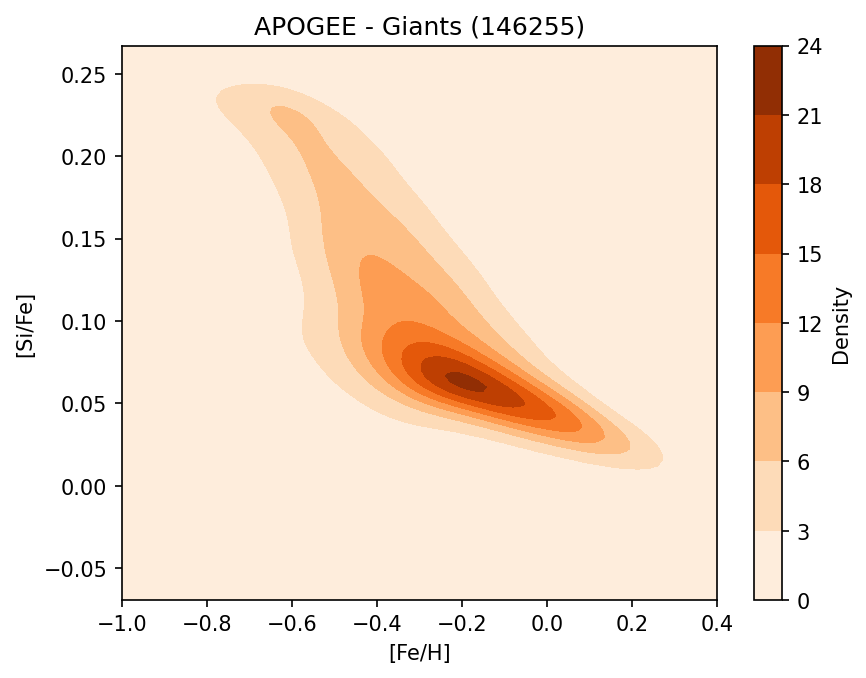} 
\includegraphics[width=0.45\textwidth,height=0.24\textheight]{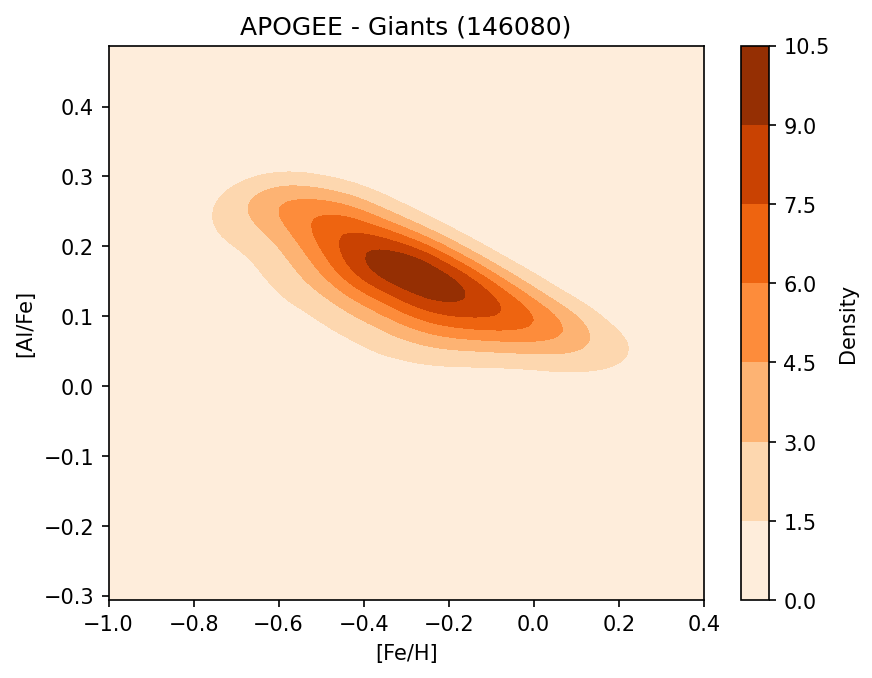} 

\caption{Density plot depicting the pairwise distribution of different atmospheric parameters and chemical abundances for a total of $\sim140\,000$ giant stars predicted in this study, based on the APOGEE based trainning set. The title of each plot indicates the exact number of stars plotted, while the color bar represents the density distribution.}
\label{fig:ParameterALLDistribution}
\end{figure*}
To handle data that fall outside the range of our training set, we apply a normalization method where values below the minimum threshold are set to zero, and values above the maximum threshold are set to one. This method can introduce biases for sources that lie outside these thresholds, so it should be used with caution. To assess the extent of this issue, we introduce a flag feature (FF) that indicates the proportion of features for each source that fall outside the training set limits. The FF value ranges from 0\% to 100\%, representing the percentage of parameters that exceed the established training set limits. \\

Table \ref{table:IDR4Prediction} displays the number of sources processed using our methodology, comprising $\sim140$ thousand giant stars and $\sim4.9$ million dwarf stars. Remarkably, the abundance ratios and stellar parameters of about $\sim90\%$ for giant stars and $\sim75\%$ of dwarf stars fall within the constraints of the training sets. While the optimal outcome is achieved when all features are within the training set limits, cases where missing features are deemed unimportant for determination may still yield acceptable results, particularly for sources with lower flag numbers.

\subsection{Catalog description}\label{Sec:CatalogDesciption}

Appendix \ref{appendixG} describes the stellar parameters and abundances ratios provided in this work and instructions on how to access them. The catalog primarily consists of the 12 S-PLUS magnitudes along with their respective errors, in addition to the abundance ratios estimated, each accompanied by a flag indicating the number of features inside the training set limits (see Section \ref{Sec:EstimationsSPLUS}). Figure \ref{fig:ParameterALLDistribution} displays the distribution of the predicted parameters for APOGEE data pertaining to giant stars. 
$T_\mathrm{eff}$ and $\log g$ do not exhibit any correlation with metallicity. Since these parameters are mainly determined by stellar characteristics unrelated to metallicity, such as stellar evolution and surface properties, the lack of correlation between $T_{\text{eff}}$ and metallicity is not expected. The apparent lack of correlation between $T_{\text{eff}}$, $\log(g)$, and metallicity in the giant star sample may be explained by the inclusion of both red clump and red giant branch stars. Red giant branch stars exhibit a metallicity dependence, particularly from one side of the branch to the other \citep{Salaris_2002}, but this trend can be obscured by the more stable properties of red clump stars \citep{Pietrinferni_2004}, resulting in a diminished overall correlation. \\
    
The plot of [Mg/Fe] as a function of [Fe/H] displays a bimodality. The bimodal distribution observed in the [Mg/Fe] plot may indicate distinct stellar populations or formation mechanisms. One possible explanation is the presence of two populations of stars with different enrichment histories: one formed from pristine gas with low [Fe/H] (stars at the MW thick disk) and another formed from gas enriched by previous generations of stars (stars at the MW tin disk). The weaker bimodality observed, particularly for the APOGEE sample, can be attributed to the fact that the S-PLUS survey primarily observed stars from the thin disk, with a lower representation of thick disk stars. Furthermore, only stars with observations in all 12 filters were included, which further reduces the sample size, particularly for thick disk populations. Such bimodality was recently revealed using APOGEE data \citep[][]{RojasArriagada-2019,Horta:2023}, indicating the consistency of our results.\\

$[\alpha/M]$, [Si/Fe], and [O/Fe] exhibit a weak correlation with [Fe/H]. The weak correlation between these elements and [Fe/H] may be attributed to their diverse formation processes and sensitivity to environmental conditions. For instance, \text{H{$\alpha$}} emission is influenced by factors such as stellar activity and gas dynamics in the interstellar medium, which may not directly correlate with [Fe/H]. This lack of correlation, as noted, could be related to different formation processes. Similarly, [Si/Fe] and [O/Fe] are produced through different nucleosynthesis pathways leading to varying degrees of sensitivity to [Fe/H] across different stellar populations \citep[e.g.,][]{Snaith:2014,Amarsi:2019,Rossi:2024}.\\

[C/Fe] displays a very weak or no correlation with [Fe/H]. Similarly to the previous topic, the lack of correlation between these elements and [Fe/H] can be attributed to their complex nucleosynthetic origins and dependence on multiple astrophysical processes. [C/Fe], for example, can be produced through both stellar nucleosynthesis and nucleosynthesis in supernova explosions, leading to a wide range of carbon abundances across different stellar populations.

\begin{figure*}
\centering
\includegraphics[width=0.48\textwidth,height=0.22\textheight]{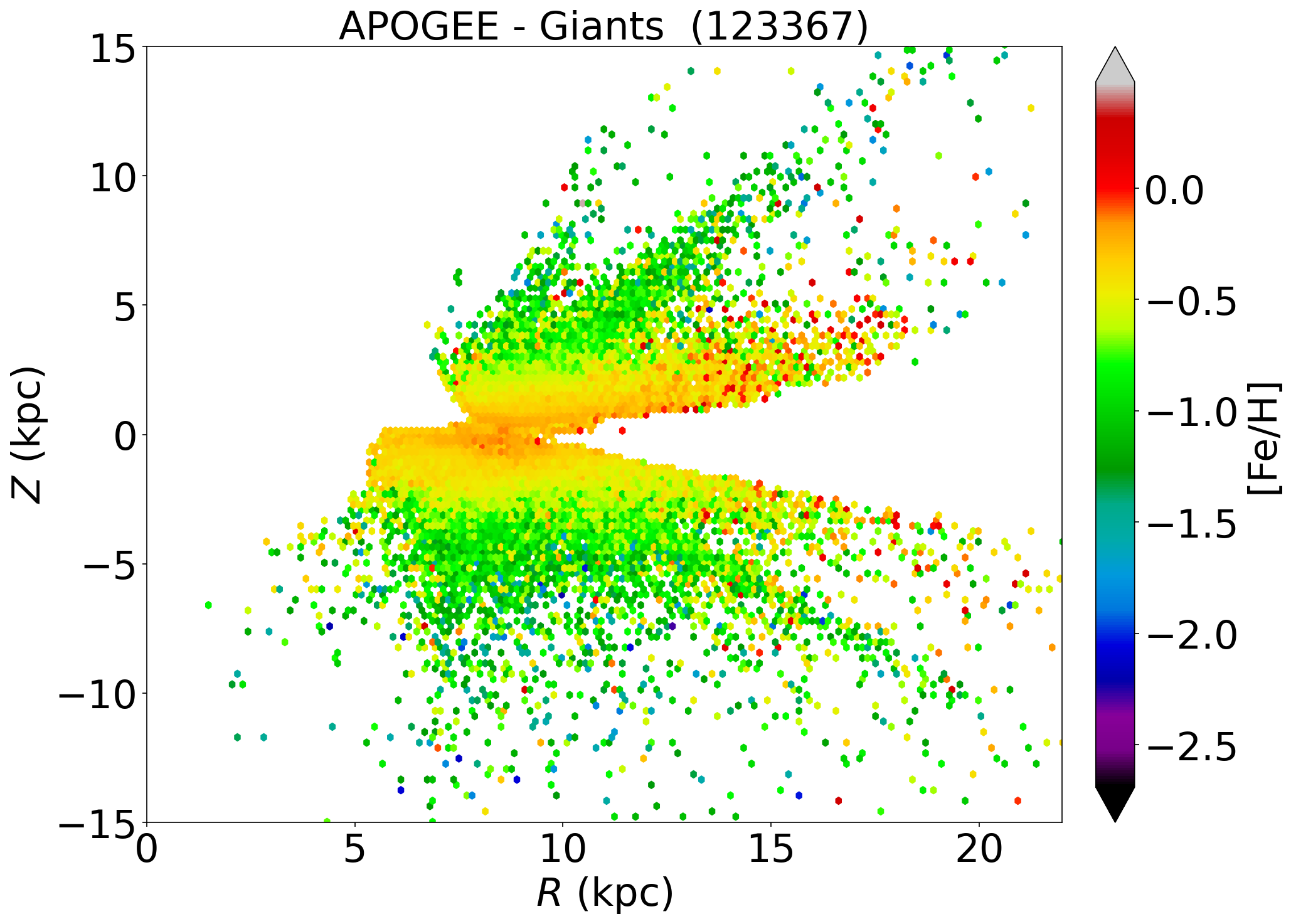} 
\includegraphics[width=0.48\textwidth,height=0.22\textheight]{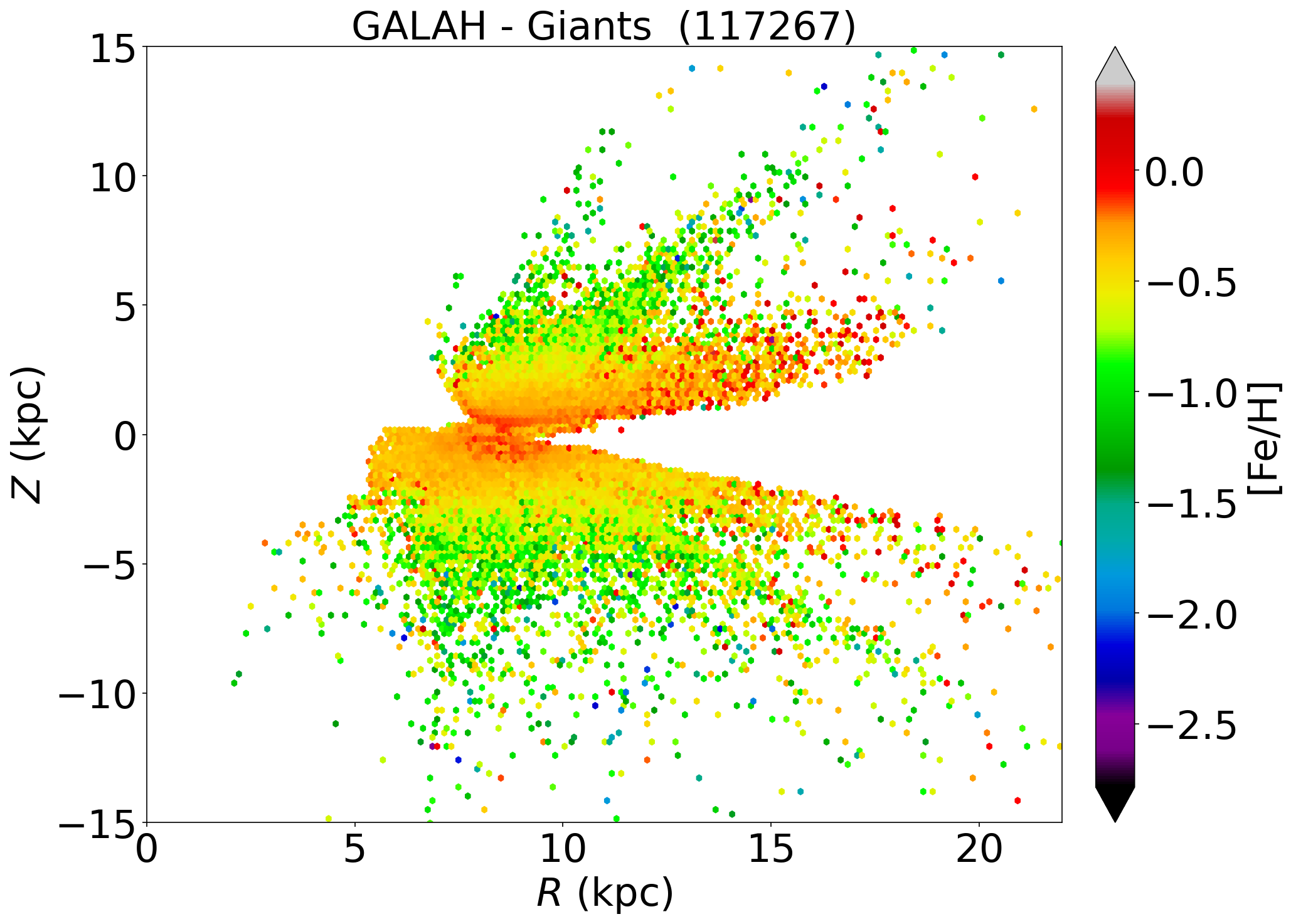} 
\includegraphics[width=0.48\textwidth,height=0.22\textheight]{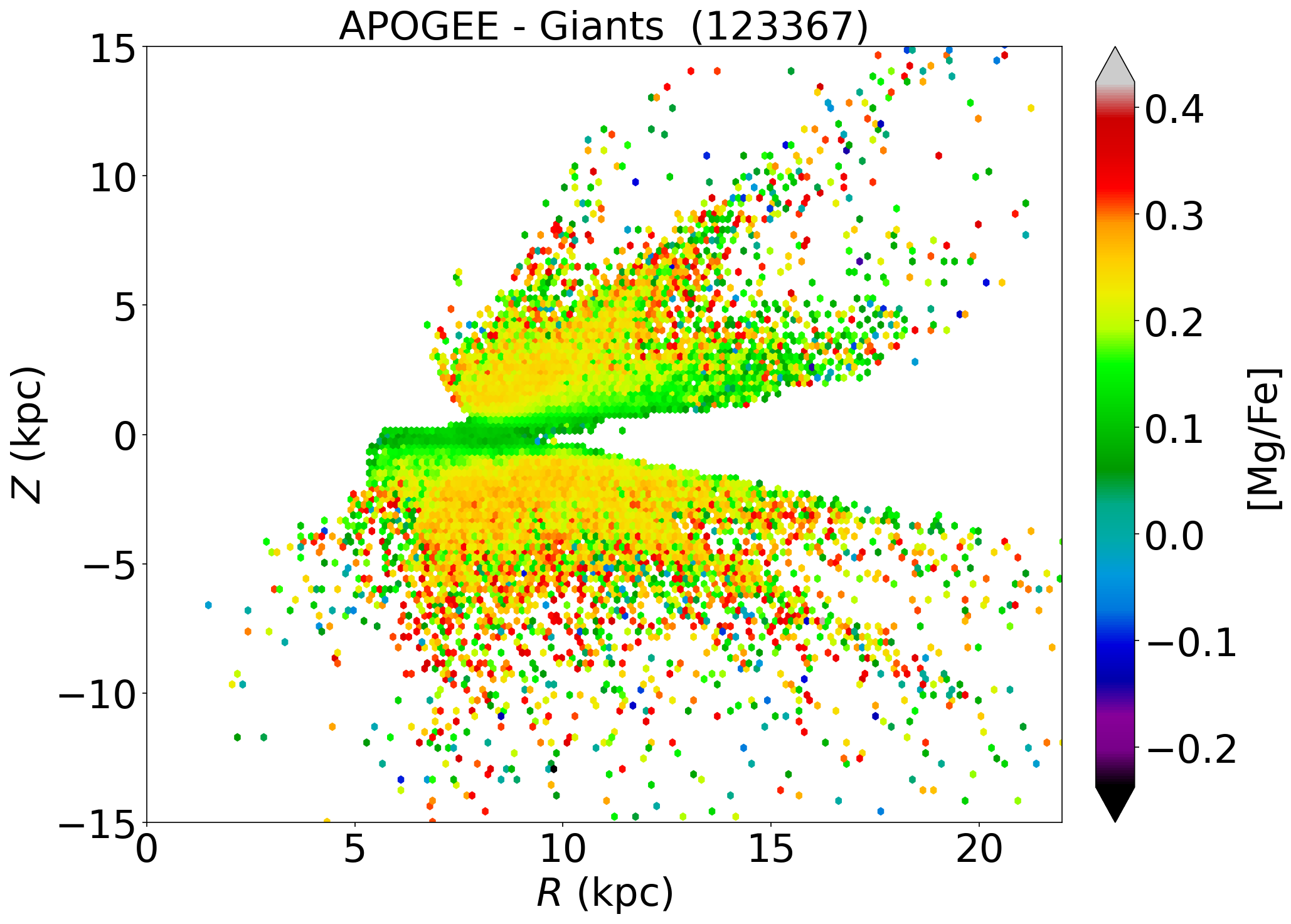} 
\includegraphics[width=0.48\textwidth,height=0.22\textheight]{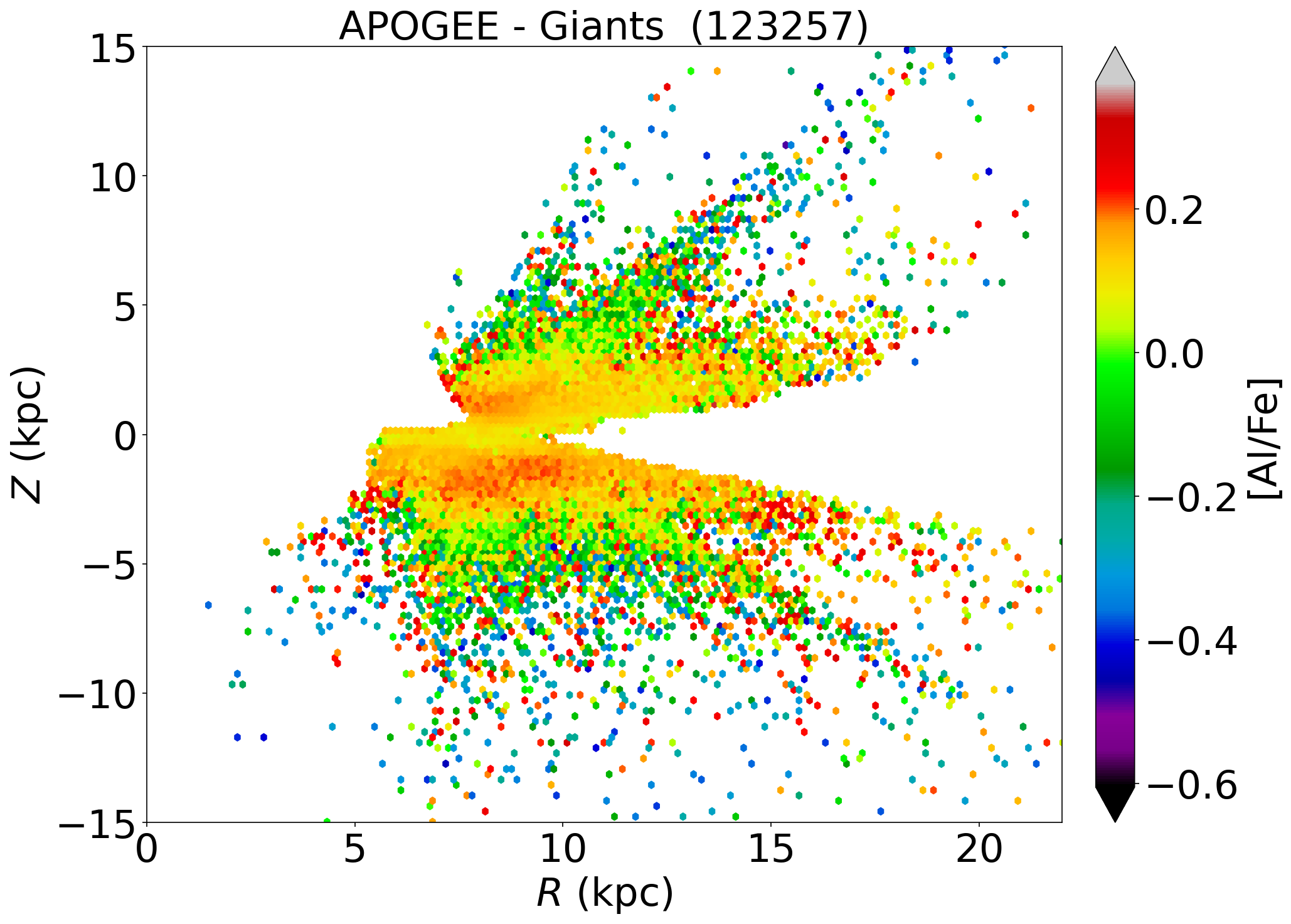} 
\includegraphics[width=0.48\textwidth,height=0.22\textheight]{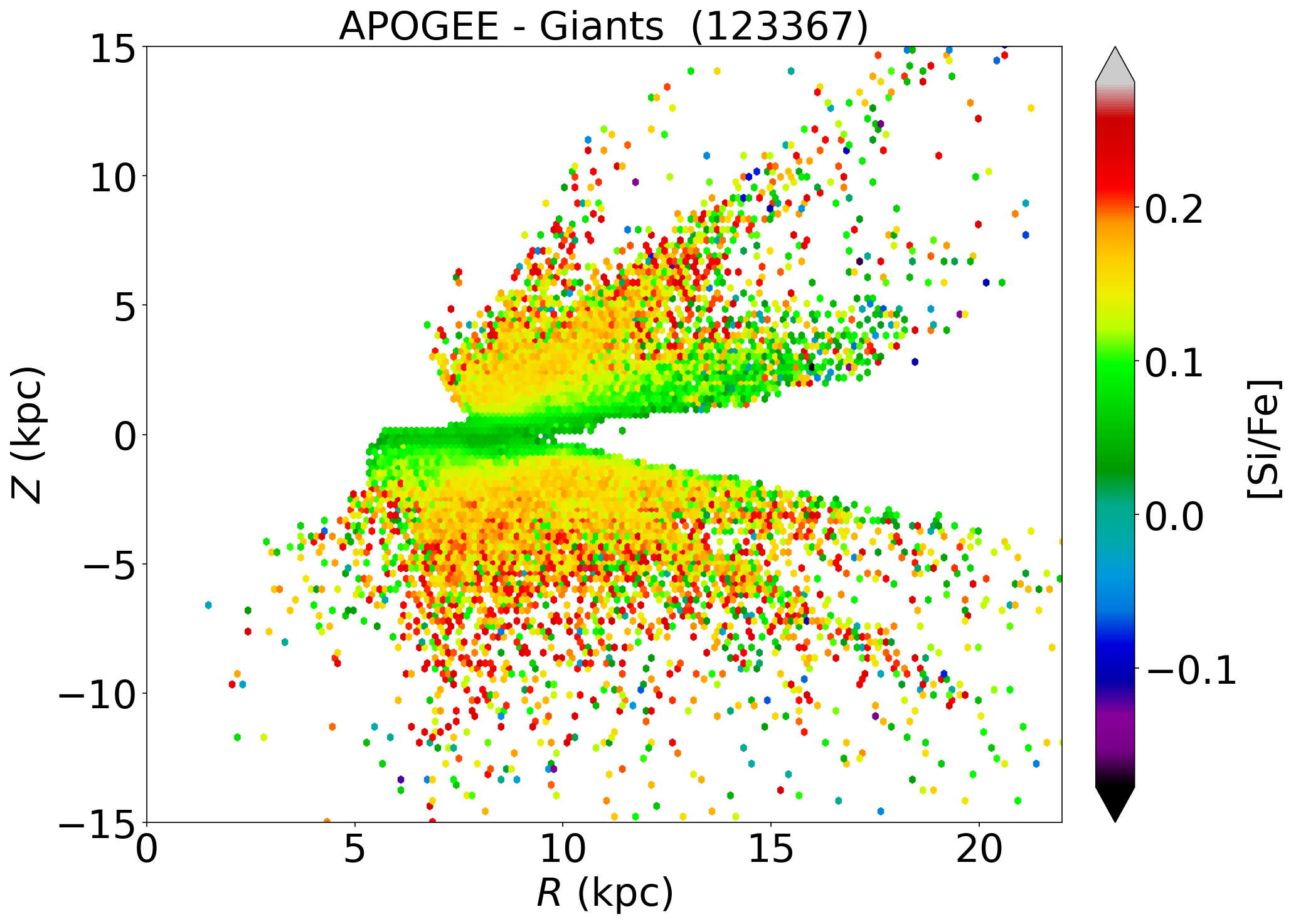} 
\includegraphics[width=0.48\textwidth,height=0.22\textheight]{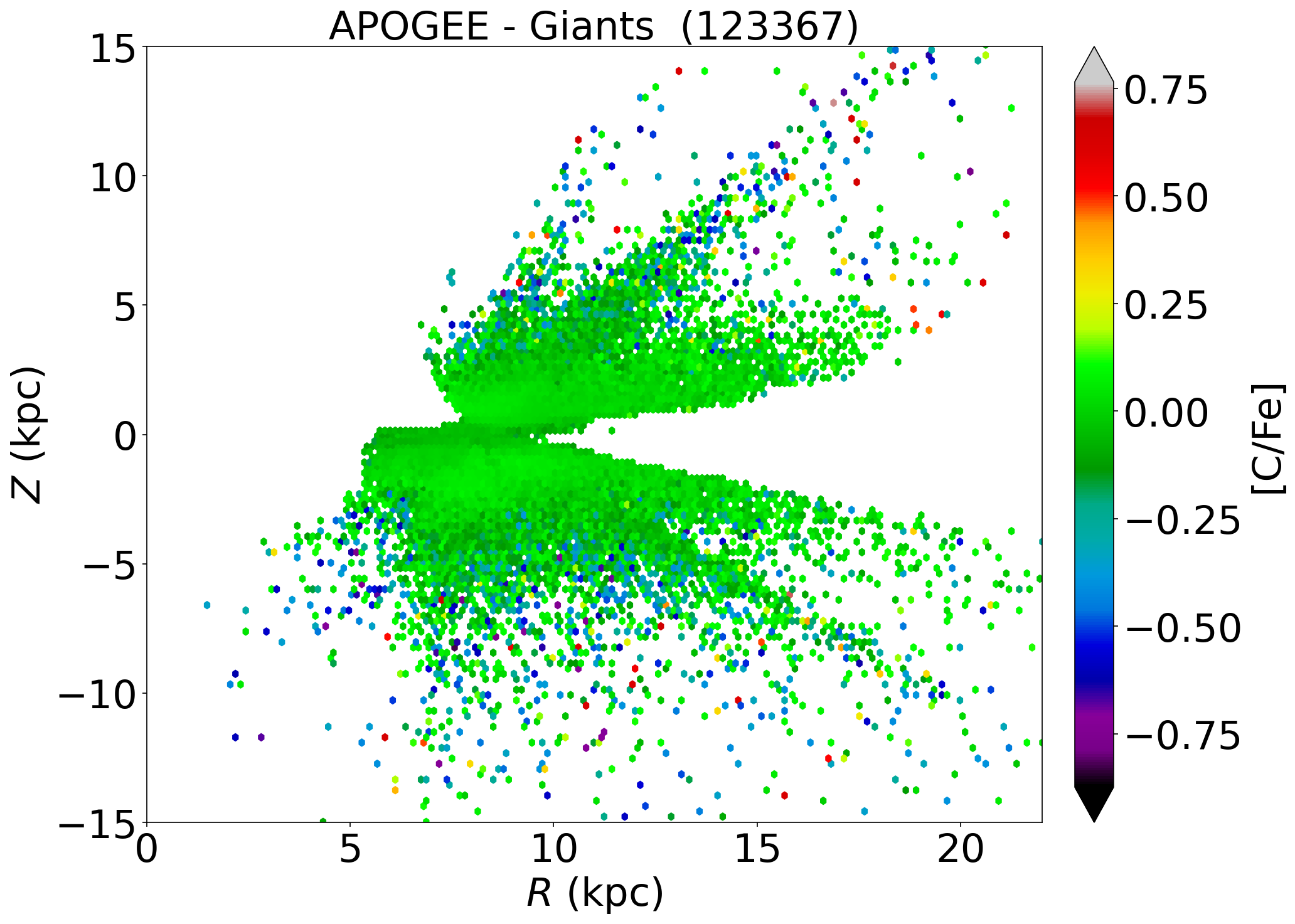} 
\includegraphics[width=0.48\textwidth,height=0.22\textheight]{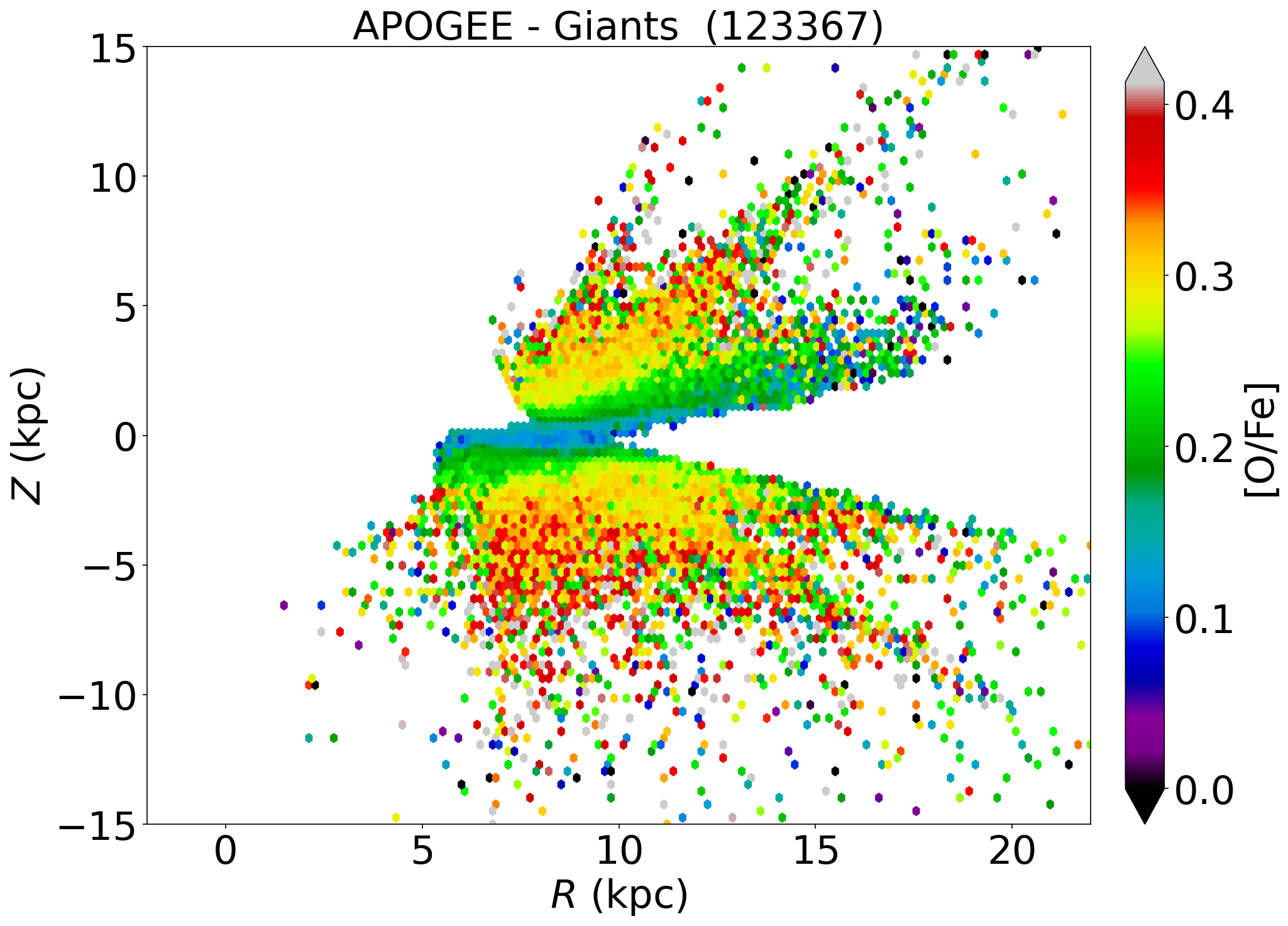}
\includegraphics[width=0.48\textwidth,height=0.22\textheight]{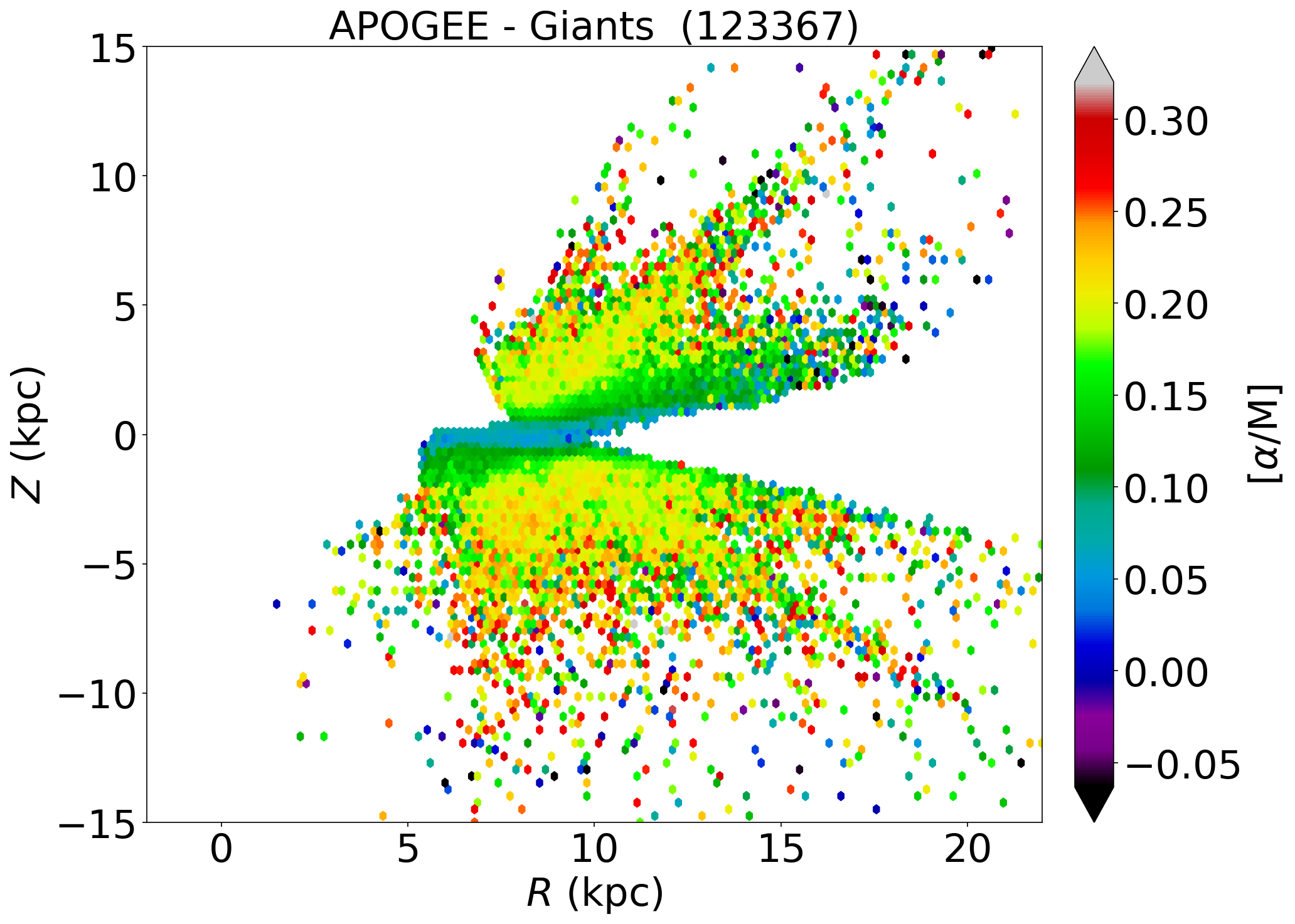}

\caption{Spatial distribution of Milky Way giant stars and their various chemical abundances calculated in this study. In each plot, radial distance (R) in kpc from the center of the Milky Way is plotted along the abscissa, and the Galactic vertical height (z) is plotted along the ordinate. The colorbar represents the range of various calculated chemical abundances obtained in this work, illustrating how these abundances vary across different Galactic positions. The title of each plot indicates the training data set used to calculate the respective chemical abundance and the number of giant stars plotted.}
\label{fig:DistanceDistributionSky}
\end{figure*}

A thorough analysis of abundance ratio correlations requires addressing the literature where these trends have been extensively discussed. Such a discussion is beyond the scope of this work that aims to release the parameters. A forthcoming paper will address a chemo-dynamical description of the Milky Way. We also utilize {\it Gaia} DR3 distances to obtain an overview of the distribution of the parameters across the sky. Figure \ref{fig:DistanceDistributionSky} displays the sky distribution of our results where the cylindrical galactocentric distance is denoted as $R$, and the distance to the Galactic plane is $Z$:

\begin{itemize}
    \item $\mathrm{[Fe/H]}$: The iron abundance reveals a clear gradient in the Milky Way. The [Fe/H] values range from approximately $0.5$ to values below $-2.5$, indicating a decrease in metallicity from the disk to the outer regions of the Galaxy. This metallicity gradient is consistent with previous studies that have shown the prevalence of metal-rich stars in the inner regions of the Galaxy and metal-poor stars in the outer regions \citep[e.g.,][]{Hayden-2015,GaiaCollaborationQuemical:2023} \\
    \item $\mathrm{[Mg/Fe]}$: Trends for this abundance range change with galactocentric distance. In the disk region, it remains relatively constant at around $0.1$, indicating a balance between Type Ia and Type II supernova enrichment. However, as we move towards the outer regions (higher Z values), [Mg/Fe] shows a gradual increase, reaching approximately $0.3$. This trend suggests a higher contribution from Type II supernovae or a lower contribution from Type Ia supernovae in the outer regions, resulting in enhanced magnesium enrichment \citep[e.g.,][]{Nidever-2014,Hayden-2015}.\\
    \item $\mathrm{[Al/Fe]}$: The distribution of aluminum abundance exhibits complex behavior across different galactocentric distances. At the Galactic center, [Al/Fe] is approximately $0.1$, indicating a relatively low abundance compared to iron. However, higher Z values around $3$ kpc, [Al/Fe] increases to around $0.2$. Interestingly, for even higher Z values, drops below $-0.1$, suggesting a decrease in aluminum abundance in the outer regions of the Galaxy. These variations may reflect different enrichment processes and stellar populations in different regions of the Milky Way \citep[e.g.,][]{Magrini-2021}. Indeed, the current assumptions are based on a broad overview but the current sample primarily represents stars from the thin and thick disks, with other regions being less represented. A more detailed analysis, particularly focusing on halo stars and streams, will be addressed in future work.\\
    \item $\mathrm{[Si/Fe]}$: Close to the Galactic center, the silicon abundance compared to iron is around 0.05. For Z around 1 kpc, it increases to approximately 0.15, trend which is accentuated in a way that  for even higher Z values, where  [Si/Fe] exceeds 0.25. This behavior suggests a higher contribution from Type II supernovae or silicon-rich stellar populations in the outer regions \citep[e.g.,][]{Hogg-2016}.\\
    \item $\mathrm{[C/Fe]}$: This abundance ratio displays significant variations with galactocentric distance, ranging from $-0.1$ at the center to around $0.05$ at a $Z$-height of about 1 kpc. Remarkably, at even higher $Z$ values, [C/Fe] drops below $-0.15$, indicating a decrease in carbon abundance in the outer regions of the Galaxy. These variations may be attributed to different enrichment processes and stellar populations \citep[e.g.,][]{Rossi:2024}.\\
    \item $\mathrm{[\alpha/M]}$: The abundance ratio of $\alpha$ elements to iron displays consistent trends across different galactocentric distances. Initially, close to the Galactic center, [$\alpha$/M] is approximately 0.05. However, as we move towards outer regions, [$\alpha$/M] exhibits a linear increase, reaching values around $0.2$. Beyond Z values of $5$ kpc, [$\alpha$/M] shows scatter. These trends reflect the varying contributions of different nucleosynthesis processes and stellar populations in different regions of the Milky Way \citep[e.g.,][]{Anders-2014}.\\
    \item $\mathrm{[O/Fe]}$: As expected, the distribution of oxygen abundance exhibits similar trends to [$\alpha$/Fe], albeit with a different range of values. Like [$\alpha$/Fe], [O/Fe] shows an increase from the Galactic center towards outer regions, reflecting the enrichment history of the Galaxy. However, [O/Fe] values may differ due to the specific nucleosynthesis processes involved in oxygen production and the different chemical evolution pathways of oxygen compared to other $\alpha$ elements \citep[e.g.,][]{Nissen-2014}.
\end{itemize}

In summary, the observed distributions of elemental abundances align with the chemical evolution and enrichment history of the Milky Way Galaxy. However, further investigations are essential, including targeted spectroscopic follow-up on selected objects to refine parameter determinations to assure the higher accuracy.  Potential spectroscopic follow-up could greatly enhance our understanding of specific stellar populations, including halo stars, globular clusters, and metal-poor stars. This would provide more detailed insights into their chemical compositions and evolutionary histories. The current study's framework also allows for expansion into these regions using photometric data, with broader studies including the halo being feasible without immediate spectroscopy. The parameters estimated in this study reproduce the global trends for the MW, in well agreement with the existing knowledge. This also holds the potential of expanding the method, without spectroscopy, to other studies. However, the integrating our with proper motion measurements from surveys like Gaia can enhance target selection for further studies of star clusters and other stellar populations.

\begin{figure}
\centering
\includegraphics[width=0.45\textwidth]{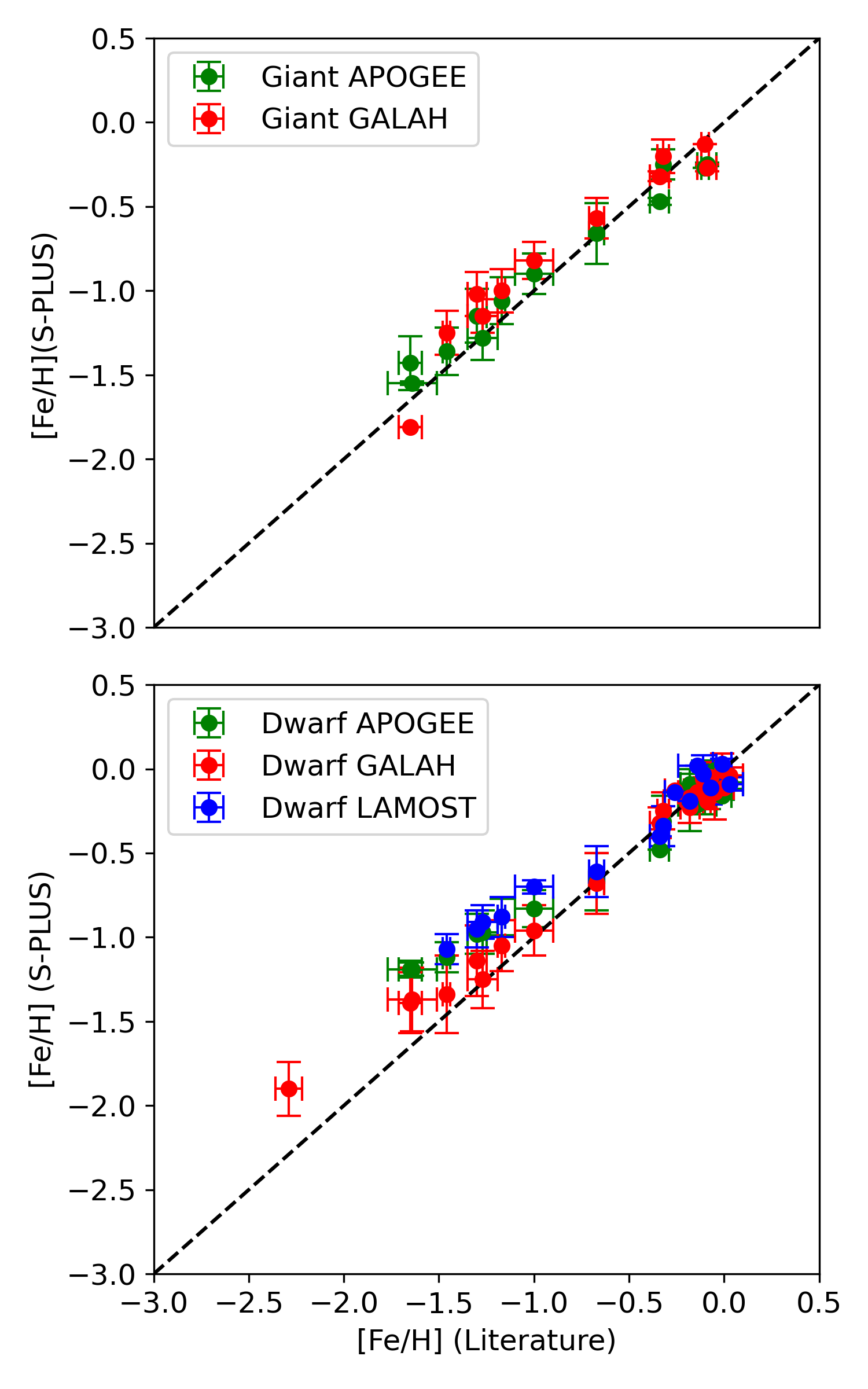} 

\caption{Comparison of metallicity ([Fe/H]) values from the literature (x-axis) with the estimations obtained in this work (y-axis) for the selected star clusters.}
\label{fig:FeHClustersComparisson02}
\end{figure}

\subsection{Star clusters}
\label{Sec:GlobularCluster}

We performed a crossmatch with star clusters to evaluate the accuracy of our derived parameters. The catalogs compiled by \citet{Vasiliev-2021} and \citet{Hunt-2023} include 170 globular clusters and 7167 open clusters, providing membership probabilities that indicate the likelihood of a star being a member of a specific cluster. These studies generate updated lists based on refined models and parameters, meaning that the membership probabilities may vary with new data or methodologies \citep[e.g.][]{Galleti-2006,AlonsoGarcia2021,Dias-2021}. For our analysis, we initially considered only stars with a membership probability greater than $0.95$, using a search radius of one arcsecond to identify cluster members. \\

A more detailed analysis was conducted, taking into account that star cluster members have been observed by GALAH, APOGEE, or LAMOST and should exhibit similar metallicity. For example, we identified two stars, 6816573828484824192 and 6817326448488737152, with metallicities of [Fe/H] $\simeq -0.04$ and [Fe/H] $\simeq -0.48$, respectively, as computed from GALAH spectra. These stars were previously considered members of NGC 7099, which has an estimated metallicity of [Fe/H] $\simeq -2.29$. Our own estimations yielded similar metallicities for these stars, i.e., [Fe/H] $\simeq -0.05$ and [Fe/H] $\simeq -0.41$. This discrepancy raises several issues, such as the possibility that our data may identify stars that do not belong to the cluster or highlight photometric problems that result in inaccurate metallicity estimations.\\

To analyze how closely the photometric metallicities align with the spectroscopic values, we focus only on stars whose metallicities are close to those reported in the literature, applying an error tolerance of $0.2 \times (1 + |[Fe/H]_{\text{Literature}}|)$. This method imposes a minimum error of 0.2 dex for stars with metallicity equal to zero. Such an assumption helps ensure the best membership assignment for the clusters. However, stars not included in this analysis may either not belong to the cluster, or have been accreted. Alternatively, our measurements may be under- or overestimated due to issues such as crowding. Appendix \ref{appendixH} presents the literature values, along with the total number of giants and dwarfs with membership probabilities greater than 0.95 $(N_{\text{all}})$. It also shows the number of giants $(N_G)$ and dwarfs $(N_D)$ used to compute our metallicity estimates, based on the criteria outlined above. The key findings are summarized below:

\begin{figure*}
\centering
\includegraphics[width=0.49\textwidth]{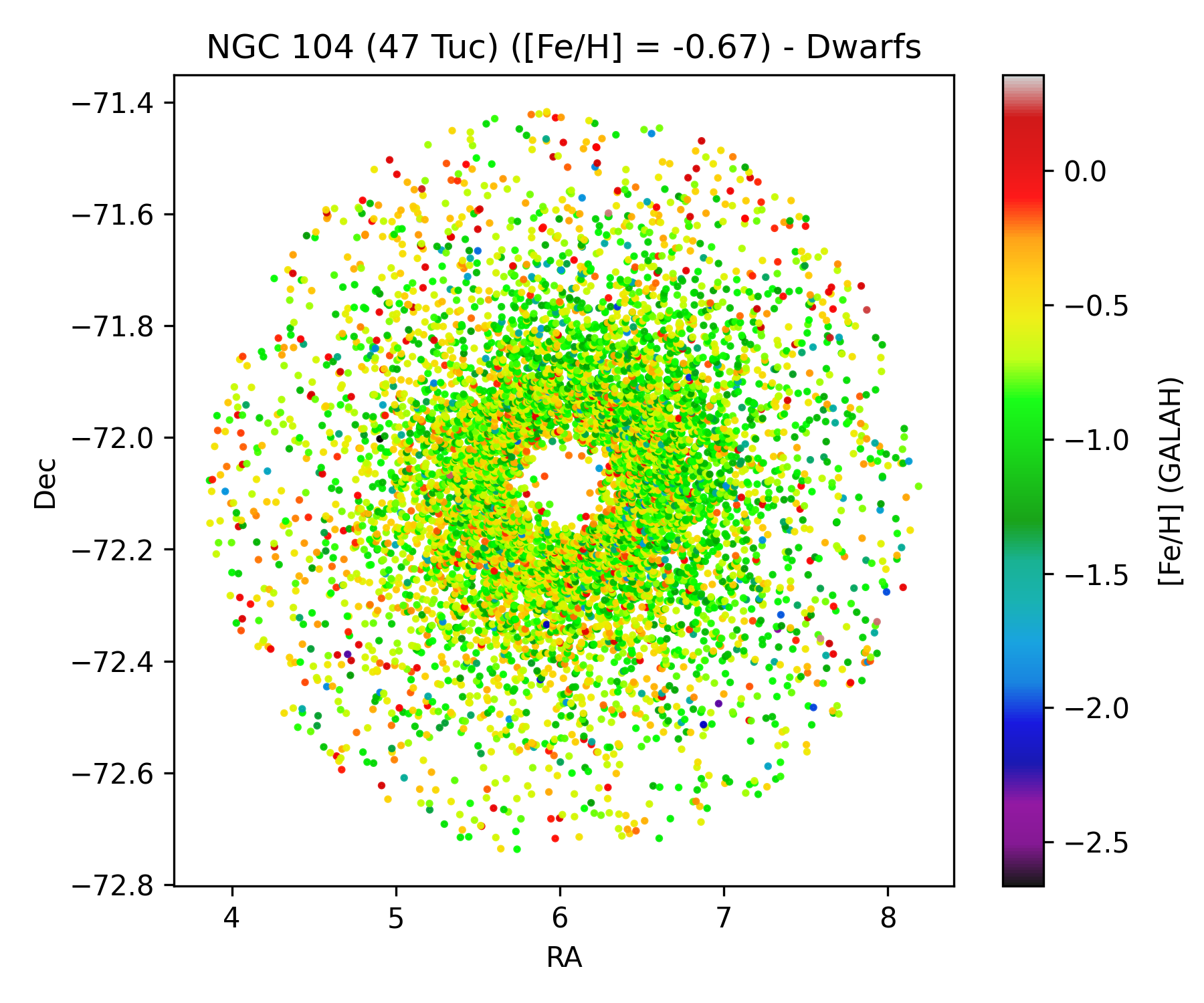} 
\includegraphics[width=0.49\textwidth]{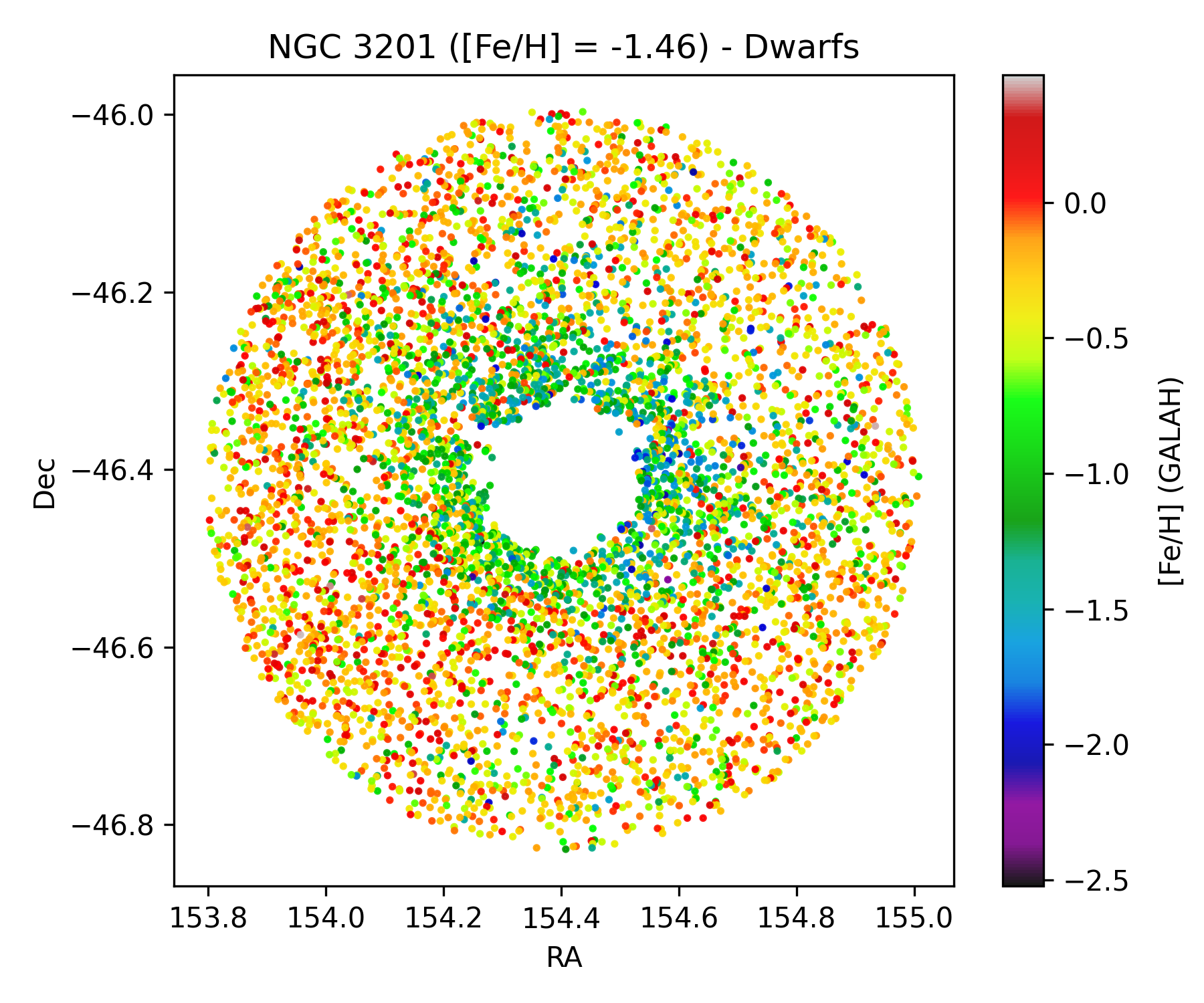} 
\caption{Sky plots for globular clusters NGC 104 (left panel) and NGC 3201 (right panel). The color bar in each plot indicates the [Fe/H] values determined in this study for each member star of the cluster, utilizing the GALAH data as training set.}
\label{fig:Clusters}
\end{figure*}

Figure \ref{fig:FeHClustersComparisson02} presents a comparison between the literature metallicities and those derived from our data. Nevertheless, the sigma values obtained from our estimates are smaller than those assumed in our criteria for selecting cluster members. The largest discrepancies are observed for metal-poor stars; however, the limited number of sources in this region prevents a deeper analysis.  Our results indicate that our estimates can reliably be used to identify member stars in star cluster analyses based on their metallicity values.\\

The proportion of stars with membership probabilities consistent with our estimated metallicity ranges varies from approximately 15\% to 100\%. This variation may arise from several factors, including the possibility that many stars do not belong to the cluster, as well as issues related to our estimations, such as photometric contamination, reddening, and stellar brightness. A more detailed analysis, focusing specifically on star clusters, will be presented in a forthcoming paper.\\
    
NGC 7099 (M30) is the star cluster with the lowest metallicity among the tested samples ($\sim-2.29$ dex), and 81 out of 339 stars ($\sim24\%$) classified as members have metallicity values consistent with our estimations \citep[][]{OMalley:2018}. Indeed, our estimated value is $\sim$0.3 dex higher than that found in the literature, suggesting that our method may overestimate the metallicity of metal-poor stars. Further investigation is required to understand the factors contributing to this discrepancy and to improve the accuracy of our estimations.\\
    
NGC 7492 and NGC 7089 have literature metallicity values of about $-1.65$ dex. While the results for giant stars from APOGEE align with these values, the metallicities for dwarf stars appear to be underestimated. On the other hand, this trend is also observed in other star clusters in our sample, suggesting that metallicity estimates for giant stars are generally more reliable than those for dwarf stars. Empirical adjustments between spectroscopic and photometric metallicities \citep[e.g.,][]{Calamida-2009, Lianou-2011, Bellazzini-2023} can be made to bring the results closer. However, applying such corrections could bias the estimates in cases where the original results are already in agreement (see the discussion on NGC 104 and NGC 3201).\\

NGC 3201 and NGC 288 star clusters have literature metallicities of $-1.46$, and $-1.30$, respectively. Our results can be used to assess membership probability based on the metallicity criterion. The values derived using the GALAH giant and APOGEE dwarf training sets are in agreement with previous estimations inside the error bars. However, our estimations indicate slightly higher metallicities for both clusters. The values derived using the APOGEE giant and GALAH dwarf training sets are more reliable compared to other methods. Based on these estimations, we find metallicities of $-1.35$ to $-1.15$ for NGC 3201 and NGC 288 respectively, both with similar sigma values of approximately  $0.15$~dex. Despite the consistency in these estimations, further research and analysis are necessary to fully understand the underlying mechanisms driving the observed differences in metallicity.\\
     
The star clusters NGC 1261, NGC 362, and Pal 12 show metallicity values consistent with those computed by us, primarily using GALAH and APOGEE as training sets. The results from LAMOST are in agreement but tend to fall at the lower limit of the literature values. Similar patterns are observed in other globular clusters when comparing with literature metallicities.\\
    
In NGC 104 (47 Tucanae), the literature value for [Fe/H] ($-0.67\pm0.04$) aligns well with the values derived from the three data samples.     NGC 104 has member stars with probabilities and metallicities close to the expected values for star clusters, ranging from 360 to 5,582. This means approximately 80\% of the stars are identified based only on the probability values. We observed a standard deviation of [Fe/H] smaller than 0.18 dex when compared to the literature. Remarkably, this cluster boasts the highest number of members among those studied, likely due to its proximity and one of the most massive GCs in the Milky Way, which facilitated a more comprehensive analysis in forthcoming work.\\

For star clusters with metallicities higher than $-0.5$ dex, we found good agreement between our estimates and values from the literature, with sigma values smaller than $-0.5$ dex. However, Ruprecht 4 stands out with a larger discrepancy: the literature reports a metallicity of about $-0.09$ dex, while our estimate is $-0.27$ dex. \citet{Carraro-2007} analyzed five stars in Ruprecht 4, finding that three of the brighter stars had metallicities around $0.00$ dex, while two fainter stars had values near $-0.35$ dex. The brighter stars were better aligned with the isochrones used by the author, leading to the adoption of the higher metallicity for the cluster. Since our study includes different stars than those analyzed in the literature, a re-evaluation of this cluster is necessary to better understand its true metallicity.\\

In general, our approach demonstrates that one can select member stars based on our outputs (see Fig. \ref{fig:FeHClustersComparisson02}, although the LAMOST models exhibit the least accuracy, with sigma values around 0.2 dex. As discussed earlier, this was necessary because some stars flagged as cluster members showed metallicities significantly different from the expected values for their clusters. To be cautious, we limited our analysis to stars with metallicities not too divergent from the cluster’s average, which introduced some bias into the results, even though the sigma values are smaller than those found in other studies. A more thorough investigation focusing exclusively on cluster stars will be conducted to clarify this issue. We also observed a notable discrepancy for metal-poor stars. Unfortunately, we were unable to find a counterpart for any of the star clusters analyzed in the LAMOST data, which prevents us from verifying whether these discrepancies are due to the signal-to-noise ratio limit we applied (SNR$ > 20$) or another factor. Indeed, the training set with the largest sample of metal-poor stars belongs to the GALAH data, in contrast to the LAMOST data (see Table \ref{table:MetalityInterval}). Consequently, the lower accuracy of [Fe/H] for LAMOST may be attributed to its limited sample of metal-poor stars, as well as the fiber size of 3 arcseconds, which makes it ill-suited for globular cluster analysis.  Based on these findings, we recommend that users rely on the metallicity estimates derived from GALAH data, as they are likely to be more accurate. \\

\begin{figure*}
\centering
\includegraphics[width=0.33\textwidth]{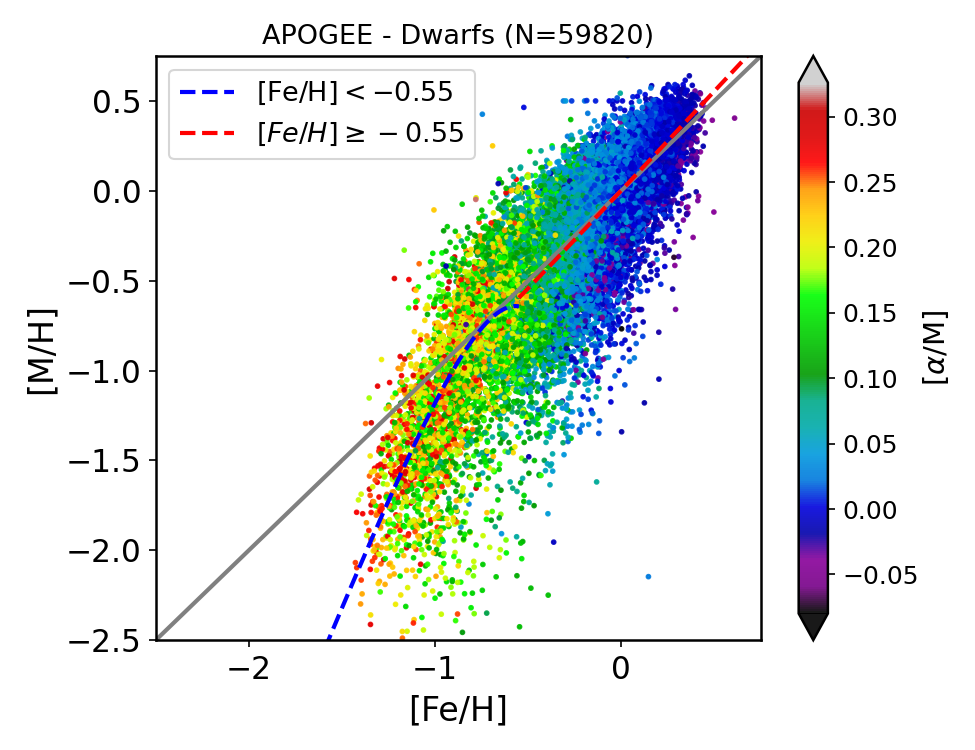} 
\includegraphics[width=0.33\textwidth]{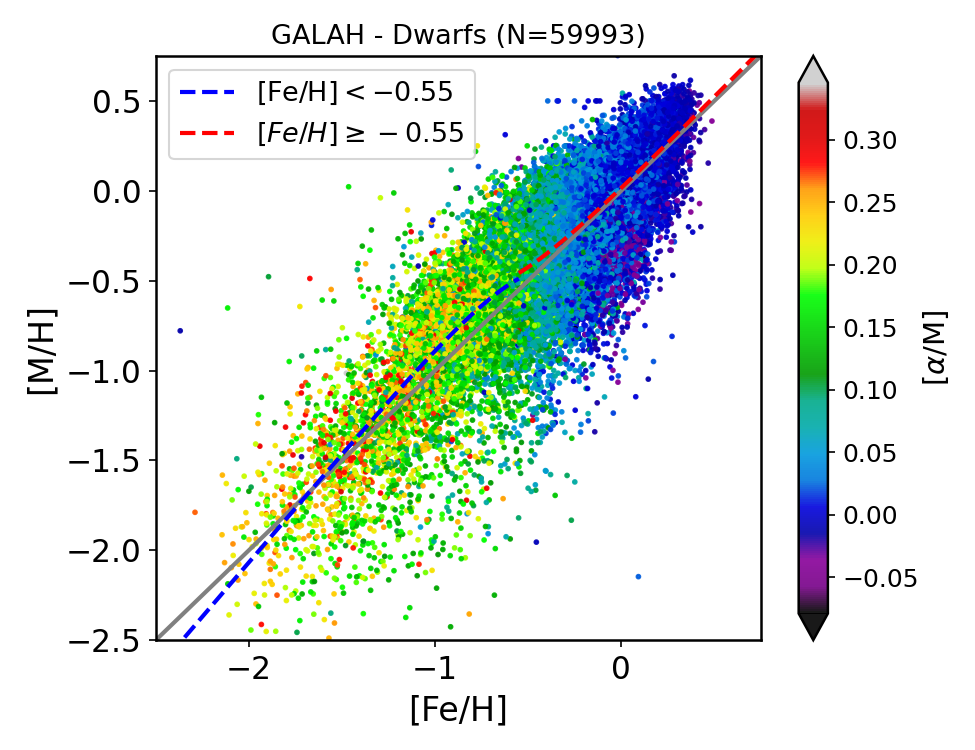} 
\includegraphics[width=0.33\textwidth]{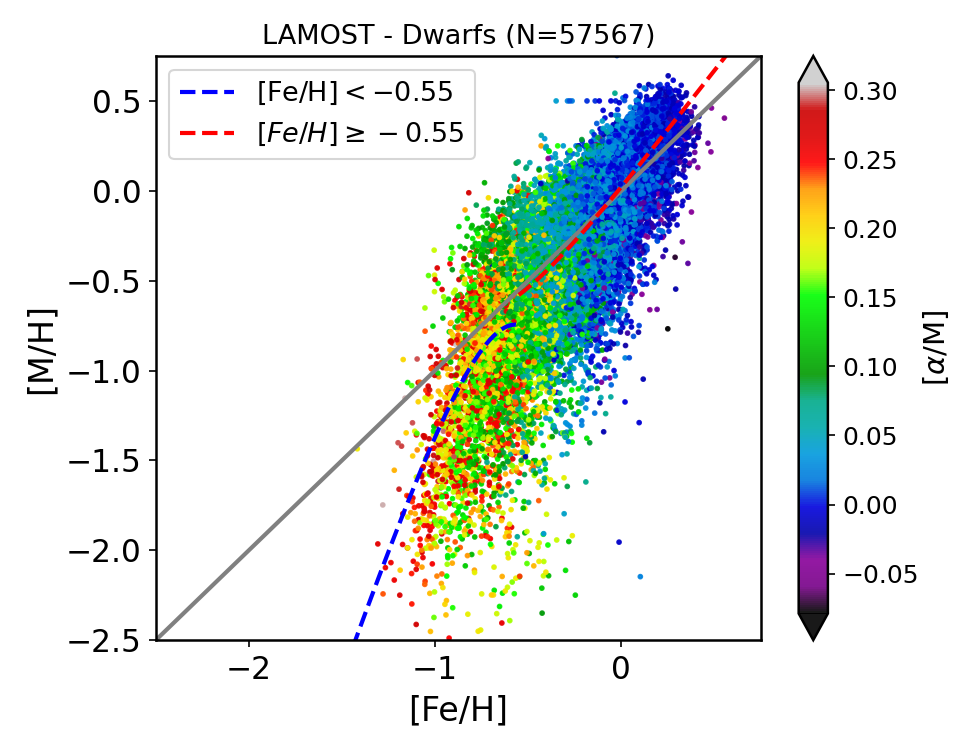} 

\includegraphics[width=0.33\textwidth]{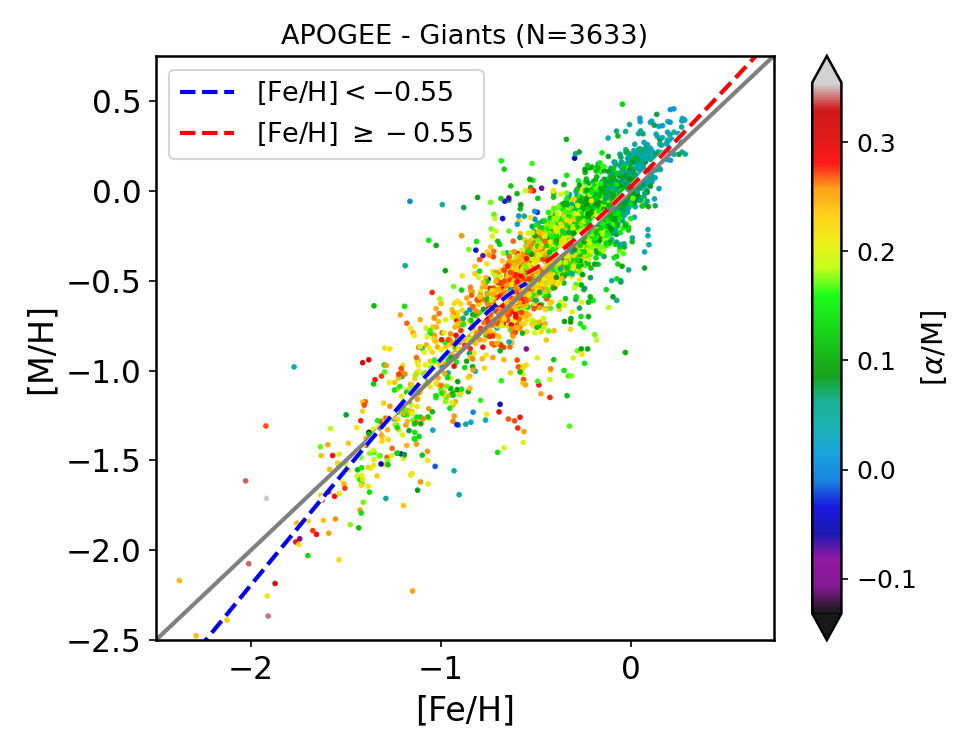} 
\includegraphics[width=0.33\textwidth]{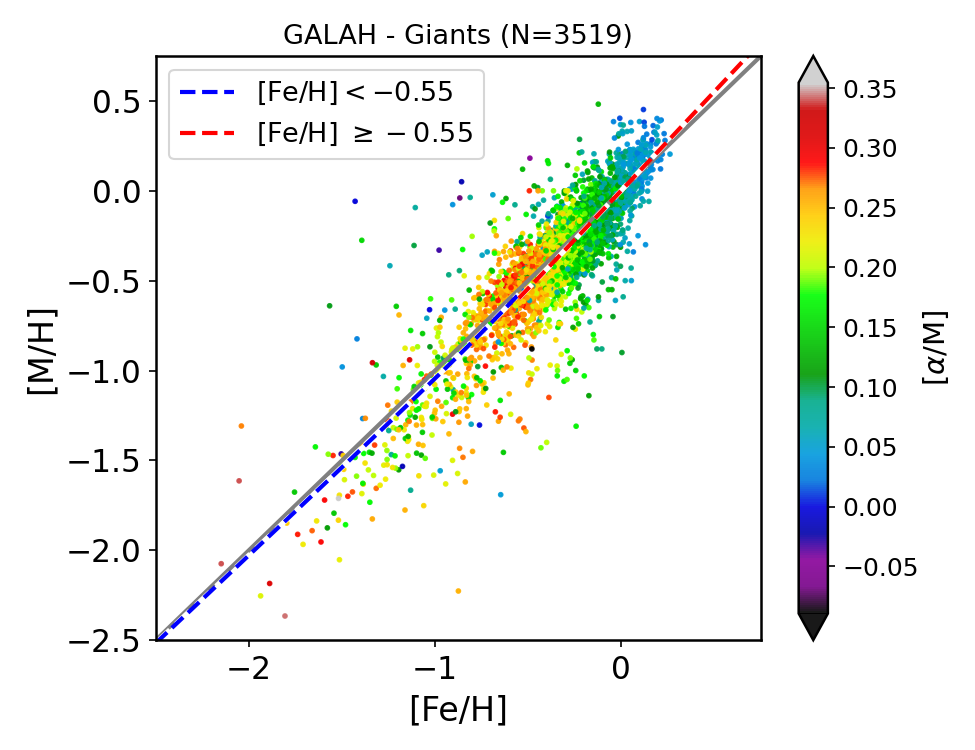}

\caption{Comparison of [M/H] vs. [Fe/H] for dwarf stars (top panels) and giant stars (bottom panels). The color gradient represents [$\alpha$/M], while the dashed and solid grey lines depict the linear fits and one-to-one relationship, respectively.}
\label{fig:tessinputcat}
\end{figure*}

The results for giant and dwarf stars derived from the GALAH and APOGEE training sets show better agreement compared to those from the LAMOST estimates. 
Typically, globular clusters are monometallic, meaning they exhibit a well-defined metallicity with narrow distributions of [Fe/H] \citep[][]{Harris:1996}. However, a particularly notable globular cluster is Omega Centauri ($\omega$ Cen), which is unique due to its significant spread in metallicity, ranging from approximately -2.2 to -0.6 dex \citep[][]{Gratton-2011}. Research into $\omega$ Cen has revealed multiple stellar populations, apart from typical globular clusters, making it a key focus in understanding the formation and evolution of globular clusters \citep[][]{Kraft-1994,Bedin-2004}. The metallicity estimates for entire clusters highlight intriguing targets for further investigation, particularly regarding potential metallicity variations and the structural properties of these clusters. For instance, in Fig. \ref{fig:Clusters} (left panel), the sky plot illustrates the distribution of [Fe/H] for NGC 104. Notably, the metallicity distribution of NGC 104 is quite homogeneous for all member sources. The no matched metallicity values suggests that some stars within these clusters may not belong to the cluster, or may have been accreted, or may have poor spectroscopic and/or photometric estimates. \\

On the other hand,  the metallicity distribution of NGC 3201 near the center of each cluster appears to be lower compared to the outer regions. Conversely, NGC 3201 exhibits a distinct metallicity gradient, with stars near the center being notably metal-poor, while those towards the outer regions show a scattering of more metal-rich stars.  The observed [Fe/H] spread may be influenced by several systematic factors. Aperture photometry, affected by crowding, likely introduces inaccuracies in flux measurements, which can be mitigated by reprocessing the data with PSF photometry. To address potential field star contamination, a crossmatch with Gaia DR3 is recommended, leveraging its precise astrometry to isolate true cluster members. Additionally, ensuring accurate photometric calibration and robust SED fitting procedures is crucial to avoid biases. By addressing these factors, we aim to determine if the observed abundance spread is genuine or an artifact of the analysis process. For the true multiple population star cluster, such results serves as a mere example of how available data can be leveraged to explore various scenarios and identify potential targets. However, comprehensive studies are imperative to substantiate these observations further \citep{Bonatto-2019}. \\

\subsection{TESS input catalog}\label{Sec:TESS}

The TESS Input Catalog (TIC) is a comprehensive catalog of stellar parameters and photometric data for stars observed by the TESS mission \citep[][]{Stassun-2019}. The creation of the TIC involves several steps where an initial catalog is cross-matched and integrated to create a unified catalog with consistent stellar parameters and photometry. Rigorous quality control measures are applied to ensure the reliability and accuracy of the data in the TIC. This includes filtering out spurious detection's, eliminating duplicate entries, and flagging uncertain measurements. Figure \ref{fig:tessinputcat} illustrates the comparison between the [Fe/H] values derived from our analysis and the [M/H] values provided by the TIC. While [Fe/H] specifically measures the abundance of iron relative to hydrogen, [M/H] encompasses all elements heavier than helium in a star's atmosphere. Therefore, comparing the two measures requires careful consideration. The fact that [$\alpha$/M] increases for metal-poor stars is expected, as it reflects the higher relative abundances of $\alpha$ elements (like O, Ne, Mg, Si, S, Ar, Ca, and Ti) in older, metal-poor stars due to their formation in core-collapse supernovae. A complete relation between [Fe/H] and [M/H] requires knowledge of both [$\alpha$/M] and [$\alpha$/Fe], which is not available in our current analysis. \\

In Figure \ref{fig:tessinputcat}, we observe a systematic discrepancy between [M/H] from the TIC catalog and [Fe/H] derived from the machine learning model based on S-PLUS colors, particularly when using the APOGEE and LAMOST datasets. This discrepancy can be attributed to several factors. APOGEE, which operates in the near-infrared, typically measures higher metallicity values compared to optical surveys like GALAH, especially at lower $T_{\mathrm{eff}}$ values \citep[][]{Hegedus:2023}. Moreover, APOGEE tends to report lower $\alpha$-element abundances [$\alpha$/Fe] than optical measurements, as noted by \citep[][]{Jonsson-2020}. Similar issues also affect LAMOST, which has lower resolution compared to GALAH. The reduced sensitivity of machine learning models using APOGEE and LAMOST data, relative to GALAH, can be explained by the role of $\alpha$ elements in shaping the spectral energy distribution. $\alpha$ elements are more prominent in optical spectra and less sensitive in the near-infrared. Studies have shown that optical surveys like GALAH, with higher resolution, provide more accurate measurements of $\alpha$-element abundances, especially for metal-poor stars and halo populations \citep[][]{Buder-2018, Ness-2015}. The S-PLUS colors, based on optical narrowband filters, are more adept at capturing these variations, leading to better agreement when trained on optical data from GALAH. In contrast, the near-infrared range of APOGEE and the low resolution of LAMOST  not capture the subtle differences in $\alpha$-element abundances as effectively, resulting in discrepancies when used for machine learning predictions.

The relation between [M/H] and [Fe/H] can vary depending on the correlation with stellar parameters and chemical abundances \citep[e.g.,][]{Salaris-1993}. In particular, \citep[][]{Stassun-2019} proposed a straightforward relation between [M/H] and [Fe/H], given by

\begin{equation}
    [\mathrm{M}/\mathrm{H}] = a \times[\mathrm{Fe}/\mathrm{H}] + b \times \left( 1 \pm \left( 1 - \exp\left( -3.6 \left| [\mathrm{Fe}/\mathrm{H}] + 0.55 \right| \right) \right) \right)
    \label{EQ:mh_feh}
\end{equation}

\begin{table}[h!]
\caption{Values of the fitted coefficients from Equation \ref{EQ:mh_feh}}.
\resizebox{\columnwidth}{!}{%
\begin{tabular}{@{}ccc|cc@{}}
\toprule \midrule
	               & \multicolumn{2}{c}{{[}Fe/H{] $< - $ 0.55}} & \multicolumn{2}{|c}{{[}Fe/H{] $\geq - $ 0.55}} \\ \midrule
               & $a\pm\sigma a$ & $b\pm\sigma b$  & $a\pm\sigma a$ & $b\pm\sigma b$ \\ \midrule
Giant &                         &                         &                           &                          \\
APOGEE         & 1.29 $\pm$ 0.04         & 0.20 $\pm$ 0.02         & 1.14 $\pm$ 0.05           & 0.17 $\pm$ 0.03          \\
GALAH          & 0.98 $\pm$ 0.07         & -0.04 $\pm$ 0.04        & 1.10 $\pm$ 0.05           & 0.01 $\pm$ 0.03          \\
               &                         &                         &                           &                          \\
Dwarf &                         &                         &                           &                          \\
APOGEE         & 2.53 $\pm$ 0.05         & 0.75 $\pm$ 0.03         & 1.11 $\pm$ 0.01           & 0.01 $\pm$ 0.01          \\
GALAH          & 1.21 $\pm$ 0.02         & 0.18 $\pm$ 0.01         & 1.04 $\pm$ 0.01           & 0.12 $\pm$ 0.01          \\
LAMOST         & 2.96 $\pm$ 0.14         & 0.88 $\pm$ 0.07         & 1.33 $\pm$ 0.01           & 0.15 $\pm$ 0.01          \\ \midrule \bottomrule
\end{tabular}}
\label{Table:CoefFit}
\end{table}

\noindent where the plus sign applies for [Fe/H]$ < -0.55$ and the minus sign otherwise. The parameters a and b were estimated to be $1$ and $0.11$, respectively, based on the dataset analyzed by the authors. The change in the [M/H] to [Fe/H] relation around [Fe/H] $< -0.55$ results from differences in element synthesis and enrichment in stellar populations \citep{McWilliam-1997, Nissen2015}. Stars with [Fe/H] $< -0.55$ are typically older and formed in a less enriched interstellar medium, leading to a higher relative abundance of $\alpha$-elements (like oxygen, magnesium, and silicon) compared to iron \citep{Thomas2005, Bensby2014}. The value $-0.55$ dex marks a transition where $[\alpha/Fe]$ increases due to the dominance of Type II supernovae (which produce $\alpha$-elements) in early galaxy enrichment, while Type Ia supernovae (which produce iron) contribute later \citep{Tinsley1979, Woosley1995}. Consequently, below [Fe/H]$ \approx -0.55$, the $[M/H]$ relation becomes non-linear, reflecting the enhanced contribution of $\alpha$-elements relative to iron \citep{Gratton2000, Schuster2004}. 

The key points observed in  Figure \ref{fig:tessinputcat} are as follows. For giant and dwarf slightly metal-rich stars (SMRS - [Fe/H] $> 0$), the value of [M/H] typically exceeds [Fe/H]. This is because [M/H] incorporates additional contributions from $\alpha$-elements and other metals, resulting in a higher overall metallicity. Therefore, for stars in the SMRS, [M/H] is expected to be greater than [Fe/H], reflecting the increased metallicity of these stars \citep[e.g.,][]{Casagrande-2011,Mikolaitis-2017}. \\
       
For stars with near-solar metallicities (\([Fe/H] = 0.0 \pm 0.5\)), the contributions from $\alpha$ elements are relatively minor compared to iron. In these stars, iron lines dominate the spectra, leading to \([Fe/H]\) values that closely reflect the iron abundance. As a result, \([M/H]\) and \([Fe/H]\) are approximately equal, since the overall metallicity measured by \([M/H]\) is heavily influenced by iron, and the additional contributions from $\alpha$ elements do not significantly alter this balance \citep[e.g.,][]{Beers-2005,Nomoto-2013}. \\
    
The distribution of [Fe/H] as a function of [M/H] for dwarfs and giants, determined using GALAH data as a training set, are nearly equal. They also follow the relation [Fe/H] $\simeq$ [M/H], which suggests that the determination of metallicity for MPS could be biased by the presence of $\alpha$ elements. This indicates that further analysis is needed to accurately disentangle the contributions of various elements to the overall metallicity. \\

As discussed above, APOGEE, which operates in the near-infrared, generally reports higher metallicity values compared to optical surveys like GALAH. On the other hand, LAMOST, with its lower spectral resolution, faces similar challenges when compared to higher-resolution optical data like GALAH. The systematic difference observed in both surveys in comparisson with those found for GALAH data can be related to the underestimation of $\alpha$-element abundances that are more important for stars having smaller metallicity values. \\

The model provided by Eq. \ref{EQ:mh_feh} offers a way to improve the agreement between [Fe/H] and [M/H] values. The inconsistency found for the LAMOST and APOGEE training sets, where a notable break occurs at metallicities around $\sim-0.5$, suggests that the [Fe/H] and [M/H] relationship is influenced by the different spectral ranges and resolutions of these surveys with their correlations with the S-PLUS colors. Since the model lacks a physical basis when [M/H] is lower than [Fe/H], the [Fe/H] values from APOGEE and LAMOST, particularly for [Fe/H]$ < -0.5$, should be interpreted with caution. An additional flag (FlagTIC) was introduced to identify stars that do not align with the analysis based on the TIC dataset, with a value of 1 assigned to such cases and 0 otherwise. \\

\begin{figure*}
\centering

\includegraphics[width=0.33\textwidth]{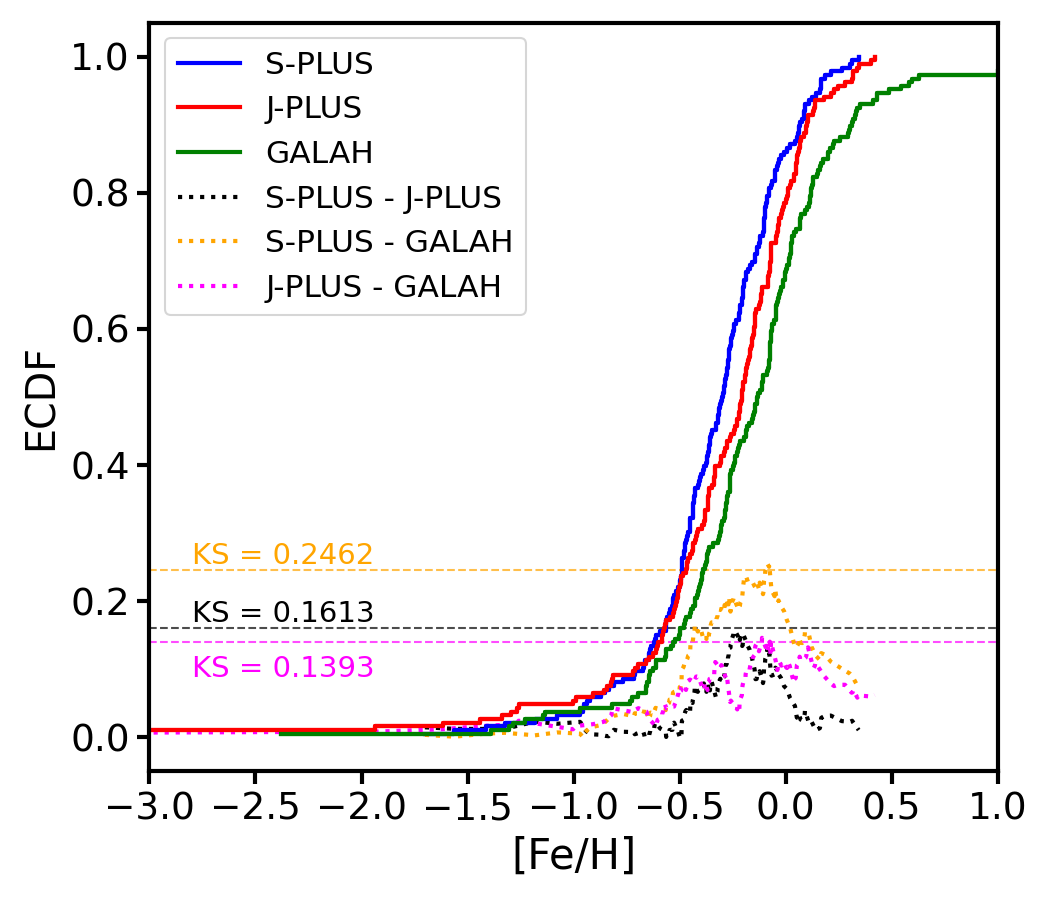} 
\includegraphics[width=0.33\textwidth]{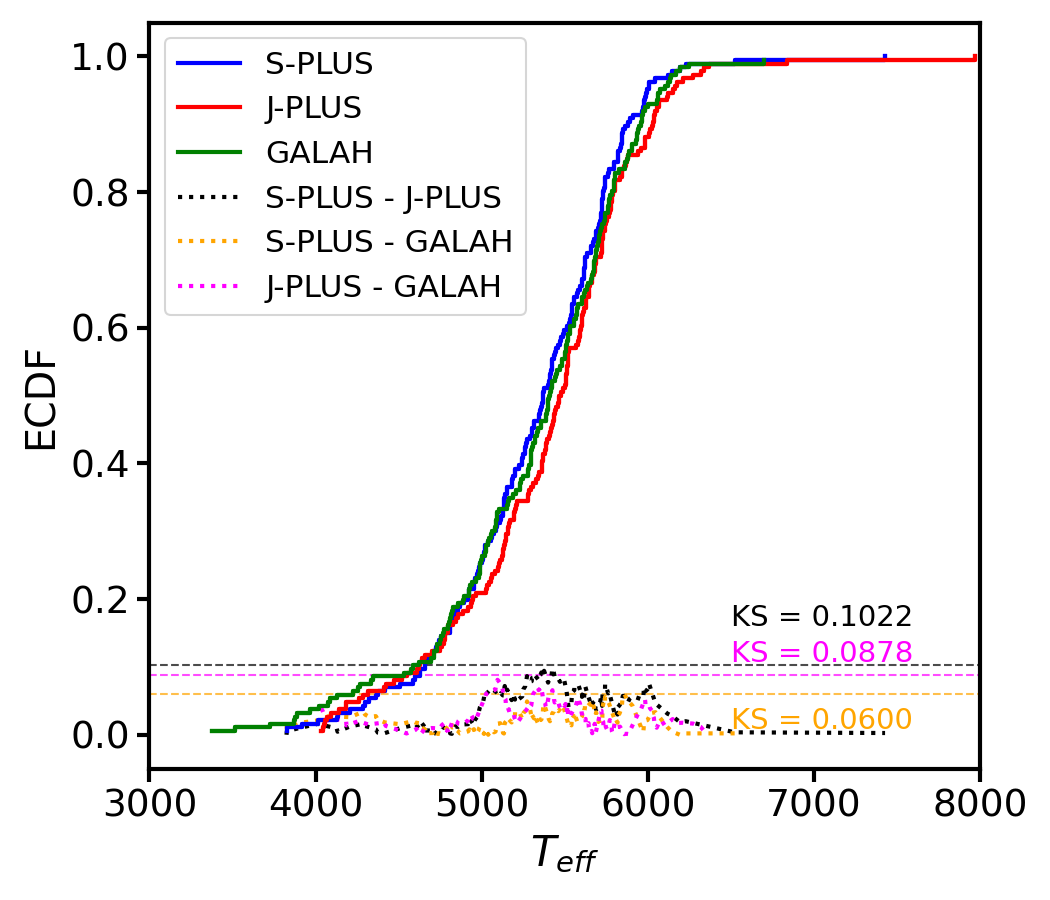} 
\includegraphics[width=0.33\textwidth]{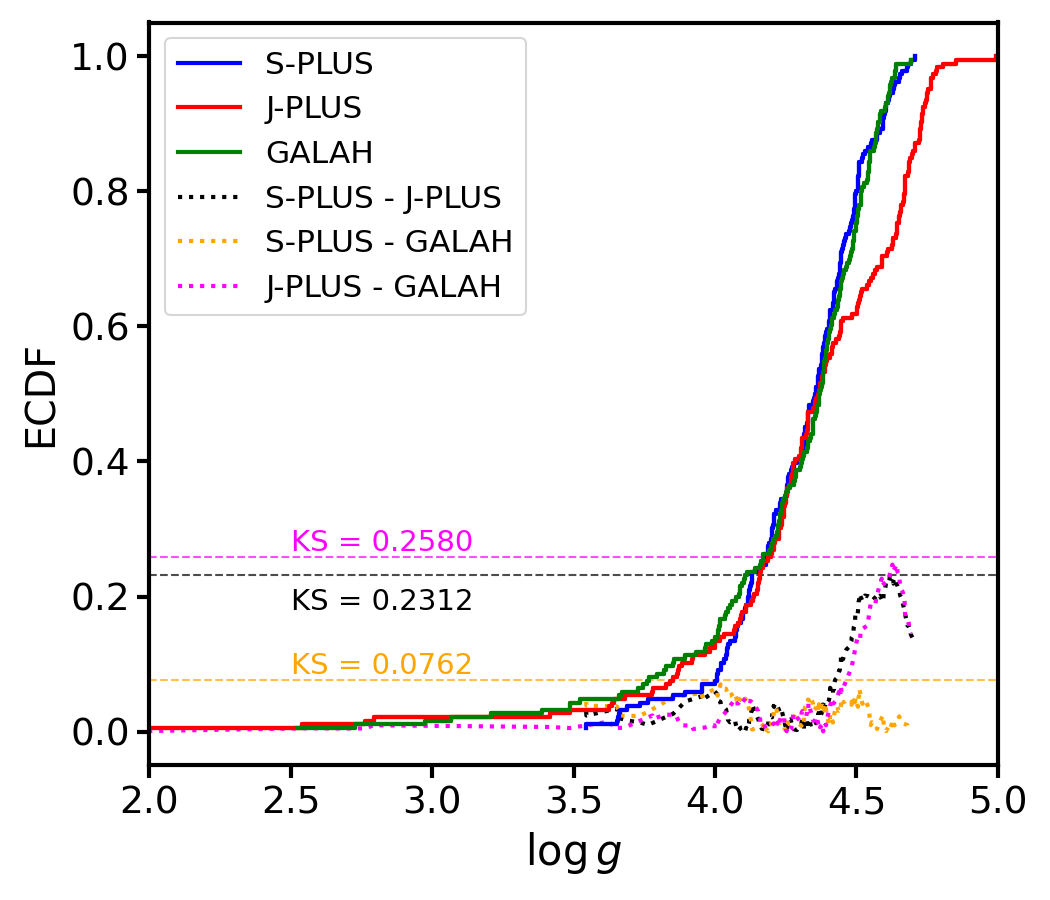} 

\caption{Empirical Cumulative Distribution Functions (ECDFs) comparing predicted [Fe/H], $T_\mathrm{eff}$, and log g  values from the S-PLUS (this work) with the predicted values of J-PLUS, and spectroscopic values from GALAH for 186 dwarf stars. The dotted lines represent the absolute differences between the ECDFs of the survey pairs, and the dashed horizontal lines indicate the Kolmogorov-Smirnov (KS) statistic for each pair of surveys, with lower KS values meaning closer agreement between the respective surveys.}
\label{fig:ecdf_3survey}
\end{figure*}

\begin{figure*}
\centering
\includegraphics[width=0.33\textwidth]{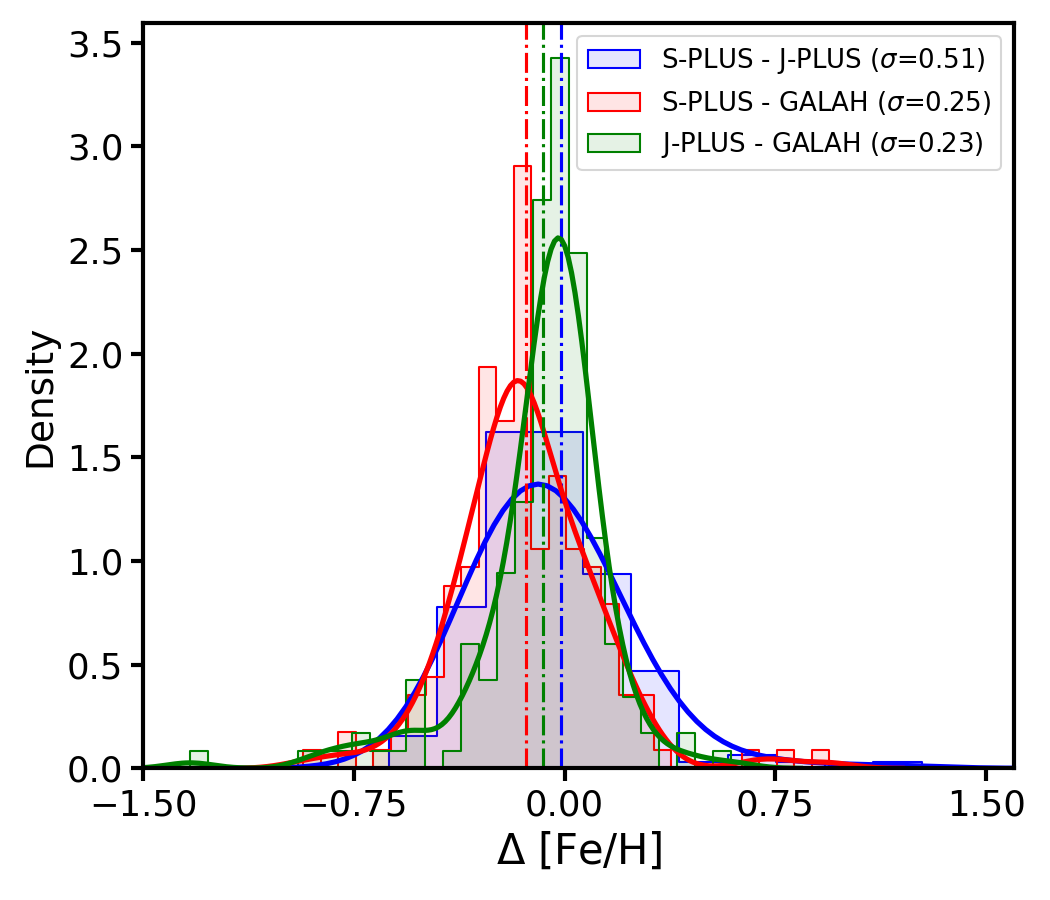} 
\includegraphics[width=0.33\textwidth]{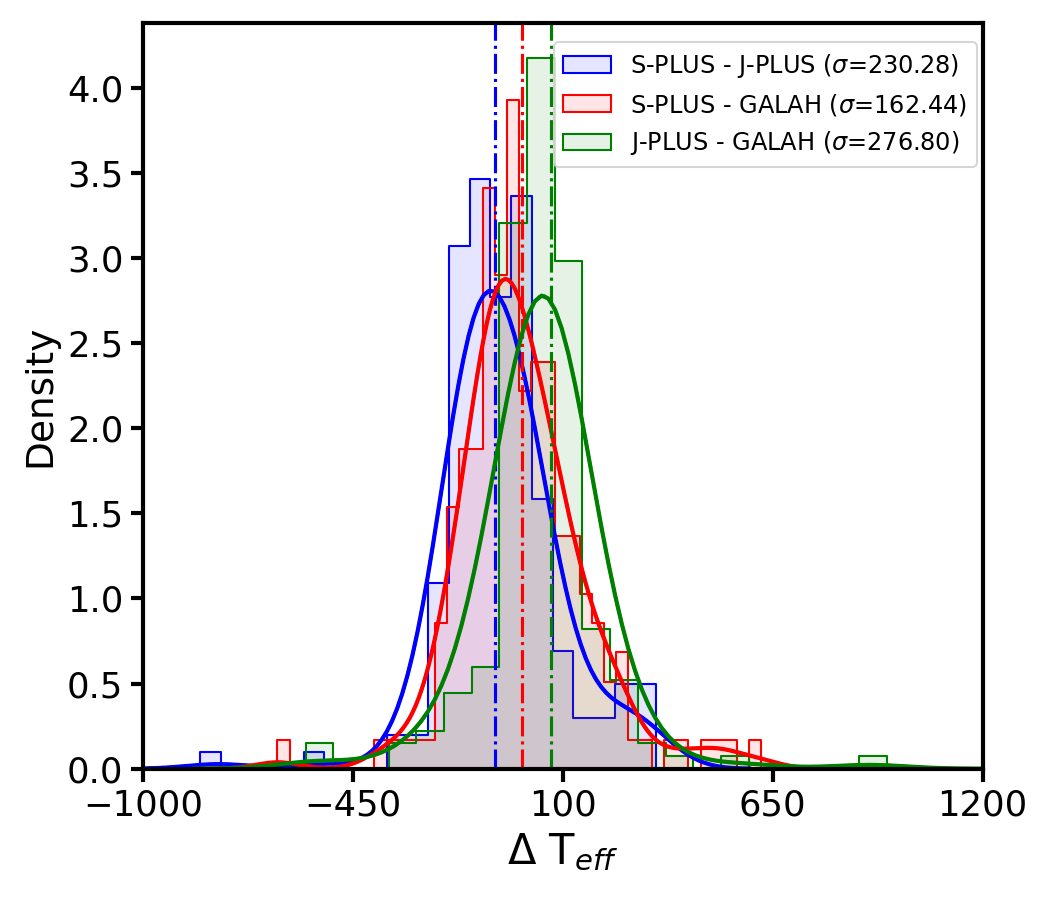} 
\includegraphics[width=0.33\textwidth]{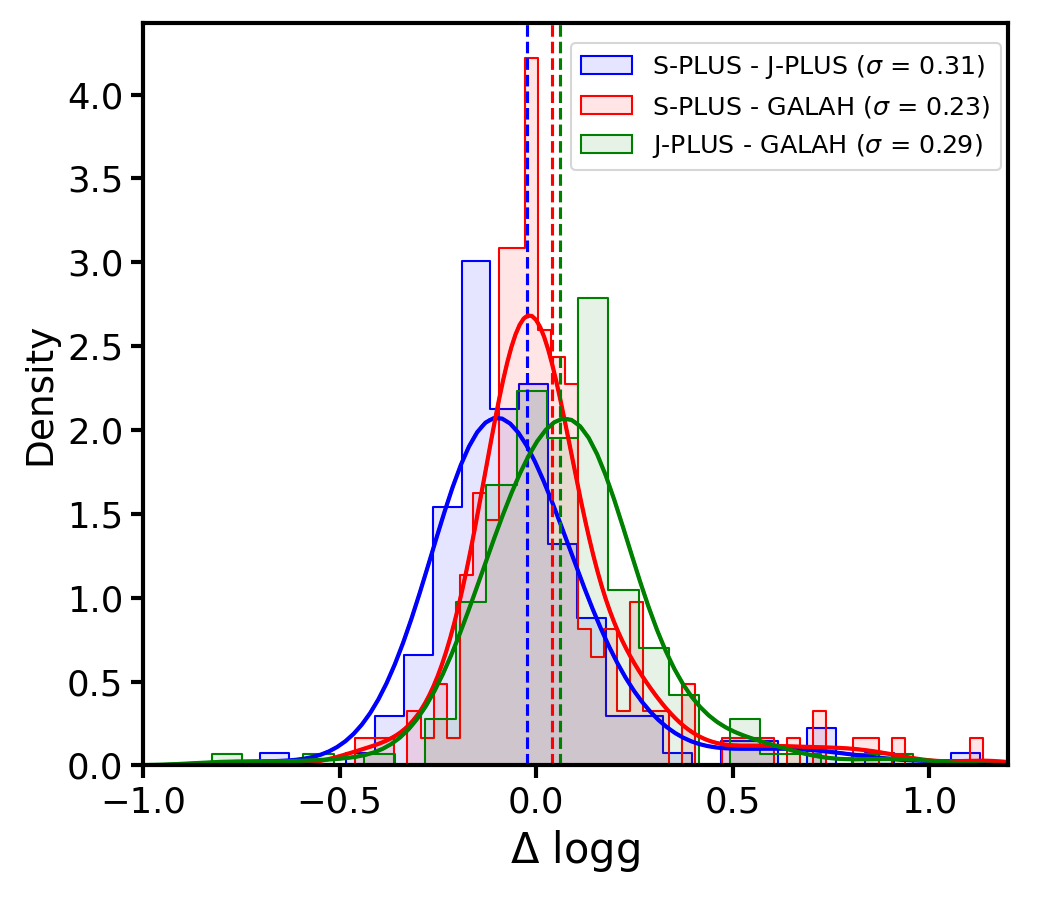} 

\caption{Comparison of the pairwise differences between the predicted values from the S-PLUS (this work), J-PLUS and spectroscopic values from GALAH for [Fe/H], $T_\mathrm{eff}$, and log g, for 186 dwarf stars. The dashed-dotted lines represent mean value of each distribution. The 1$\sigma$ value of the difference of each pair of surveys are displayed in the legend.}
\label{fig:hist_3survey}
\end{figure*}

For the GALAH training sets, the coefficients in both parts of Eq. \ref{EQ:mh_feh} are nearly equal, indicating a consistent correlation between [Fe/H] and [M/H] for both giant and dwarf stars (see Table \ref{Table:CoefFit}). This suggests that high-resolution optical spectroscopic surveys are better suited for developing models based on photometric colors for the elements analyzed in the current study.

A note of caution, TIC is compiled from various catalogs, some of which may include sources that overlap with our training set. However, the total number of entries in the TIC is at least three times larger than that of our training set. Additionally, [M/H] is not directly measured from [Fe/H]. The variation in coefficients governing the relationship between [Fe/H] and [M/H] for dwarfs and giants across different surveys underscores the complexity of determining stellar metallicities. While some results align with theoretical expectations, caution is warranted when interpreting the parameters provided. Notably, APOGEE and GALAH for giant stars, and GALAH for dwarf stars, exhibit good agreement between [M/H] and [Fe/H]. However, disparities arise between APOGEE and LAMOST for dwarf stars. A potential explanation for these differences lies in the varying spectral resolutions among surveys, with GALAH having a resolution up to seven times higher than APOGEE and LAMOST, as well as differences in the spectral range in which these surveys operate.

\begin{table*}[]
\centering
\caption{Standard deviation values for the comparison of cross-matched S-PLUS, J-PLUS, and GALAH samples for giant and dwarf stars.}
\resizebox{\textwidth}{!}{%
\label{tab:sigma_values}
\begin{tabular}{l|c|c|c|c|c|c}
\toprule
\midrule
     & \multicolumn{3}{c|}{Giant (N=8)} & \multicolumn{3}{c}{Dwarf (N=186)} \\ 
     \midrule
     & S-PLUS - J-PLUS & S-PLUS - GALAH & J-PLUS - GALAH & S-PLUS - J-PLUS & S-PLUS - GALAH & J-PLUS - GALAH \\ \midrule
$\sigma_{\Delta \mathrm{[Fe/H]}}$     & 0.09   & 0.07  & 0.09   & 0.51   & 0.25  & 0.23  \\
$\sigma_{\Delta T_\mathrm{eff}}$ & 47     & 75    & 91     & 230    & 162   & 277   \\
$\sigma_{\Delta \log g}$      & 0.19   & 0.12  & 0.20   & 0.31   & 0.23  & 0.29  \\
\midrule
\bottomrule
\end{tabular}%
}
\label{Table:SigmaStarClusters}
\end{table*}

\subsection{Comparision with J-PLUS and GALAH}
\label{Sec:JPLUSPLUS}

We crossmatched our predicted S-PLUS catalog with the predicted catalog of J-PLUS and the spectroscopic catalog of GALAH DR3; and found a total of 186 dwarfs and 8 giants common in all three catalogs. To assess the similarity and differences between our predicted values along with the predicted values of J-PLUS and GALAH, in terms of [Fe/H], $T_\mathrm{eff}$, and \(\log g\) of the 186 dwarf stars, we performed the Kolmogorov-Smirnov test \citep[KS: ][]{Kolmogorov-1933,Massey-1951}. From this test, we compare the empirical cumulative distribution functions (ECDFs) of the datasets, providing a statistic that quantifies the maximum difference between them. A lower KS statistic indicates greater similarity between the distributions. \\

Fig. \ref{fig:ecdf_3survey} illustrates the ECDFs and KS statistics of the 186 dwarfs stars in three surveys. For [Fe/H], the KS statistic between S-PLUS and J-PLUS is 0.1613, between S-PLUS and GALAH is 0.2462, and between J-PLUS and GALAH is 0.1393, yielding the similarity between the three datasets. For the $T_\mathrm{eff}$, the KS statistics are notably lower, with S-PLUS vs J-PLUS yielding 0.1022, S-PLUS vs GALAH yielding 0.0600, and J-PLUS vs GALAH yielding 0.0878. These small values indicate a higher degree of similarity in the temperature distributions across the three surveys. For \(\log g\) the KS statistics show larger variability, with S-PLUS vs J-PLUS at 0.2312, S-PLUS vs GALAH at 0.0762, and J-PLUS vs GALAH at 0.2580. This indicates that the \(\log g\) values differ more significantly between J-PLUS and the other two surveys, particularly in comparison with GALAH, where S-PLUS vs GALAH shows much closer agreement.\\

It is important to note that the crossmatching for giant stars yielded only 8 common stars, which was insufficient for a meaningful KS test comparison, and thus the giant sample was not included in the KS analysis. Fig. \ref{fig:hist_3survey} displays the distribution of pairwise difference of the predicted values among three surveys, for 186 dwarf stars. For $T_\mathrm{eff}$ and \(\log g\) values of dwarf stars, S-PLUS surplases JPLUS, while comparing with GALAH, with 1$\sigma$ values of 162 and 0.23 for $T_\mathrm{eff}$ and \(\log g\), respectively. For [Fe/H] values of dwarf stars, the J-PLUS-GALAH has lower standard deviation, i.e. 0.23. Table \ref{Table:SigmaStarClusters} presents the standard deviation values of [Fe/H], $T_\mathrm{eff}$, and $\log g$ for comparisons between S-PLUS, J-PLUS, and GALAH, based on a sample of 8 giant stars and 186 dwarf stars. The sigma values for giant stars are typically smaller than those for dwarfs; however, due to lower statistical significance, drawing a conclusion on this topic is challenging.

\section{Conclusions}\label{Sec:conclusion}

We estimated the stellar abundances  $[$Al$/\text{Fe}]$, $[$C$/\text{Fe}]$, $[$Li$/\text{Fe}]$, $[$Mg$/\text{Fe}]$, $[$O$/\text{Fe}]$, $[$Si$/\text{Fe}]$, and $[$Cu$/\text{Fe}]$, and fundamental stellar parameters ([Fe/H], \(\log g\), $T_\mathrm{eff}$, $\alpha$/M) for approximately $\sim5$ million sources selected from the 12-band multi-filter photometry provided by the S-PLUS survey. To achieve this, we utilized three training sets (APOGEE, GALAH, and LAMOST) employing RF and NN methods. The NN including $T_\mathrm{eff}$ and \(\log g\) provides a better result than the other approaches tested.
We included only those parameters that achieved a goodness-of-fit greater than 50\% across all approaches (see Appendix \ref{appendixB}) to enhance the reliability of the estimated parameters. Additionally, we introduced a flag to inform users which stars have all input parameters within the training set limits (see Table \ref{table:IDR4Prediction}). This combination of information serves as a strong indicator of the reliability of the computed abundances.
Our findings highlight the importance of narrowband filters, such as $J$0378, $J$0395, $J$0410, and $J$0430, along with the u-band, for the determination of photometric abundances. Importantly, the frameworks developed in this study serve as a cost-effective alternative to spectroscopy and are designed for application in the S-PLUS Internal Data Release 5 (IDR5) as well, which is already available for internal use of the collaboration, as well as in the S-PLUS Ultra-Short Survey \citep[USS-][]{Perottoni:2024}. IDR5 encompasses approximately twice the data volume of DR4, providing expanded opportunities for research and analysis. Meanwhile, the USS covers the observation of bright stars across approximately 9300 square degrees of the Southern sky. We expect that the present work, which provides a means to determine detailed abundance ratios and physical parameters for both dwarf and giant stars from S-PLUS multi-band photometry, will enable a variety of astrophysical applications in the near future. \\

\section{Data Availability} The catalog is available in electronic form at the CDS via anonymous ftp to \url{cdsarc.u-strasbg.fr} (200.54.220.98). 

\begin{acknowledgements}
The authors thank Dr. Jessica Schonhut-Stasik for her insightful comments and contributions, which have greatly enhanced the quality and clarity of this paper.
C.E.F.L and this project is supported by ANID’s Millennium Science Initiative through grant ICN12\_12009, awarded to the Millennium Institute of Astrophysics (MAS); by ANID/FONDECYT Regular grant 1231637; by DIUDA 88231R11.
Support for N.M is provided by doctorado becas Chile/2021 - 21211323. The work of V.M.P. is supported by NOIRLab, which is managed by the Association of Universities for Research in Astronomy (AURA) under a cooperative agreement with the U.S. National Science Foundation. G.L. acknowledges FAPESP proc. 2021/10429-0. 
Support for M.C. is provided by ANID's FONDECYT Regular grant \#1171273; ANID's Millennium Science Initiative through grants ICN12\textunderscore 009 and AIM23-0001, awarded to the Millennium Institute of Astrophysics (MAS); and ANID's Basal project FB210003. D.H. acknowledges the support provided by ANID through doctoral fellowship grant 21232262 for pursuing Ph.D. and ANID/FONDECYT Regular grant 1231637. LAGS and AVSC acknowledges financial support from Consejo Nacional de Investigaciones Cient\'ificas y T\'ecnicas (CONICET), Agencia I+D+i (PICT 2019-03299) and Universidad Nacional de La Plata (Argentina). J.A.-G., acknowledges support from Fondecyt Regular 1201490 and by ANID – Millennium Science Initiative Program – ICN12\_009 awarded to the Millennium Institute of Astrophysics MAS. V.O. acknowledges Coordenação de Aperfeiçoamento de Pessoal de Nível Superior - Brasil (CAPES) - Finance Code 001.  M.B.F. acknowledges financial support from the National Council for
Scientific and Technological Development (CNPq) Brazil (grant number: 307711/2022-6). AAC acknowledges financial support from the Severo Ochoa grant CEX2021-001131-S funded by MCIN/AEI/10.13039/501100011033.  MEDR acknowledges support from {\it Agencia Nacional de Promoci\'on de la Investigaci\'on, el Desarrollo Tecnol\'ogico y la Innovaci\'on} (Agencia I+D+i, PICT-2021-GRF-TI-00290, Argentina). The work of E.M.-P. is supported by CAPES (proc. 88887.605761/2021-00 and 88881.846754/2023-01) and by NOIRLab, which is managed by the Association of Universities for Research in Astronomy (AURA) under a cooperative agreement with the U.S. National Science Foundation. S.D. acknowledges CNPq/MCTI for grant 306859/2022-0. PKH gratefully acknowledges the Fundação de Amparo à Pesquisa do Estado de São Paulo (FAPESP) for the support grant 2023/14272-4. F.A.-F. acknowledges funding for this work from FAPESP (procs 2018/20977-2). The work of E.M.-P. is supported by CAPES (proc. 88887.605761/2021-00 and 88881.846754/2023-01) and by NOIRLab, which is managed by the Association of Universities for Research in Astronomy (AURA) under a cooperative agreement with the U.S. National Science Foundation. D.R.G. acknowledges grants from FAPERJ (E-26/211.527/2023) and CNPq (315307/2023-4).

\end{acknowledgements}

\bibliographystyle{aa}
\bibliography{MyLIB-AA.bib}

\appendix
\section{Training sets (\href{https://doi.org/10.5281/zenodo.14226800}{see link})}\label{appendixA}
This section presents the plot containing the distribution of parameters identified in the GALAH and APOGEE datasets, the same as presented in Fig. \ref{Fig:SPLUS-TrainsetAPOGEE}

\section{Goodness-of-fit values (\href{https://doi.org/10.5281/zenodo.14226800}{see link})}\label{appendixB}
Goodness-of-fit values for the three methodologies (ABC, see Section \ref{Sec:Methodology}) employed in this study.

\section{Importance Plots (\href{https://doi.org/10.5281/zenodo.14226800}{see link})}\label{appendixC}
This section presents the importance feature plots for some of the parameters analyzed in this work. The remaining plots can be requested from the authors.

\section{Plots of RF as function of NN (\href{https://doi.org/10.5281/zenodo.14226800}{see link})}\label{appendixD}
This section presents the plots illustrating RF as a function of NN estimations. These visualizations help to understand the relationship and dependencies between RF and NN in our analysis.

\section{Test set plots (\href{https://doi.org/10.5281/zenodo.14226800}{see link})}\label{appendixF}
This section features plots illustrating the testing set results for stellar parameters and elemental abundances using the Neural Network (NN) approach C.

\section{Catalog description (\href{https://doi.org/10.5281/zenodo.14226800}{see link})}\label{appendixG}
Description of the released table containing the estimated parameters obtained in this study. 

\section{Comparison of [Fe/H] Values for Star Clusters (\href{https://doi.org/10.5281/zenodo.14226800}{see link})}\label{appendixH}
The section contains the table summarizing the comparison between [Fe/H] values from various star clusters in our predicted sample and corresponding values from the literature.

\end{document}